\documentclass[11pt]{article}
\usepackage{jheppub}
\usepackage{bm,bbm}
\usepackage{booktabs,amsmath}
\usepackage{accents}
\usepackage{multirow}
\usepackage{graphics}
\usepackage{braket}
\usepackage{mathtools}
\usepackage{comment}
\usepackage{autobreak}
\usepackage{subcaption}
\usepackage{enumerate}
\usepackage{blkarray}
\usepackage{longtable}
\usepackage{enumitem}

\usepackage{tikz}
\usetikzlibrary{patterns}
\usetikzlibrary{arrows,shapes,positioning}
\usetikzlibrary{decorations.markings}
\usetikzlibrary{decorations.pathmorphing}
\tikzset{snake it/.style={decorate, decoration=snake}}

\hypersetup{
    colorlinks=true,
    linkcolor=blue,
    filecolor=magenta,      
    urlcolor=cyan,
    pdfpagemode=FullScreen,
}

\usepackage{arydshln}
\usepackage{mathtools}

\edef\restoreparindent{\parindent=\the\parindent\relax}
\usepackage{parskip}
\restoreparindent

\usepackage{ascmac}
\usepackage{mathrsfs}
\usepackage{tabularray}
\UseTblrLibrary{booktabs}

\renewcommand{\arraystretch}{1.25}

\usepackage{amsthm}
\newtheoremstyle{break}
  {\topsep}{\topsep}%
  {\upshape}{}%
  {\bfseries}{}%
  {\newline}{}%
\theoremstyle{break}

\def\d{{\rm d}}
\def\i{{\rm i}}
\def\CC{{\cal C}}
\def\CD{{\cal D}}

\def\CN{{\cal N}}

\def\BQ{\mathbb{Q}}
\def\BR{\mathbb{R}}

\def\BZ{\mathbb{Z}}

\def\d{\mathrm{d}}
\def\O{\mathrm{O}}

\def\SU{\mathrm{SU}}

\def\Spin{\mathrm{Spin}}

\def\U{\mathrm{U}}

\title{
Fermionic CFTs from topological boundaries in abelian Chern-Simons theories
}

\author[a]{Kohki Kawabata,}
\author[b]{Tatsuma Nishioka,}
\author[c]{Takuya Okuda}
\author[a]{and Shinichiro Yahagi}

\affiliation[a]{Department of Physics, Faculty of Science,
The University of Tokyo,\\
Bunkyo-Ku, Tokyo 113-0033, Japan}

\affiliation[b]{Department of Physics, The University of Osaka,\\
Machikaneyama-Cho 1-1, Toyonaka 560-0043, Japan}

\affiliation[c]{Graduate School of Arts and Sciences, The University of Tokyo, Komaba,\\
Meguro-ku, Tokyo 153-8902, Japan}

\preprint{OU-HET-1263, UT-Komaba/25-1}

\abstract{A quantum field theory is referred to as bosonic (non-spin) if its physical quantities are independent of the spacetime spin structure, and as fermionic (spin) if they depend on it. We explore fermionic conformal field theories (CFTs) that emerge from bosonic abelian Chern-Simons theories, playing the role of a symmetry topological field theory, by imposing topological boundary conditions. Our construction includes the fermionic generalization of code CFTs. When the Chern-Simons theory is associated with the root lattice of a simply laced Lie algebra, this approach yields a fermionic CFT with a level-one affine Lie algebra symmetry. As an application, we consider the Chern-Simons theories corresponding to a class of supersymmetric vertex operator algebras studied by Johnson-Freyd and classify their fermionic topological boundary conditions that give rise to supersymmetric CFTs.
}

\begin{document} 
\maketitle
\flushbottom

\section{Introduction}

Quantum field theory (QFT) is a basic framework for describing the dynamics of particles and fields, and plays a central role in both high-energy and condensed matter theories.  A key distinction in QFTs is whether they are bosonic or fermionic, which refers to the behavior of their physical quantities in relation to the spin structure of spacetime. Specifically, a QFT is called bosonic (non-spin) if its observables are independent of the spacetime spin structure, while it is called fermionic (spin) if these quantities are dependent on the spin structure. This classification has significant implications for the topological properties of such theories.

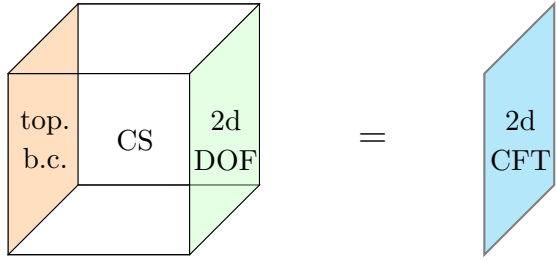
\begin{figure}
\centering
\begin{tikzpicture}[scale=1.2]

    \begin{scope}
            \coordinate (O) at (0,0,0);
            \coordinate (A) at (0,2,0);
            \coordinate (B) at (0,2,2);
            \coordinate (C) at (0,0,2);
            \coordinate (D) at (2,0,0);
            \coordinate (E) at (2,2,0);
            \coordinate (F) at (2,2,2);
            \coordinate (G) at (2,0,2);
            
            \draw (O) -- (C) -- (G) -- (D) -- cycle;% Bottom Face
            \draw (O) -- (A) -- (E) -- (D) -- cycle;% Back Face
            \draw[fill=orange!50, opacity=0.5] (O) -- (A) -- (B) -- (C) -- cycle;% Left Face
            \draw[fill=green!20,opacity=0.5] (D) -- (E) -- (F) -- (G) -- cycle;% Right Face
            \draw (C) -- (B) -- (F) -- (G) -- cycle;% Front Face
            \draw (A) -- (B) -- (F) -- (E) -- cycle;% Top Face

            \node at (1.63, 0.5) {\parbox{1cm}{\centering 2d\\ DOF}};

            \node at (-0.36, 0.5) {\parbox{1cm}{\centering top.\\ b.c.}};

            \node at (0.63, 0.5) {CS};
    \end{scope}

    \begin{scope}[xshift=3.25cm, yshift=0.5cm]
        \node at (0,0) {\Large $=$};
    \end{scope}

    \begin{scope}[xshift=5.25cm]
            \coordinate (O) at (0,0,0);
            \coordinate (A) at (0,2,0);
            \coordinate (B) at (0,2,2);
            \coordinate (C) at (0,0,2);

            \draw[thick, fill=cyan!50, opacity=0.5] (O) -- (A) -- (B) -- (C) -- cycle;

            \node at (-0.37, 0.5) {\parbox{1cm}{\centering 2d\\ CFT}};
    \end{scope}

\end{tikzpicture}

\caption{The bulk Chern-Simons (CS) theory supports 2d degrees of freedom (DOF) on the physical boundary and is subject to a topological boundary condition.
At low energies, a 2d CFT describes the combined 3d$+$2d system.
}
\label{fig:3d}
\end{figure}

In this paper, we investigate two-dimensional conformal field theories (CFTs) derived from three-dimensional bosonic abelian Chern-Simons theories by imposing topological boundary conditions. 
An abelian Chern-Simons theory, defined by the action 
\begin{equation}\label{eq:CS-action-intro}
    S = \frac{\i}{4\pi} \int  K_{IJ} A^I \wedge \d A^J
    \end{equation}
for abelian gauge fields $A^I$ $(I=1,\cdots, n)$ of the gauge group~$\mathrm{U}(1)^n$,
is bosonic when the diagonal entries $K_{II}$ of the non-degenerate symmetric integer matrix $(K_{IJ})$ are all even~\cite{Dijkgraaf:1989pz,Belov:2005ze}.
In our framework, the abelian Chern-Simons theory plays the role of a symmetry topological field theory (SymTFT) in the so-called sandwich construction~\cite{Gaiotto:2020iye, Apruzzi:2021nmk,Freed:2022qnc,Kaidi:2022cpf} (figure~\ref{fig:3d}). 
The three-dimensional spacetime has two boundaries, and we impose a topological boundary condition on one boundary and a physical boundary condition on the other.
While much of the existing literature on similar setups, such as~\cite{Kapustin:2010hk,Kapustin:2010if,Kaidi:2021gbs}, focuses on bosonic CFTs that arise from bosonic topological boundaries, we instead place particular emphasis on fermionic CFTs that emerge from fermionic topological boundaries.%
\footnote{%
For the SymTFT construction of fermionic CFTs via fermionic topological boundaries not based on Chern-Simons theories, see the recent papers~\cite{Fukusumi:2024cnl,Wen:2024udn,Huang:2024ror,Bhardwaj:2024ydc,Chen:2024ulc}.
}

Our framework is motivated by and generalizes the constructions of chiral CFTs through Euclidean lattices~\cite{frenkel1984natural,frenkel1989vertex,Dolan:1994st,Gaiotto:2018ypj,Kawabata:2023nlt,Kawabata:2023rlt,Kawabata:2024gek,Okada:2024imk} and non-chiral CFTs through Lorentzian lattices~\cite{Dymarsky:2020bps,Dymarsky:2020qom,Dymarsky:2020pzc,Dymarsky:2021xfc,Henriksson:2021qkt,Buican:2021uyp,Yahagi:2022idq,Furuta:2022ykh,Henriksson:2022dnu,Angelinos:2022umf,Henriksson:2022dml,Dymarsky:2022kwb,Kawabata:2022jxt,Furuta:2023xwl,Alam:2023qac,Kawabata:2023usr,Kawabata:2023iss,Aharony:2023zit,Barbar:2023ncl,Singh:2023mom,Ando:2024gcf,Singh:2024qjm,Mizoguchi:2024ahp} from classical and quantum codes.
In particular, reference~\cite{Kawabata:2023iss} proposed a SymTFT construction of (bosonic) Narain code CFTs (see also~\cite{Barbar:2023ncl,Dymarsky:2024frx}), which we elaborate on.
A key contribution of this work is the fermionic generalization of non-chiral Narain code CFTs, expanding on these lines of research.

The SymTFT construction naturally leads to our central application.
When the bosonic Chern-Simons theory is associated with the root lattice of a simply laced Lie algebra, our construction yields a fermionic CFT with a level-one affine Lie algebra symmetry. 
The affine symmetries for simply laced Lie algebras, namely  $A_m = su(m+1)$, $D_m=spin(2m)$, and  $E_6, E_7, E_8$, are of particular interest because they can be realized in terms of free bosons~\cite{Frenkel:1980rn,Segal1981UnitaryRO}.  By classifying fermionic topological boundary conditions,
we classify CFT (modular covariant) completions of a class of super vertex operator algebras (SVOAs) studied by Johnson-Freyd~\cite{Johnson-Freyd:2019wgb}, namely the $\mathcal{N}=1$ supersymmetric SVOAs without free fermions and with the bosonic subalgebra given by a product of level-one affine Lie algebras for simply laced Lie algebras.%
\footnote{%
We emphasize the difference between a vertex operator algebra and a full CFT.
The former specifies the characters while the latter specifies the modular invariant (in the bosonic case) or covariant (in the fermionic case, for a given spin structure) combination of the characters as the partition function.
In the terminology of~\cite{Freed:2012bs}, the former defines a relative theory while the latter defines an absolute theory.
}

This paper is organized as follows.
Section~\ref{sec:general} is mostly a review.
In section~\ref{sec:canonical-quantization}, we relate the physical (dynamical) boundary in the SymTFT construction to the canonical quantization of abelian Chern-Simons theories based on holomorphic/K\"ahler quantization approach.
We then review bosonic and fermionic topological boundary conditions in section~\ref{sec:topological-boundary}.
In section~\ref{sec:torus} we show that the dimensional reduction of an
abelian Chern-Simons theory on an interval with physical and topological boundary conditions reduces to a bosonic or fermionic two-dimensional CFT.
Section~\ref{sec:orbifolds} discusses orbifolds within our framework.
In particular, we explain the appearance of discrete torsion in bosonic orbifolds within our framework in section~\ref{sec:orb-bos}.
Section~\ref{sec:examples} illustrates our construction by interesting bosonic examples.
In~section~\ref{sec:fermionic-code-CFTs}, we discuss the fermionic generalization of non-chiral Narain code CFTs.
In~section~\ref{sec:level-one-affine}, we construct CFTs with level-one affine Lie algebra symmetries from topological boundary conditions in the associated Chern-Simons theories.
We conclude with discussion in section~\ref{sec:discussion}.
Appendices collect some technical details relevant to section~\ref{sec:level-one-affine}.

\section{Bosonic and fermionic CFTs from bosonic abelian Chern-Simons theories}
\label{sec:general}

In this section, we study how the SymTFT/sandwich construction works in the case of abelian Chern-Simons theories.
As already mentioned in the Introduction, the construction involves two kinds of boundary conditions, physical and topological boundary conditions (figure~\ref{fig:3d}).
One advantage of abelian Chern-Simons theories is that the physical boundary condition can be explicitly worked out by relating it to canonical quantization, as we review in section~\ref{sec:canonical-quantization}.
In section~\ref{sec:topological-boundary}, we will see that topological boundary conditions also admit straightforward descriptions in terms of the so-called Lagrangian subgroups.
Combining the two boundary conditions, we obtain the partition functions of two-dimensional CFTs on a Riemann surface, which we explain in section~\ref{sec:torus} in the genus-one case.
In section~\ref{sec:orbifolds}, we study how orbifold and discrete torsion arise within our formulation.

\subsection{Physical boundary and canonical quantization}
\label{sec:canonical-quantization}

In this subsection, we relate the physical boundary defined by Dirichlet boundary conditions to the canonical quantization of an abelian Chern-Simons theory~\cite{Bos:1989wa,Polychronakos:1989cd,Belov:2005ze,Porrati:2021sdc,Aharony:2023zit}.
The space of gauge fields on a Riemann surface is equipped with a symplectic structure and a metric,%
\footnote{In the notations introduced below, the symplectic structure is given by $(K_{IJ}/4\pi)\int \delta A^I\wedge \delta A^J = \sum_i (\epsilon_i/4\pi) \int \delta A^i\wedge \delta A^i$ and the metric by $\sum_i (1/2\pi) \int \d^2z\, \delta A^i_{\bar z} \delta A^i_z  $.
} 
which define a K\"ahler structure.  In K\"ahler quantization, a wave function is identified with a holomorphic section of a line bundle with a connection whose curvature is the symplectic structure (up to normalization).  In this framework, we quotient the space of gauge fields by gauge transformations, leading to a finite-dimensional Hilbert space.  Wave functions (conformal blocks of a CFT~\cite{Witten:1988hf}) are identified with the functions on the space of one-forms that transform properly under large gauge transformations.
For detailed explanations of this method, see~\cite{Axelrod:1989xt,Nair:2005iw,Belov:2005ze}.

Let us consider the bosonic abelian Chern-Simons theory defined by the action~(\ref{eq:CS-action-intro}) for the gauge fields $A^I$ ($I=1,\ldots,n$) of the gauge group $\mathrm{U}(1)^n$.
Large gauge transformations act as the shifts
\begin{equation}\label{eq:AI-large-gauge}
A^I \rightarrow A^I +  \omega^I \,,
\end{equation}
where $\omega^I/2\pi$ for each value of $I$ is a closed one-form with integral periods, i.e., we can write $\i\, \omega^I = (g^I)^{-1} \d g^I$ where $g^I$ is a $\U(1)$-valued function.
There exists a real matrix $(M_I{}^j)$ such that
\begin{equation}\label{eq:KMepsilon}
K_{IJ} =  \sum_{i=1}^n M_I{}^i M_J{}^i \epsilon_i \,,
\end{equation}
where we have $\epsilon_a =+1$ for $a=1,\ldots,n_L$ and $\epsilon_{n_L+\bar a}=-1$ for $\bar a =1,\ldots, n_R$ ($n_L+n_R=n$).
The matrix $(K_{IJ})$ is invariant under the replacement 
\begin{equation}\label{eq:O-nL-nR}
M_I{}^i \rightarrow M_I{}^j L_j{}^i
\quad \text{ for } \quad
(L_j{}^i) \in \O(n_L,n_R;\mathbb{R}) \,,     
\end{equation}
giving rise to parameters to be identified with moduli of the resulting CFT later in section~\ref{sec:torus}.
In terms of
\begin{equation}
A^i := A^IM_I{}^i \,,
\end{equation}
we have
\begin{equation}
S = \frac{\i}{4\pi} \int \sum_i\, \epsilon_i \,A^i\wedge \d A^i \,.
\end{equation}

Let us take as the spacetime manifold the product space $M=\Sigma\times I$, where $\Sigma$ is a torus defined by the identifications $z\sim z+1\sim z+\tau$ of the complex coordinate $z=x^1+\i\, x^2$ and $I=\{0\leq x^3\leq 1\}$ is an interval.
In the rest of this subsection, we focus on the boundary component $\Sigma_\text{ph}:=\Sigma\times\{x^3=1\}$, which plays the role of the physical boundary in the sandwich construction.
Following~\cite{Aharony:2023zit} we deform the action by a boundary term:
\begin{equation}
S' = S - \frac{1}{2\pi} \int_{\Sigma_\text{ph}} \d^2 z\, \sum_{i=1}^n  A^i_z\, A^i_{\bar z} \ .
\end{equation}
We will often write $A_\mu^a := A_\mu^{i=a}$ ($a=1,\ldots,n_L$), $A_\mu^{\bar a} := A_\mu^{i=n_L+\bar{a}}$ ($\bar a =1,\ldots,n_R$).%
\footnote{We will use similar notations for other vectors $\nu$, $\alpha$, $\gamma$, $\lambda$, etc. introduced later.}
Since  $\int \d^2z A^a_{\bar z}\, \delta A^a_z$ and $\int \d^2z A^{\bar a}_{ z}\, \delta A^{\bar a}_{\bar z}$ are absent in $\delta S'$ due to cancellations, we can impose Dirichlet boundary conditions on $A^a_{\bar z}$ and $A^{\bar a}_z$.
We parameterize their boundary values by $\nu^a$ and $\bar \nu^{\bar a}$ as
\begin{equation} \label{eq:nu-nubar-bc}
A^a_{\bar z}\big|_{\Sigma_\text{ph}} = \frac{\pi \, \i}{\tau_2}\, \nu^a\,,
\qquad
A^{\bar a}_z\big|_{\Sigma_\text{ph}} = -  \frac{\pi\,  \i}{\tau_2} \,\bar\nu^{\bar a} \,.
\end{equation}

In K\"ahler quantization, wave functions $\Psi[A]:=\Psi(\nu^a,\bar\nu^{\bar a})$ must transform under gauge transformations so that the inner product
\begin{equation}
\int\mathcal{D}A \exp\left(- \frac{1}{\pi} \int_{\Sigma_\text{ph}} \d^2 z\, \sum_{i=1}^n  A^i_z\, A^i_{\bar z}\right)
\Psi_1[A]^*
\Psi_2[A]
\end{equation}
between two arbitrary wave functions $\Psi_i$ ($i=1,2$) is invariant.
The invariance holds if, under the large gauge transformation~(\ref{eq:AI-large-gauge}), the wave function transforms as
\begin{equation}
\label{eq:Psi-A-transform}
\begin{aligned}
\Psi[A] & \rightarrow\Psi[A+\omega]= \Psi[A]\times 
\exp\left[\frac{\i}{4\pi}\int_{\Sigma_\text{ph}} K_{IJ}\, \omega^I\wedge A^J\right] 
\\
&
\qquad
\qquad
\qquad
\qquad
\times
\exp\left[ \frac{1}{2\pi}
\int \d^2z
\left(A^i_z\, \omega^i_{\bar z} + A^i_{\bar z}\,\omega^i_z + \omega^i_z\,\omega^i_{\bar z}\right)\right]
e^{\i\,\varphi(\omega)}
\,.
\end{aligned}
\end{equation}
The phase $\varphi(\omega)$ must satisfy the relation
\begin{equation} \label{eq:varphi-condition}
\varphi(\omega+\tilde\omega)
=\varphi(\omega)+\varphi(\tilde\omega)
-\frac{K_{IJ}}{4\pi} \int_{\Sigma_\text{ph}}
\omega^I\wedge\tilde\omega^{J}
\quad \text{ mod } 2\pi 
\end{equation}
for the equality $\Psi[A+(\omega+\tilde\omega)]= \Psi[(A+\omega)+\tilde\omega]$ to hold.
The transformation~(\ref{eq:Psi-A-transform}) can be written as
\begin{equation}\label{eq:Psi-transform}
\begin{aligned}
&\ \, \quad\Psi(\nu,\bar\nu)
\\
&\rightarrow 
\Psi(\nu+\delta\nu,\bar\nu+\delta\bar\nu)
\\
&
\, = \,
\Psi(\nu,\bar\nu)
\exp\Big[
\frac{\pi}{2\tau_2}
\big(
 2\nu^a (\delta\nu^a)^* + \delta\nu^a (\delta \nu^a)^* 
+2\bar\nu^{\bar a}(\delta\bar\nu^{\bar a})^* + \delta\bar\nu^{\bar a} (\delta\bar\nu^{\bar a})^*
\big)
+\i\, \varphi
\Big] \,, 
\end{aligned}
\end{equation}
where%
\footnote{%
In~(\ref{eq:Psi-transform}), $*$ denotes complex conjugation, which we also indicate by a bar.
}
\begin{equation}
    \delta\nu^a = (n^I+m^I\tau)\, M_I{}^a\,,
    \qquad
    \delta\bar\nu^{\bar a} = (n^I+m^I\bar\tau)\, M_I{}^{\bar a} \,,
\end{equation}
and the parameters $(n^I,m^I)$ for large gauge transformations are defined by
\begin{equation}
\omega_{\bar z}^I = 
\frac{\pi\,\i}{\tau_2}\,(n^I+m^I\tau)
\,,
\qquad
\omega_{z}^I 
= -\frac{\pi\,\i}{\tau_2}\,(n^I+m^I\bar\tau) \,,
\qquad
n^I,m^I\in\mathbb{Z} \,.
\end{equation}

Let us write $q=e^{2\pi \i \tau}$.
We define $\Gamma$ to be the even integral lattice generated by the basis vectors $e_I
$ equipped with the inner product $e_I\oslash e_J = K_{IJ}$.%
\footnote{We thus have $v\oslash w = K_{IJ} v^I w^J = \sum_i \epsilon_i v^i w^i = \sum_a v^a w^a - \sum_{\bar a} v^{\bar a} w^{\bar a}$.}
Let $\Gamma^*$ be the dual of $\Gamma$.
For $\gamma\in\Gamma^*$, we write
$\gamma^2_L:=\gamma^a\gamma^a$ and $\gamma^2_R:=\gamma^{\bar a}\gamma^{\bar a}$.
Let~$\eta(\tau)= q^{1/24}\prod_{j=1}^\infty(1-q^j)$ be the Dedekind eta function.
A direct calculation shows that for $[\alpha]\in\Gamma^*/\Gamma$,%
\footnote{%
Throughout the paper, the notation~$[\alpha]=\alpha+\Gamma\in\Gamma^*/\Gamma$ denotes the equivalence class which is an element of the quotient group $\Gamma^*/\Gamma$ and is represented by $\alpha\in\Gamma^*$.}
the functions
\begin{equation}
\label{eq:wave-function}
    \begin{aligned}
    \Psi_{[\alpha]}(\tau,\bar\tau; \nu,\bar\nu)
        &:= \frac{1}{\eta(\tau)^{n_L} \overline{\eta(\tau)}{}^{n_R}} \sum_{\gamma\in \Gamma} 
        q^{\frac{1}{2} 
        (\gamma+\alpha)^2_L
        }\,
        \bar q^{\frac{1}{2} 
        (\gamma+\alpha)^2_R}
            \\
        &\qquad\qquad\qquad\qquad\times 
        e^{2\pi \i \,\left( (\gamma+\alpha)^a \nu^a -(\gamma+\alpha)^{\bar a} \bar\nu^{\bar a} \right)} \,
        e^{\frac{\pi}{2\tau_2}(\nu^a \nu^a +\bar\nu^{\bar a} \bar\nu^{\bar a})} 
    \end{aligned}
\end{equation}
transform as in~(\ref{eq:Psi-transform}) with the phase
\begin{equation}
    \varphi= \pi\, K_{IJ}\, m^I n^J \,,
\end{equation}
which satisfies the condition~(\ref{eq:varphi-condition}).
Their modular transformations are given by
\begin{equation}\label{eq:Psi-CS-T}
    \Psi_{[\alpha]}(\tau+1,\bar\tau+1;\nu,\bar\nu) 
        = 
        e^{-\frac{\pi \i}{12}(n_L-n_R)} e^{\pi \i\,\alpha\oslash\alpha}\,\Psi_{[\alpha]}(\tau,\bar\tau;\nu ,\bar\nu) \,,
\end{equation}
\begin{equation}\label{eq:Psi-CS-S}
    \Psi_{[\alpha]}\left(-\frac{1}{\tau},-\frac{1}{\bar\tau}\,;
\frac{\nu}{\tau},\frac{\bar\nu}{\bar\tau}
    \right) 
        =
        \frac{1}{\sqrt{|\Gamma^*/\Gamma|}}
        \sum_{[\beta]\in\Gamma^*/\Gamma}
         e^{-2\pi \i\, \alpha\oslash \beta} \,
        \Psi_{[\beta]}(\tau,\bar\tau;\nu ,\bar\nu) \,.
\end{equation}
The functions~$\Psi_{[\alpha]}$ form a basis of the Hilbert space for the Chern-Simons theory quantized on the space~$\Sigma_\text{ph}$.
We emphasize that the dependence of the wave functions on the complex structure~$\tau$ arises from the boundary conditions~(\ref{eq:nu-nubar-bc}).
In the bra-ket notation, we can write
\begin{equation}
    \Psi_{[\alpha]}(\tau,\bar\tau;\nu,\bar\nu) 
        = 
        \langle
   \tau,\bar\tau ; \nu,\bar\nu
        \, \big|
        \Psi_{[\alpha]}
        \rangle \,.
\end{equation}

Let us consider the 1-cycles $\mathsf{A}=\{z=s\,|\, 0\leq s\leq 1\}$ and $\mathsf{B}=\{z=s\, \tau\,|\, 0\leq s \leq 1\}$ on the torus, and 
the Wilson loop operators $W_{[\alpha]}(\mathsf{X})=\exp \left(\mathrm{i}\oint_{\,\mathsf{X}} \alpha_I A^I\right)$ ($\mathsf{X} =\mathsf{A}$, $\mathsf{B}$) along them.
They act on the wave functions as
\begin{equation}\label{eq:Wilson-block-non-spin}
    \begin{aligned}
        W_{[\beta]}(\mathsf{A}) 
        |\Psi_{[\alpha]} \rangle 
            & =  
                e^{2\pi\mathrm{i}\,\alpha\oslash \beta}\, 
                |\Psi_{[\alpha]}\rangle
                \,,
                \\   
        W_{[\beta]}(\mathsf{B}) | \Psi_{[\alpha]} \rangle 
            & = 
            |\Psi_{[\alpha+\beta]} \rangle
            \,,
    \end{aligned}
\end{equation}
as one can see by expressing $\oint A^I$ in terms of $\nu^a, \partial/\partial \nu^a$, $\bar\nu^{\bar a}$, and $\partial/\partial\bar\nu^{\bar a}$.
The relations~(\ref{eq:Wilson-block-non-spin}) admit the natural interpretation~\cite{Witten:1988hf} that $ |\Psi_{[\alpha]} \rangle $ is the state obtained by a Chern-Simons path integral on a solid torus such that the cycle $\mathsf{A}$ on the boundary $\Sigma$ is contractible in the bulk and the Wilson line $W_{[\alpha]}$ is
inserted along cycle $\mathsf{B}$:
\begin{equation} \label{eq:Psi-W}
    |\Psi_{[\alpha]} \rangle 
= 
W_{[\alpha]}(\mathsf{B})\, |\Psi_{[0]} \rangle \,,
\end{equation}
where $ |\Psi_{[0]} \rangle $ is the state without Wilson line insertion.

\subsection{Topological boundary condition}\label{sec:topological-boundary}

In this subsection, we review the topological boundary conditions, which constitute another ingredient necessary for the construction of 2d CFTs from abelian Chern-Simons theories.

We will consider bosonic and fermionic topological boundary conditions.
Both types are characterized by how the Wilson line behaves on the topological boundary~$\Sigma_\text{top}:=\Sigma\times\{x^3=0\}$, where we recall from section~\ref{sec:canonical-quantization} that $x^3\in [0,1]$ labels an interval.
The discriminant group~$\mathcal{D} \equiv \Gamma^*/\Gamma$ labels distinct Wilson lines as
\begin{equation}
    W_{[\alpha]}(\mathsf{X})
        =    
        \exp \left(\mathrm{i}\oint_{\,\mathsf{X}} \alpha_I A^I\right)  \,,
\end{equation}
where $[\alpha]\in\Gamma^*/\Gamma$.
There is a natural quadratic form%
\footnote{%
The quadratic form $Q$ is non-degenerate and gives~$\mathcal{D}$ the structure of a metric group~\cite{2009arXiv0909.3140E,MR3242743}.
}  $Q: \mathcal{D} \rightarrow \mathbb{Q}/\mathbb{Z}$
defined by
\begin{equation}
[\alpha] \mapsto Q([\alpha]):= \frac{1}{2}  \alpha \oslash\alpha 
    \text{ mod }  \mathbb{Z}     
\end{equation} 
and the related inner product $\mathcal{D}\times \mathcal{D} \rightarrow 
\mathbb{Q}/\mathbb{Z}$
\begin{equation}
    ([\alpha],[\beta])\mapsto 
    [\alpha]\oslash[\beta] := \alpha\oslash\beta 
    \text{ mod } \mathbb{Z}
\end{equation}
induced from $Q$ as
\begin{equation}
    [\alpha]\oslash [\beta] = Q([\alpha]+[\beta])-Q([\alpha])-Q([\beta])\,.
\end{equation}
The first relation in~(\ref{eq:Wilson-block-non-spin}) implies that the braiding phase between the two anyons, whose worldlines are the Wilson lines $W_{[\alpha]}$ and $W_{[\beta]}$, is $\exp\left(2\pi\i[\alpha]\oslash[\beta]\right)$.
A subgroup $\mathcal{C}\subset \mathcal{D}$ is called Lagrangian when the two conditions
\begin{align}
     [\alpha],\, [\beta] \in \mathcal{C} \quad &\Longrightarrow \quad [\alpha] \oslash [\beta] = 0\,,
     \label{eq:self-orthogonality}
    \\
    [\alpha] \oslash [\beta] = 0  \text{ for all }[\beta]\in\mathcal{C} \quad
    &\Longrightarrow \quad [\alpha] \in \mathcal{C}
     \label{eq:coisotropy}
\end{align}
are satisfied.
In other words, $\mathcal{C}$ is Lagrangian if $\mathcal{C}=\mathcal{C}^\perp$,%
\footnote{%
When~$\mathcal{C}$ is regarded as a classical code, $\mathcal{C}^\perp$ is called the dual of~$\mathcal{C}$.}
where 
\begin{equation}\label{eq:C-dual}
\CC^\perp         :=  \left\{ [\alpha] \in \Gamma^\ast/\Gamma\, \big|\,  [\alpha] \oslash [\beta] = 0 ~~\text{for all}~[\beta] \in \CC\right\}\,.    
\end{equation}
Kapustin and Saulina argued in~\cite{Kapustin:2010hk} that a Lagrangian subgroup $\mathcal{C}$ gives rise to a topological boundary condition by specifying the Wilson lines that can end on and be absorbed by the boundary.

We will consider the case when the Wilson lines in $\mathcal{C}$ are all bosonic, i.e., they have integer spins,%
\footnote{%
The spin of~$W_{[\alpha]}$ is $Q([\alpha])=\alpha\oslash\alpha/2$ mod $\mathbb{Z}$.}
and the case when some of the Wilson lines are fermionic, i.e., they have half-integer spins.
We call the Lagrangian subgroup~$\mathcal{C}$ even if
\begin{equation}\label{eq:evenness-condition}
\alpha \oslash \alpha \in 2\, \mathbb{Z}
\quad
\text{ for all } [\alpha] \in \mathcal{C} \,,
\end{equation}
and odd if%
\footnote{%
When the Lagrangian subgroup~$\mathcal{C}$ is identified with a binary code, the evenness and the oddness of the Lagrangian subgroup are equivalent to the double evenness and the single evenness of the code~\cite{conway2013sphere}, respectively.
}
\begin{equation}\label{eq:oddness-condition}
\alpha \oslash \alpha \in 2\, \mathbb{Z}+1
\quad
\text{ for some } [\alpha] \in \mathcal{C} \,.
\end{equation}
We note that for given  $[\alpha]\in\mathcal{C}\subset\Gamma^*/\Gamma$, $\alpha\oslash\alpha$ mod 2 is independent of the representative~$\alpha$ because $\Gamma$ is even.
In terms of the quadratic form $Q$, $\mathcal{C}$ is even if $Q([\alpha])=0 \text{ mod } \mathbb{Z}$ for all $[\alpha]\in\mathcal{C}$, and is odd otherwise.

\subsubsection{Bosonic boundary}
\label{sec:bosonic-boundary}

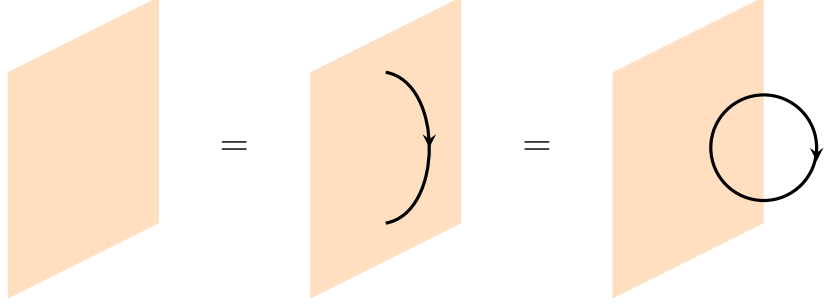
\begin{figure}
    \centering
    
\begin{tikzpicture}[thick, >=stealth]
    
        \begin{scope}
            \draw[fill=orange!50, opacity=0.5, draw=none] (-1,-2) -- (1, -1) -- (1, 2) -- (-1, 1) -- cycle;
        \end{scope}
        
        \node at (2, 0) {\Large $=$};

        \begin{scope}[xshift=4cm, decoration={markings, mark=at position 0.5 with {\arrow[very thick]{>}}}]
            \draw[fill=orange!50, opacity=0.5, draw=none] (-1,-2) -- (1, -1) -- (1, 2) -- (-1, 1) -- cycle;
            \draw[very thick, postaction={decorate}] (0,1) to[out=-10,in=10] (0,-1);
        \end{scope}
        
        \node at (6, 0) {\Large $=$};
        
        \begin{scope}[xshift=8cm, decoration={markings, mark=at position 0.0 with {\arrow[very thick]{<}}}]
            \draw[fill=orange!50, opacity=0.5, draw=none] (-1,-2) -- (1, -1) -- (1, 2) -- (-1, 1) -- cycle;
            \draw[very thick, postaction={decorate}] (1,0) circle(0.7);
        \end{scope}
    \end{tikzpicture}
    \caption{A bosonic topological boundary condition corresponding to an even Lagrangian subgroup $\mathcal{C}\subset \mathcal{D}$ specifies a maximal set of mutually local Wilson lines that can end on and be freely absorbed by a topological boundary. 
    }
    \label{fig:topological-boundary}
\end{figure}

A bosonic topological boundary condition is specified by the choice of an even Lagrangian subgroup~$\mathcal{C}$.
The Wilson lines whose charges are in $\mathcal{C}$ become trivial on the topological boundary:
\begin{equation}\label{eq:bosonic-condensation}
    W_{[\gamma]}\big|_{\Sigma_\text{top}} = 1  \quad \text{ if ~$[\gamma]\in\mathcal{C}$.}
\end{equation}
See figure~\ref{fig:topological-boundary}.
We say that the anyon of charge $[\gamma]\in\mathcal{C}$ condenses on the boundary.
The condition~(\ref{eq:bosonic-condensation}) can be consistently imposed because the braiding phase is trivial between the two anyons whose charges are in $\mathcal{C}$.

For a given level matrix $K_{IJ}$ with even diagonal elements, an even Lagrangian subgroup~$\mathcal{C}$ of $\mathcal{D}\simeq \mathbb{Z}^n/K \mathbb{Z}^n$ may or may not exist, i.e., the Chern-Simons theory may or may not admit a bosonic topological boundary condition.
The condition was studied in detail in~\cite{Kaidi:2021gbs} and was characterized by the so-called higher central charges.
Alternatively, \cite{Kaidi:2021gbs} found that the Chern-Simons theory admits a topological boundary condition if and only if it is equivalent to a Dijkgraaf-Witten theory~\cite{Dijkgraaf:1989pz} for an abelian group $G$, which is isomorphic to $\mathcal{D}/\mathcal{C}$.
In turn, the bosonic topological boundary conditions of the latter are labeled by a subgroup $H\subset G$ and an element $\psi\in H^2(H,\U(1))$~\cite{Bhardwaj:2017xup,Gaiotto:2020iye}.
This means that different bosonic topological conditions for a given Chern-Simons theory give rise to CFTs that are related to each other by gauging/orbifolding by a subgroup $H\subset G$, possibly with a discrete torsion corresponding to $\psi$.
We will explain how discrete torsion appears within our construction in section~\ref{sec:orb-bos}.

\subsubsection{Fermionic boundary}
\label{sec:fermionic-boundary}

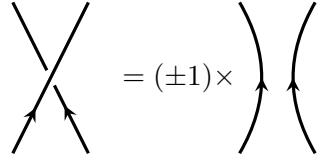
\begin{figure}
    \centering
    
    \begin{tikzpicture}[thick, >=stealth]

        \begin{scope}[xshift=4cm, decoration={markings, mark=at position 0.3 with {\arrow[very thick]{>}}}]
            \draw[very thick, postaction={decorate}] (0,0) to (1,2);
        \end{scope}

        \begin{scope}[xshift=4cm, decoration={markings, mark=at position 0.7 with {\arrow[very thick]{>}}}]
            \draw[very thick, postaction={decorate}] (1,0) to (0.55,0.9);
        \end{scope}        
        
        \begin{scope}[xshift=4cm]
            \draw[very thick, postaction={decorate}] (0.45,1.1) to (0,2);
        \end{scope}           

        \node at (6.2, 1) {$= (\pm 1) \times$};

        \begin{scope}[xshift=4cm, decoration={markings, mark=at position 0.5 with {\arrow[very thick]{>}}}]
            \draw[very thick, postaction={decorate}] (3,0) to [out=60,in=-60] (3,2);
            \draw[very thick, postaction={decorate}] (4,0) to [out=120,in=-120] (4,2);
        \end{scope}

      \end{tikzpicture}
    \caption{The skein relation (half braiding) for a  Wilson line~$W_d$ with $(-1)^{d\oslash d}=\pm 1$.}
    \label{fig:skein}
\end{figure}

Let $\mathcal{C}$ be an odd Lagrangian subgroup of $\mathcal{D}$.
By definition, there exists an element $\alpha_\mathrm{F} \in \Gamma^*$ such that $[\alpha_\mathrm{F}]  \in\mathcal{C}$ and $\alpha_\mathrm{F} \oslash \alpha_\mathrm{F} \in 2\mathbb{Z}+1$.
For $\beta \in\Gamma^*$ such that $[\beta]\in\mathcal{C}$ and $\beta\oslash\beta \in 2\mathbb{Z}$, $\gamma = \alpha_\mathrm{F} + \beta$ satisfies $\gamma\oslash \gamma \in 2\mathbb{Z}+1$.
Conversely, any $\gamma \in \Gamma^*$,  such that $[\gamma]\in\mathcal{C}$ and $\gamma\oslash\gamma\in 2\mathbb{Z}+1$, is of the form $\gamma = \alpha_\mathrm{F} + \beta$ with $\beta\oslash\beta \in 2\mathbb{Z}$.
For such $\gamma$, due to the skein relation in figure \ref{fig:skein}, we have
\begin{equation}\label{eq:W-relation-fermionic}
    W_{[\gamma]}(\mathsf{C}_1 + \mathsf{C}_2 ) 
        =  
        W_{[\gamma]}(\mathsf{C}_1) \,W_{[\gamma]}(\mathsf{C}_2) \,(-1)^{\#( \mathsf{C}_1 \cap \mathsf{C}_2)} \,,
\end{equation}
where $\mathsf{C}_i$ ($i=1,2$) and $\mathsf{C}_1+\mathsf{C}_2$ are (possibly non-connected) non-self-intersecting closed curves, and $\mathsf{C}_1+\mathsf{C}_2$ is obtained from $\mathsf{C}_1$ and $\mathsf{C}_2$ by applying the skein relation repeatedly.

\begin{figure}
    \centering
    
\begin{tikzpicture}[thick, >=stealth]
   
        \begin{scope}
            \draw[fill=orange!50, opacity=0.5, draw=none] (-1,-2) -- (1, -1) -- (1, 2) -- (-1, 1) -- cycle;
        \end{scope}

        \node at (0, 0) {\Large $(\pm 1)$};
         
        \node at (2, 0) {\Large $=$};

        \begin{scope}[xshift=4cm, decoration={markings, mark=at position 0.5 with {\arrow[very thick]{>}}}]
            \draw[fill=orange!50, opacity=0.5, draw=none] (-1,-2) -- (1, -1) -- (1, 2) -- (-1, 1) -- cycle;
            \draw[very thick, postaction={decorate}] (0,0.7) to[out=-10,in=10] (0,-0.7);
            \draw[dashed] (0,1.5) to[out=-90,in=90] (0,0.7);  \draw[dashed] (0,-0.7) to[out=-90,in=90] (0,-1.5);
        \end{scope}
        
        \node at (6, 0) {\Large $=$};

        \begin{scope}[xshift=8cm, decoration={markings, mark=at position 0.5 with {\arrow[very thick]{>}}}]
            \draw[fill=orange!50, opacity=0.5, draw=none] (-1,-2) -- (1, -1) -- (1, 2) -- (-1, 1) -- cycle;
            \draw[very thick, postaction={decorate}] (1.5,1.5) to[out=-90,in=90] (1.5,-1.5);
        \end{scope}
    \end{tikzpicture}
    \caption{A fermionic topological boundary condition corresponding to an odd Lagrangian subgroup $\mathcal{C}\subset \mathcal{D}$ specifies a maximal set of mutually local Wilson lines that can end on and be absorbed by a topological boundary, up to a sign dependent on the spin structure and the homology class as in~(\ref{eq:fermionic-condensation}).
    The dashed segments in the middle represent a transparent topological line defect that becomes the sign factor when the Wilson line is fully absorbed.
    }
    \label{fig:fermionic-topological-boundary}
\end{figure}
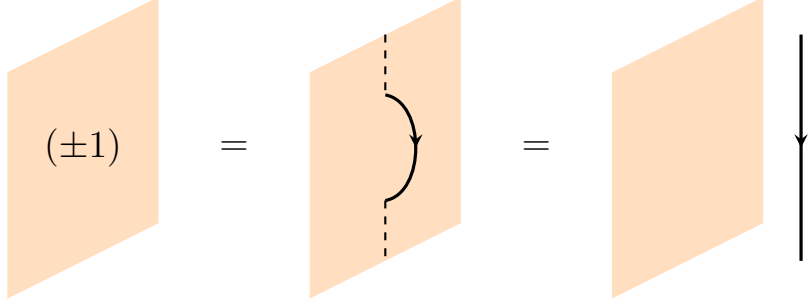

When considering Wilson lines coincident on the topological boundary~$\Sigma_\text{top}$,  $\mathsf{C}_i$ become closed curves on a two-dimensional surface~$\Sigma$.
A topological boundary condition is specified by a maximal set~$\mathcal{C}$ of Wilson line charges $[\gamma]=\gamma+\Gamma$ such that $W_{[\gamma]}$ restricted to the boundary condense, i.e., become non-zero values.
The relation~(\ref{eq:W-relation-fermionic}) implies that the values cannot be simply 1 as in the case when the relevant Wilson lines are all bosonic.
This issue can be resolved with the help of a spin structure as follows.
(See for example~\cite{Tachikawa-TASI}.)
Recall from~\cite{Atiyah1971,MR588283} that a spin structure on the surface~$\Sigma$ is in a  one-to-one correspondence with a function $\rho: H_1(\Sigma,\mathbb{Z}_2) \rightarrow \{\pm 1\}$ such that
\begin{equation}\label{eq:spin-structure-relation}
\rho(\mathsf{C}_1 + \mathsf{C}_2 ) =  \rho(\mathsf{C}_1) \rho(\mathsf{C}_2) (-1)^{\#( \mathsf{C}_1 \cap \mathsf{C}_2)} \,.
\end{equation}
(It is implicit that $\mathsf{C}_i$ is to be replaced by the corresponding homology class.)
We will simply use the symbol $\rho$ to denote the spin structure.
Then we can consistently impose the  boundary condition
\begin{equation}\label{eq:fermionic-condensation}
W_{[\gamma]}(\mathsf{C}) \big|_{\Sigma_\text{top}} = \rho(\mathsf{C})^{\gamma\oslash\gamma}
\end{equation}
for $[\gamma] \in \mathcal{C}$, at the same time introducing a dependence on the spin structure~$\rho$ on the surface~$\Sigma$.
See figure~\ref{fig:fermionic-topological-boundary}.
Equation~(\ref{eq:fermionic-condensation}) dictates how a fermionic anyon of charge $[\gamma]$ condenses on the boundary.

\subsection{Torus partition functions}
\label{sec:torus}

We now put together the results of sections~\ref{sec:canonical-quantization} and \ref{sec:topological-boundary} to describe the bosonic and fermionic CFTs constructed from abelian Chern-Simons theories.

\subsubsection{Bosonic CFTs}

A bosonic CFT is a CFT whose partition function does not depend on the spin structure.
On the torus, the partition function (without chemical potentials $\nu^a$ and $\bar{\nu}^{\bar{a}}$) can be written as a trace:
\begin{equation}\label{eq:torus-partition-function-bosonic}
{\rm Tr}_{\mathcal{H}} \left[
q^{L_0- \frac{n_L}{24}} \,\bar q^{\bar L_0- \frac{n_R}{24}}
\right] \,,
\end{equation}
where $L_0$ and $\bar L_0$ are the left- and right-moving Virasoro zero-modes.
Invariance under $\tau\rightarrow\tau+1$ would require that the chiral central charge $n_L-n_R$ is a multiple of 24 and that the spin $L_0-\bar{L}_0$ is an integer.%
\footnote{%
A bosonic CFT arising from a bosonic topological boundary condition has $n_L-n_R$ that is a multiple of~8~\cite{Kaidi:2021gbs}.
(An even self-dual lattice exists only if $n_L-n_R\in 8 \mathbb{Z}$~\cite{Narain:1985jj}.)
Non-invariance under $\tau\rightarrow\tau+1$ can be canceled by multiplying the theory with copies of the $(E_8)_1$ CFT.
}
We now study the bosonic CFTs that arise from abelian Chern-Simons theories via dimensional reduction.

We saw in section~\ref{sec:bosonic-boundary} that a bosonic topological boundary condition in the abelian Chern-Simons theory, defined by the level matrix $K_{IJ}$ with even diagonal elements, is specified by an even Lagrangian subgroup~$\mathcal{C}$ of $\mathcal{D}=\Gamma^*/\Gamma$.
Let $|\mathcal{C}\rangle$ be the boundary state defined by the bosonic topological boundary condition~(\ref{eq:bosonic-condensation}):%
\footnote{%
The state $|\mathcal{C}\rangle$ may be interpreted as spanning the one-dimensional code subspace of a quantum error-correcting code defined by the stabilizers $\{W_c(\mathsf{C})\,|\, c\in\mathcal{C} \,, \mathsf{C}\in H_1(\Sigma,\mathbb{Z}_2)\}$, where  $\{W_d(\mathsf{C})\,|\, d\in\mathcal{D}\,, \mathsf{C}\in H_1(\Sigma,\mathbb{Z}_2)\}$ is the Pauli group.
}
\begin{equation}
W_c(\mathsf{C}) |\mathcal{C}\rangle = |\mathcal{C}\rangle
\end{equation}
for any closed curve $\mathsf{C}$ on the torus and for any $c\in\mathcal{C}$.
For $d\notin \mathcal{C}$, on the other hand, the first line of~(\ref{eq:Wilson-block-non-spin}) implies that the state $|\Psi_d\rangle$ is an eigenstate of $W_c(\mathsf{A})$ with an eigenvalue different from $+1$ for some $c\in\mathcal{C}$.
Thus, using also~(\ref{eq:Psi-W}), we get
\begin{equation}
\langle\Psi_d| \mathcal{C}\rangle 
=
\left\{
\begin{array}{cc}
1 & \text{ if } d \in \mathcal{C} \,, \\
0 &  \text{ if } d \notin \mathcal{C} \,.
\end{array}
\right. \,.
\end{equation}
We chose the normalization such that $\langle \Psi_0 | \mathcal{C}\rangle=1$.
The partition function of the 3d Chern-Simons theory on $\Sigma\times [0,1]$, with the physical boundary condition~(\ref{eq:nu-nubar-bc}) along $\Sigma\times \{1\}=\Sigma_\text{ph}$ and the bosonic topological boundary condition~(\ref{eq:bosonic-condensation}) along $\Sigma\times \{0\}=\Sigma_\text{top}$, can be expressed as a sum of the conformal blocks compatible with the boundary condition:
\begin{equation}\label{eq:bosonic-partition-function-reduction}
    Z_\mathcal{C} (\tau,\bar\tau;\nu,\bar\nu ) 
=
\langle \tau,\bar\tau ; \nu,\bar\nu
\big|
\left(\sum_{d\in \mathcal{D}}
|\Psi_d\rangle \,\langle \Psi_d|
\right)
\big| \mathcal{C}   \rangle 
    = \sum_{[\alpha]\in \mathcal{C} } 
    \Psi_{[\alpha]}(\tau,\bar\tau;\nu,\bar\nu)  \,.
\end{equation}

We now show that the partition function~(\ref{eq:bosonic-partition-function-reduction}) can be identified with that of a 2d bosonic CFT.
The group~$\mathcal{C}$ gives rise to a lattice
\begin{equation}\label{eq:Lambda-Gamma-mathcalC}
    \Lambda(\mathcal{C}) 
        = 
        \Lambda(\Gamma;\mathcal{C} ) 
        := 
        \{\gamma\in\Gamma^* \, | \, \gamma + \Gamma \in \mathcal{C}  \} \,.
\end{equation}
Note that $\mathcal{C}= \Lambda(\mathcal{C}) /\Gamma$ by definition.
The lattice~$ \Lambda(\mathcal{C})$ is even, i.e., 
\begin{equation}\label{eq:lambda-even}
\lambda\oslash\lambda\in2\mathbb{Z} \quad\text{ for} \quad \lambda \in \Lambda(\mathcal{C})\ , 
\end{equation}
because $\mathcal{C}$ is even in the sense defined in~(\ref{eq:evenness-condition}).
The lattice is self-dual because $\mathcal{C}$ is Lagrangian, i.e., $\mathcal{C}=\mathcal{C}^\perp$, where the dual~$\mathcal{C}^\perp$ is defined in~(\ref{eq:C-dual}).
Since $\Lambda(\mathcal{C})$ is an even self-dual lattice, we can identify it with the momentum lattice of a bosonic CFT denoted as~$T[\mathcal{C}]$.
In particular, $T[\mathcal{C}]$ is a Narain CFT when $n_L=n_R$~\cite{Narain:1985jj,Narain:1986am}.
All states have integer spins because of~(\ref{eq:lambda-even}) and are bosonic.
The corresponding torus partition function is
\begin{equation}
\label{eq:torus-partition-function}
        \frac{1}{\eta(\tau)^{n_L} \overline{\eta(\tau)}{}^{n_R}} \sum_{\lambda\in\Lambda(\mathcal{C})} 
        q^{\frac{1}{2}
        \lambda^2_L
        }
        \bar q^{\frac{1}{2} 
        \lambda^2_R}
        e^{2\pi \i\, (\lambda^a \nu^a -\lambda^{\bar a} \bar\nu^{\bar a})} \,,
\end{equation}
which coincides with~(\ref{eq:bosonic-partition-function-reduction}) up to normalization%
\footnote{%
The normalization factor is the last exponential in~(\ref{eq:wave-function}), which is necessary for the modular transformation property~(\ref{eq:Psi-CS-S}) to hold.
}
and takes the form~(\ref{eq:torus-partition-function-bosonic}) if we set $\nu^a=\bar\nu^{\bar a}=0$.
We can write
\begin{equation}
    \lambda^a \nu^a -\lambda^{\bar a}\bar\nu^{\bar a}
    = \lambda_I \nu^I
\end{equation}
in terms of $\nu^I := \nu^a\,  (M^{-1})_a{}^I + \bar\nu^{\bar a}\,  (M^{-1})_{\bar a}{}^I $, allowing us to interpret $\nu_I$ as the chemical potentials for the $\U(1)$ symmetries whose charges form the Narain lattice~$\Lambda(\mathcal{C})$.
The partition function~(\ref{eq:torus-partition-function}) is invariant under the separate actions of $\mathrm{O}(n_L;\mathbb{R})$ (acting on the index $a$) and $\mathrm{O}(n_R;\mathbb{R})$ acting on $\bar{a}$, which therefore do not change the 2d CFT.
Combined with the group action in~(\ref{eq:O-nL-nR}), the quotient space
\begin{equation}\label{eq:moduli-space}
\frac{\mathrm{O}(n_L,n_R;\mathbb{R})
}{
\mathrm{O}(n_L;\mathbb{R})
\times
\mathrm{O}(n_R;\mathbb{R})
}
\end{equation}
parameterizes the moduli of the CFT, which are subject to further discrete identifications due to T-dualities.
We note that there is no moduli
in the chiral case $n_R=0$ or $n_L=0$.
Note that the moduli space~(\ref{eq:moduli-space}) does not depend on the choice of topological boundary condition~$\mathcal{C}$.

We note that the Lagrangian properties~(\ref{eq:self-orthogonality}) and (\ref{eq:coisotropy}) imply the modular invariance of the partition function.
First, we have
% See Takuya's 01-18 note.
\begin{equation}
    Z_\mathcal{C} (\tau+1,\bar\tau+1;\nu,\bar\nu) =
    e^{-\frac{\pi \i}{12}(n_L-n_R)} Z_\mathcal{C} (\tau,\bar\tau;\nu,\bar\nu)  
\end{equation}
for the T-transformation.
The phase factor is trivial if we require that $n_L-n_R\in 24\mathbb{Z}$.
For the S-transformation, we get
\begin{equation}
    \begin{aligned}    
    Z_\mathcal{C} \left(-\frac{1}{\tau},-\frac{1}{\bar\tau};
    \frac{\nu}{\tau},\frac{\bar\nu}{\bar\tau}
    \right)
    &=
    \frac{1}{\sqrt{|\Gamma^*/\Gamma|}}
    \sum_{[\beta]\in\Gamma^*/\Gamma}
    \left( \sum_{[\alpha]\in \mathcal{C} }  e^{-2\pi \i \,\alpha\oslash \beta} \right)
    \Psi_{[\beta]}(\tau,\bar\tau;\nu,\bar\nu)
    \\
    &=
    \frac{|\mathcal{C} |}{\sqrt{|\Gamma^*/\Gamma|}}
     \sum_{[\beta] \in \mathcal{C} ^\perp} \Psi_{[\beta]}(\tau,\bar\tau;\nu,\bar\nu)
    \,.
    \end{aligned}
\end{equation}
We note that $|\mathcal{C} |\, |\mathcal{C} ^\perp| = |\Gamma^*/\Gamma|$.
Since $\mathcal{C} ^\perp=\mathcal{C} $, we have invariance under the S-transformation
\begin{equation}
Z_\mathcal{C} \left(-\frac{1}{\tau},-\frac{1}{\bar\tau};
\frac{\nu}{\tau},\frac{\bar\nu}{\bar\tau}
\right)
=
Z_\mathcal{C} (\tau,\bar\tau;\nu,\bar\nu)
\,.
\end{equation}

\subsubsection{Fermionic CFTs}
\label{sec:ferm-cft}

A fermionic CFT is defined as a CFT whose partition function depends on the spin structure.
For the torus, let us specify the spin structure by $(S_1,S_2) \in \{0,1\}^2$ as follows.
\begin{center}
\begin{tabular}{c|cccc}
&$\mathrm{NS}$ &$\widetilde{\mathrm{NS}}$ &$\mathrm{R}$ &$\widetilde{\mathrm{R}}$ \\
\hline
$(S_1,S_2)$& $(0,0)$ & $(0,1)$ & $(1,0)$ & $(1,1)$
\end{tabular}
\end{center}
Here, NS and R indicate that the spinor is respectively anti-periodic and periodic along the spatial direction.
On the other hand, the presence and the absence of a tilde respectively indicate that the spinor is periodic and anti-periodic along the time direction.
The labels $(S_1,S_2)$ can be related to the functions  $\rho: H_1(\Sigma,\mathbb{Z}_2) \rightarrow \{\pm 1\}$ introduced in section~\ref{sec:fermionic-boundary} by setting%
\footnote{%
By these conditions, we fix the ambiguity from the redefinition $\rho(\mathsf{C}) 
\rightarrow \rho(\mathsf{C})(-1)^{\mathsf{x}\cdot \mathsf{C}}$ that  maintains the relation~(\ref{eq:spin-structure-relation}), where $\mathsf{x}$ is a cocycle representing a cohomology class in $ H^1(\Sigma,\mathbb{Z}_2)$.}
$\rho_{(S_1,S_2)}(\mathsf{A}) =(-1)^{S_1}$ and $\rho_{(S_1,S_2)}(\mathsf{B}) =(-1)^{S_2}$.
When $\nu^a=\bar\nu^{\bar a}=0$, the spin structure-dependent partition functions can be expressed in terms of the trace as
\begin{equation}
\label{eq:fermionic-partition-functions-trace}
\begin{aligned}
Z_{(0,0)} &= Z_\mathrm{NS}  =
{\rm Tr}_{\mathcal{H}_\text{NS}} \left[
q^{L_0- \frac{n_L}{24}} \bar q^{\bar L_0- \frac{n_R}{24}}
\right]
\,, 
\\
Z_{(0,1)} &= Z_\mathrm{\widetilde{NS}}  
= {\rm Tr}_{\mathcal{H}_\text{NS}} \left[
(-1)^F
q^{L_0- \frac{n_L}{24}} \bar q^{\bar L_0- \frac{n_R}{24}}
\right]
\,, \\
Z_{(1,0)} & = Z_\mathrm{R}
=  {\rm Tr}_{\mathcal{H}_\text{R}} \left[
q^{L_0- \frac{n_L}{24}} \bar q^{\bar L_0- \frac{n_R}{24}}
\right]
\,,  \\
Z_{(1,1)} &= Z_\mathrm{\widetilde{R}}
= {\rm Tr}_{\mathcal{H}_\text{R}} \left[
(-1)^F
q^{L_0- \frac{n_L}{24}} \bar q^{\bar L_0- \frac{n_R}{24}}
\right]
\,,
\end{aligned}
\end{equation}
where $F$ is the fermion number.
Let us now focus on the fermionic CFTs that arise from abelian Chern-Simons theories.

When $\mathcal{C}$ is an odd Lagrangian subgroup, the lattice $\Lambda:=\Lambda(\mathcal{C})$ is odd and self-dual.
The set $\Lambda_0 := \{\lambda\in\Lambda \, | \,  \lambda\oslash\lambda \in 2\mathbb{Z}\}$ is a sublattice of index 2, i.e., $|\Lambda/\Lambda_0|=2$.
Let us consider the dual $\Lambda_0^*$ of $\Lambda_0$.
The set $S:=\Lambda_0^* \, \backslash \, \Lambda$
is called the shadow of $\Lambda$~\cite{conway2013sphere}.
Any element $\chi\in 2S =\{2s\,|\, s\in S\}$ satisfies 
\begin{equation}\label{eq:characteristic-vector}
\chi \oslash \lambda = \lambda \oslash \lambda 
\text{ mod } 2 \quad \text{ for all } \lambda\in \Lambda \,.
\end{equation}
% See the iPad note 2024-06-17_QEC-CS.pdf shared on Discord for a proof.
Such $\chi$ is called a characteristic vector.
Let us fix $s\in S$ and $\chi = 2s$.
The torus($=\Sigma$) partition function of the Chern-Simons theory on $\Sigma\times [0,1]$, with the physical boundary condition~(\ref{eq:nu-nubar-bc}) along $\Sigma\times \{1\}=\Sigma_\text{ph}$ and the fermionic topological boundary condition~(\ref{eq:fermionic-condensation}) along $\Sigma\times \{0\}=\Sigma_\text{top}$,
can be written as
\begin{equation}
    Z_{\mathcal{C},\rho} (\tau,\bar\tau;\nu,\bar\nu ) 
=
\langle \tau,\bar\tau ; \nu,\bar\nu
| \mathcal{C} ,\rho  \rangle
=
\langle \tau,\bar\tau ; \nu,\bar\nu
|
\left(\sum_{d\in \mathcal{D}}
|\Psi_d\rangle \,\langle \Psi_d|
\right)
| \mathcal{C} ,\rho  \rangle
      \,,
\end{equation}
where $|\mathcal{C},\rho\rangle$ is the topological boundary state defined by~(\ref{eq:fermionic-condensation}).
Using~(\ref{eq:characteristic-vector}), we can characterize the state~$|\mathcal{C},\rho\rangle$ by
\begin{equation}\label{eq:characterize-C-rho}
W_c (\mathsf{C})
|\mathcal{C},\rho\rangle
=
\rho(\mathsf{C})^{\chi\oslash c} 
|\mathcal{C},\rho\rangle 
\quad\text{ for }\quad c\in\mathcal{C}
\,.
\end{equation}
Boundary states with different spin structures are related by the action of appropriate Wilson lines.
We have
\begin{equation} \label{eq:C-rho-WW}
|\mathcal{C},\rho_{S_1,S_2} \rangle 
=
W_{[s]}(\mathsf{B})^{S_1 } 
W_{[s]}( \mathsf{A})^{S_2}
|\mathcal{C},\rho_{0,0} \rangle 
\end{equation}
since the right-hand side satisfies~(\ref{eq:characterize-C-rho}).%
\footnote{%
We have normalized the states $|\mathcal{C},\rho_{S_1,S_2} \rangle$ so that the left-hand side is equal, rather than just proportional, to the right-hand side in~(\ref{eq:C-rho-WW}).
Further, we normalize $|\mathcal{C},\rho_{0,0} \rangle$ so that $\langle \Psi_c | \mathcal{C},\rho_{0,0}\rangle = 1$ for $c\in\mathcal{C}$.
}
Using~(\ref{eq:Wilson-block-non-spin}), we can now recast the partition functions as
\begin{equation}\label{eq:CFT-part-ferm-bc}
Z_{\mathcal{C},(S_1,S_2)}(\tau,\bar\tau;\nu,\bar\nu)
= \sum_{c\in\mathcal{C}}
  (-1)^{S_2[\chi]\oslash 
  c
  } \Psi_{c+S_1[s]}(\tau,\bar\tau;\nu,\bar\nu)
\end{equation}
up to overall phases.
We note that the odd self-dual lattice~$\Lambda=\Lambda(\mathcal{C})$ and its shadow~$S$ constitute the NS and R spectra, respectively.
We can also write the torus partition functions as sums over lattice vectors.
For simplicity, we set $\nu^a = \bar\nu^{\bar a} =0$.
Then,
\begin{align}\label{eq:ZC0011}
    \begin{aligned}
        Z_{\mathcal{C},(0,0)} &=  Z_{\mathcal{C},\mathrm{NS}} =  
        \frac{1}{\eta^{n_L} \overline{\eta}{}^{n_R}}
        \sum_{\lambda\in\Lambda(\mathcal{C})}  q^{\frac{1}{2} \lambda^2_L}         \bar q^{\frac{1}{2} \lambda^2_R}
        \,, 
        \\
        Z_{\mathcal{C},(0,1)} & 
        =   Z_{\mathcal{C},\widetilde{\mathrm{NS}}} =
        \frac{1}{\eta^{n_L} \overline{\eta}{}^{n_R}}
        \sum_{\lambda\in\Lambda(\mathcal{C})} 
        (-1)^{\chi\oslash\lambda}
        q^{\frac{1}{2} \lambda^2_L}         \bar q^{\frac{1}{2} \lambda^2_R}
        \,, \\
        Z_{\mathcal{C},(1,0)} &
        =  Z_{\mathcal{C},\mathrm{R}} =
        \frac{1}{\eta^{n_L} \overline{\eta}{}^{n_R}}
        \sum_{\lambda\in\Lambda(\mathcal{C})}  q^{\frac{1}{2} (\lambda+s)^2_L}         \bar q^{\frac{1}{2} (\lambda+s)^2_R}
        \,, \\
        Z_{\mathcal{C},(1,1)} &
        =  Z_{\mathcal{C},\widetilde{\mathrm{R}}} =
        \frac{1}{\eta^{n_L} \overline{\eta}{}^{n_R}}
        \sum_{\lambda\in\Lambda(\mathcal{C})} 
        (-1)^{\chi\oslash\lambda}
        q^{\frac{1}{2} (\lambda+s)^2_L}         \bar q^{\frac{1}{2} (\lambda+s)^2_R}
        \,,
        \end{aligned}
\end{align}
where 
$\eta$ and $\overline{\eta}$ denote $\eta(\tau)$ and $\overline{\eta(\tau)}$, respectively.
Let us identify $(-1)^{\chi\oslash\lambda}$ with $(-1)^F$, so that vertex operators with odd $\chi\oslash\lambda (= \lambda\oslash\lambda$ mod 2) are fermionic.%
\footnote{\label{footnote:R-F}%
For the R sector ($S_1=1$), the definition of the fermion number depends on the choice of $s\in S$, which determines $\lambda=\lambda_\mathrm{R}-s\in\Lambda$ from a given vector $\lambda_\mathrm{R}\in  S$ specifying an R state.
}
Then the partition functions~(\ref{eq:ZC0011}), obtained as those of the 3d Chern-Simons theory on $\Sigma\times[0,1]$, take the forms of the torus partition functions~(\ref{eq:fermionic-partition-functions-trace}) for the 2d CFT~$T[\mathcal{C}]$.
We note that fermionic operators arise from fermionic Wilson lines stretching in the bulk between the physical and topological boundaries.
In addition, we observe from the partition functions~(\ref{eq:ZC0011}) that both the NS spectrum $\Lambda(\mathcal{C})$ and the R spectrum $\Lambda(\mathcal{C})+s = S$ are determined by the odd Lagrangian subgroup~$\mathcal{C}=\Lambda(\mathcal{C})/\Gamma$.

\subsubsection{Shadow of odd Lagrangian subgroup}
\label{sec:shadow-lagrangian}

We have seen that the operator content of the NS sector is determined by the odd self-dual lattice $\Lambda(\CC)$ for an odd Lagrangian subgroup $\CC\subset\CD$.
On the other hand, the R sector is given by the shadow $S=\Lambda(\CC) + s$ of the lattice.
Here, we introduce the shadow $S(\CC)$ of an odd Lagrangian subgroup $\CC$, which determines the shadow of the lattice and the R sector.

We consider the discriminant group $\CD = \Gamma^*/\Gamma$ and an odd Lagrangian subgroup $\CC\subset\CD$.
We divide the odd Lagrangian subgroup into $\CC = \CC_0\sqcup \CC_2$ where
\begin{align}
\begin{aligned}
    \CC_0 &= \{\,[\alpha]\in\CC\mid \alpha\oslash \alpha\in 2\BZ\,\}\,,\\
    \CC_2 &= \{\,[\alpha]\in\CC\mid \alpha\oslash \alpha\in 2\BZ+1\,\}\,,
\end{aligned}
\end{align}
and consider the dual $\CC_0^\perp$ of $\CC_0$.
Then, any element $[s]\in \CC_0^\perp\backslash\CC$ satisfies
\begin{align}
\label{eq:shadow_grad}
    \alpha\oslash \alpha = \alpha\oslash (2s)\mod 2\BZ\,,
\end{align}
for $[\alpha]\in \CC$.
This can be shown as follows.
Because $[2\alpha]\in \CC_0$ and $[s]\in \CC_0^\perp$, we have $\alpha\oslash (2s)= (2\alpha)\oslash s \in \mathbb{Z}$.
Thus modulo $2\mathbb{Z}$, $\alpha\oslash (2s)$ is either $0$ or $1$.
If $[\alpha]\in \CC_0$, we have $ \alpha \oslash s \in \mathbb{Z}$ and hence $\alpha\oslash (2s) \in 2\mathbb{Z}$ because $[s]\in \CC_0^\perp$.
Now suppose that $[\alpha]\in \CC_2$.
Because $[s]$ is not in $\CC$, there must be an element $[t]\in \CC$ such that $s\oslash t \notin \mathbb{Z}$.
Since $[t]$ cannot be in $\CC_0$, $[t]
$ is in $\CC_2$.
Noting that $[\alpha-t]$ is in $\CC_0$, we obtain that $s\oslash \alpha \notin \mathbb{Z}$ and that $\alpha\oslash (2s) = 1$  mod $2\mathbb{Z}$, completing the proof. 
 
In analogy with the shadow of a code~\cite{dougherty2000shadow,dougherty2001shadow,bonnecaze2003splitting,dougherty2003generalized}, we define the shadow $S(\CC)$ of an odd Lagrangian subgroup $\CC$ by
\begin{align}
    S(\CC ) = \CC + [s]\subset \CD\,.
\end{align}
Then, $S(\CC) = \CC_1\sqcup \CC_3$ where $\CC_1 = \CC_0 + [s]$ and $\CC_3 = \CC_2+[s]$.
We define the construction similar to~\eqref{eq:Lambda-Gamma-mathcalC} to a subset $K\subset \CD$ by
\begin{align}
    \Lambda(K) := \{\gamma\in\Gamma^*\mid \gamma + \Gamma\in K\}\,.
\end{align}
When $K = S(\CC)$, this construction results in the shadow of the lattice: $\Lambda(S(\CC)) = \Lambda(\CC) + s = S$.
The torus partition functions shown in~\eqref{eq:ZC0011} can be rewritten as
\begin{align} \label{eq:Z_by_C0123}
    \begin{aligned}
        Z_{\CC,\mathrm{NS}} = \chi_{\CC_0} + \chi_{\CC_2}\,,\qquad & Z_{\CC,\widetilde{\mathrm{NS}}} = \chi_{\CC_0} - \chi_{\CC_2}\,, \\
        Z_{\CC,\mathrm{R}} = \chi_{\CC_1} + \chi_{\CC_3}\,,\qquad &\;Z_{\CC,\widetilde{\mathrm{R}}} = \chi_{\CC_1} - \chi_{\CC_3}\,,
    \end{aligned}
\end{align}
where we fix the ambiguity in the sign of the $\widetilde{\mathrm{R}}$ partition function.
Here, the symbol $\chi_K$ for a subset $K\subset\CD$ denotes
\begin{align}
    \chi_K = \frac{1}{\eta^{n_L} \overline{\eta}{}^{n_R}}
        \sum_{\lambda\in\Lambda(K)}  q^{\frac{1}{2} \lambda^2_L}        \, \bar q^{\frac{1}{2} \lambda^2_R}\,.
\end{align}
Thus, one can determine the spectrum of both NS and R sectors from the odd Lagrangian subgroup $\CC$ and its shadow $S(\CC)$, respectively.

\subsection{Orbifolds}
\label{sec:orbifolds}

An abelian Chern-Simons theory may or may not admit a bosonic topological boundary condition.%
\footnote{%
See~\cite{Kaidi:2021gbs} for the characterizations of abelian Chern-Simons theories that admit bosonic topological boundary conditions.
}
When it does, there can be more than one such boundary condition, leading to multiple bosonic CFTs.
In this subsection, we will first study the relations among the bosonic CFTs that arise from different bosonic boundary conditions on a single bulk abelian Chern-Simons theory.
Then, we will generalize our discussion to the fermionic CFTs arising from different fermionic boundary conditions.
We assume that, for $\gamma_1,\gamma_2\in\Gamma^*$, the vertex operators $V_{\gamma_1}$ and $V_{\gamma_2}$ have an OPE (operator product expansion)
$
V_{\gamma_1}(z) V_{\gamma_2}(0) \sim 
z^{\gamma_1^a\gamma_2^a} 
\bar{z}^{\gamma_1^{\bar{a}}\gamma_2^{\bar{a}}}
V_{\gamma_1+\gamma_2}(0)
$.

\subsubsection{Bosonic CFTs}
\label{sec:orb-bos}

Let $\mathcal{C}$ and $\mathcal{C}'$ be two even Lagrangian subgroups of the discriminant group $\mathcal{D}=\Gamma^*/\Gamma$.
The group $\mathcal{D}$ naturally acts on the vertex operators $V_\gamma$ of the bosonic CFT $T[\mathcal{C}]$ as $V_\gamma \mapsto e^{2\pi \i\, d\oslash c }\, V_\gamma$ for $d\in\mathcal{D}$ and $ c = \gamma +\Gamma$.%
\footnote{%
The vertex operator $V_\gamma$ represents 
the Wilson line $W_{[\gamma]}$ stretching between the physical and topological boundaries.
Another Wilson line $W_d$ acts on $W_{c}$ by linking with it, giving the phase $e^{2\pi\i d\oslash c}$ as the action of a 1-form symmetry in 3d, which descends to a 0-form symmetry in 2d.}
The subgroup $\mathcal{C}\subset \mathcal{D}$ acts trivially on $T[\mathcal{C}]$.
Thus we treat the quotient $G:= \mathcal{D}/\mathcal{C}$ as the global symmetry.
Let $\varphi: \mathcal{D}\rightarrow \mathcal{D}/\mathcal{C}=G$ be the projection and set $H:=\varphi(\mathcal{C}')$.
It can be checked that $H$ is non-anomalous.%
\footnote{%
The anomaly is given by the F-symbol restricted to $\mathcal{C}'$~\cite{Bhardwaj:2017xup}, which is known to be trivial when $\mathcal{C}'$ is a Lagrangian subgroup.
See also~\cite{Kaidi:2021gbs}.
}
We wish to consider orbifolding $T[\mathcal{C}]$ by $H$.
We need to understand how to describe the twisted sectors.
An operator $\mathcal{O}_h$ in the sector twisted by $h\in H$ must satisfy
\begin{equation}
    V_\gamma(e^{2\pi \i } z, e^{-2\pi\i}\bar z)\,\mathcal{O}_h(0,0) = e^{2\pi\i\, c\oslash h}\, V_\gamma(z, \bar z)\,\mathcal{O}_h(0,0) 
\end{equation}
for $\gamma\in \Lambda(\mathcal{C})$ and $c=\gamma+\Gamma$.
This means that the twisted sector is spanned by the descendants of the vertex operators $V_\lambda$ such that $\varphi(\lambda+\Gamma)=h$. 
We note that $\lambda+\Gamma \in\mathcal{C}+\mathcal{C}'$.
There is a natural (given the choice of $\mathcal{C}'$) action of $h_1\in H$ on the twisted sector:
\begin{equation}
    \label{eq:orb-action-C'}
    V_\lambda \mapsto e^{2\pi \i \, \lambda'_1\oslash\lambda}\, V_\lambda \,,
\end{equation}
where $c'_1 = \lambda'_1+\Gamma \in\mathcal{C}'$ and $\varphi(\lambda'_1+\Gamma) = h_1$.
With this $H$-action, the physical spectrum of the orbifold theory consists of the $V_\lambda$'s such that $c'_1\oslash (\lambda+\Gamma) =0$ for all $c'_1\in \mathcal{C}'$.
Because $\mathcal{C}'$ is Lagrangian, we have that $\lambda+\Gamma\in\mathcal{C}'$.
This shows that the orbifold theory is precisely the bosonic CFT $T[\mathcal{C}']$.
Noting that $H\simeq \mathcal{C}'/(\mathcal{C}\cap\mathcal{C}')$, we write the relation as
\begin{equation}\label{eq:orbifold-relation}
T[\mathcal{C}'] = \frac{T[\mathcal{C}]}{\mathcal{C}'/(\mathcal{C}\cap\mathcal{C}')} \,,
\end{equation}
where it is understood that the action of $\mathcal{C}'/(\mathcal{C}\cap\mathcal{C}')$ on the twisted sectors is given by~(\ref{eq:orb-action-C'}).

Next, suppose that there is another even Lagrangian subgroup $\mathcal{C}''$ such that $\varphi(\mathcal{C}'')=\varphi(\mathcal{C}')=H$.
When orbifolding $T[\mathcal{C}]$ by $H$, we can use as the orbifold action of $h_1 \in H$
\begin{equation}
    \label{eq:orb-action-C''}
    V_\lambda \mapsto e^{2\pi \i\, \lambda''_1\oslash\lambda}\, V_\lambda 
\end{equation}
instead of~(\ref{eq:orb-action-C'}), where
$c''_1 = \lambda''_1+\Gamma \in\mathcal{C}''$ and $\varphi(\lambda''_1+\Gamma) = h_1$.
This gives
$T[\mathcal{C}''] = T[\mathcal{C}]/ [\mathcal{C}''/(\mathcal{C}\cap\mathcal{C}'')] $.
Although $\mathcal{C}'/(\mathcal{C}\cap\mathcal{C}')\simeq \mathcal{C}''/(\mathcal{C}\cap\mathcal{C}'')$, the different orbifold actions lead to different Narain CFTs.%
\footnote{%
The two orbifold theories share the common untwisted sector consisting of $V_\mu$ with $\mu+\Gamma \in \mathcal{C}\cap\mathcal{C}' = \mathcal{C}\cap\mathcal{C}''$.
}
Such a difference is accounted for by discrete torsion~\cite{Vafa:1986wx}. 
In our case, it is given by the relative phase 
\begin{equation} \label{eq:discrete-torsion-def}
    \epsilon(h,h_1) 
        : = 
        \frac{e^{2\pi \i\, \lambda'_1\oslash\lambda}}{e^{2\pi \i\, \lambda''_1\oslash\lambda}} \,.
\end{equation}

As we now show, the phase~(\ref{eq:discrete-torsion-def}) as a function on $H\times H$ satisfies the standard relations
\begin{align}
    \epsilon(h_1,h_2) 
        &= 
            \epsilon(h_2,h_1)^{-1} \,, \label{eq:disc-tor-1} \\
    \epsilon(h_1,h_2)\, \epsilon(h_1,h_3)  
        & = 
            \epsilon(h_1,h_2 +h_3)  \,,  \label{eq:disc-tor-2}\\
    \epsilon(h_1,h_1)  
        &= 
            1  
            \label{eq:disc-tor-3}
\end{align}
for discrete torsion~\cite{Vafa:1986wx,Polchinski:1998rr}.
Here, $h_1$, $h_2$, and $h_3$ are arbitrary elements of $H$.
First, let us compute 
\begin{equation}
\epsilon(h_1,h_2) 
= \frac{e^{2\pi \i \lambda'_2\oslash\lambda_1}}{e^{2\pi \i \lambda''_2\oslash\lambda_1}}
= \frac{e^{2\pi \i c'_2\oslash (c_1+c'_1)}}{e^{2\pi \i c''_2\oslash (\hat{c}_1+c''_1)}} 
= \frac{e^{2\pi\i c_1\oslash c'_2}}{e^{2\pi\i\hat{c}_1\oslash c''_2}} 
= e^{2\pi\i (c_1-\hat{c}_1)\oslash c'_2}
= e^{2\pi\i c''_1 \oslash c'_2} \,,
\label{eq:disc-tor-rewite}
\end{equation}
where $\varphi(c'_1)=\varphi(c''_1) = h_1 $, $c'_1 = \lambda'_1+\Gamma \in\mathcal{C}'$, $c''_1 = \lambda''_1 +\Gamma \in \mathcal{C}''$, $c_1+c'_1 = \hat{c}_1 + c''_1 = \lambda_1+\Gamma \in \mathcal{D}$, and $c_1,\hat c_1 \in\mathcal{C}$.
The equality 
\begin{equation}
0=(c_1-\hat c_1) \oslash (c_2-\hat c_2) = (c''_1-c'_1)\oslash (c''_2-c'_2) = - c'_1\oslash c''_2 - c'_2 \oslash c'_2
\end{equation}
implies~(\ref{eq:disc-tor-1}).
Equation~(\ref{eq:disc-tor-2}) immediately follows from the last expression in~(\ref{eq:disc-tor-rewite}).
Finally, the evenness of the Lagrangian subgroups $\mathcal{C}$, $\mathcal{C}'$, and $\mathcal{C}''$ implies that
\begin{equation}
2\mathbb{Z} \ni c'_1\oslash c'_1 + c''_1 \oslash c''_1-(c_1-\hat c_1) \oslash (c_1-\hat c_1) = 2 c'_1\oslash c''_1 \,.
\end{equation}
Thus $c'_1\oslash c''_1 \in\mathbb{Z}$ and~(\ref{eq:disc-tor-3}) follows.

\subsubsection{Fermionic CFTs}
Let us consider the fermionic CFTs that arise from the fermionic boundary conditions specified by the two odd Lagrangian subgroups, $\mathcal{C}$ and $\mathcal{C}'$, of $\mathcal{D}$.
Most of the discussion above~(\ref{eq:orbifold-relation}) in the bosonic case goes through without a change if we replace  $\mathcal{O}_h$ by an operator $\mathcal{O}_h^\mathrm{NS}$ in the NS sector twisted by $h\in H=\varphi(\mathcal{C}')$, showing that the NS spectrum of the orbifold theory $T[\mathcal{C}]/H$ coincides with the NS spectrum~$\Lambda(\mathcal{C}')$ of $T[\mathcal{C}']$.
Every topological operation involving symmetry~$H$ should be realizable by a topological boundary of a symmetry topological field theory for $H$,%
\footnote{%
See~\cite{Kawabata:2024gek} for examples of orbifolding fermionic CFTs by symmetries outside $H$, where the R spectrum of the orbifold theory is determined by the consistency between the NS spectrum and modular covariance.
}
namely the abelian Chern-Simons theory we have been considering.
Thus the orbifold theory $T[\mathcal{C}]/H$ should arise from the Chern-Simons theory through our construction, where the NS spectrum~$\Lambda(\mathcal{C}')$ determines the Lagrangian subgroup~$\mathcal{C'}=\Lambda(\mathcal{C}')/\Gamma$ and hence also R spectrum~$\Lambda(\mathcal{C})+s$.
(See the observation made at the end of section~\ref{sec:ferm-cft}.)
Therefore, the orbifold of the fermionic CFT $T[\mathcal{C}]$ by $H\simeq \mathcal{C}'/(\mathcal{C}\cap\mathcal{C}')$ is another fermionic CFT $T[\mathcal{C}']$, i.e., the relation~(\ref{eq:orbifold-relation}) holds in the fermionic case as well.

\section{Bosonic examples}
\label{sec:examples}
\subsection{CFTs from the $\mathbb{Z}_k$ toric code}

Here we consider the $\mathbb{Z}_k$ toric code, of which the standard $\mathbb{Z}_2$ toric code is a special case.
The low-energy theory is an abelian Chern-Simons theory with gauge group $\U(1)\times \U(1)$ with the level matrix
\begin{equation}\label{eq:Zk-toric-K}
K =
\begin{pmatrix}
~0~&~k~\\
~k~&~0~
\end{pmatrix} \,.
\end{equation}
The discriminant is $\mathcal{D} = \Gamma^*/\Gamma \simeq \mathbb{Z}_k\times\mathbb{Z}_k $ with the inner product $\oslash$ given by
\begin{equation}
(m_1,m_2)\oslash(m'_1,m'_2) 
=
\frac{m_1\, m'_2 + m_2\, m'_1}{k}
\ \text{ mod } \mathbb{Z}
\end{equation}
for $(m_1,m_2)\in \mathbb{Z}_k \times \mathbb{Z}_k$.
We parameterize the matrix $M_I{}^i$ in~(\ref{eq:KMepsilon}) as
\begin{equation}
    (M_I{}^i) 
        = 
        \sqrt{\frac{k}{2}}
            \begin{pmatrix}
            \frac{1}{r} & ~\frac{1}{r} \\
            r & -r
            \end{pmatrix}    
\end{equation}
with $r>0$.
For simplicity, we set $\nu^a = \bar\nu^a=0$.
The partition function takes the form
\begin{equation} \label{eq:Z-pL-pR}
    Z_\mathcal{C} 
        = 
            \frac{1}{|\eta(\tau)|^2} 
            \sum_{\lambda\in\Lambda(\mathcal{C})}
            q^{\frac{1}{2}p_L^2}
            \bar{q}^{\frac{1}{2}p_R^2} \,,
\end{equation}
where we write $p_L = \lambda^{a=1}$, $p_R = \lambda^{\bar{a}=1}$.\footnote{
We note that $\lambda=\lambda^I e_I=\lambda^i (M^{-1})_i{}^I e_I$, $\lambda^{a=1}=\lambda^{i=1}$, $\lambda^{\bar{a}=1} =\lambda^{i=2}$.
}

The theory has $k^2$ distinct Wilson lines
\begin{equation}
    W_{m_1,m_2} 
        = 
        \exp\left[ \i\oint \left(m_1 A^1 + m_2 A^2\right) \right]
\end{equation}
with $(m_1,m_2)\in \mathbb{Z}_k\times\mathbb{Z}_k$.
The theory has two basic topological boundary conditions:
\begin{equation}
    (\text{i}) \  A^1| 
        = 
            \text{pure gauge} \,,
            \qquad
    (\text{ii}) \ A^2|
        = 
            \text{pure gauge} \,.
\end{equation}
The boundary condition (i) implies that for any value of $m\in\mathbb{Z}_k$,  the operator $W_{m,0}$  wrapping a loop becomes the identity operator when the loop approaches and becomes coincident with the boundary.
It follows that the Wilson line $W_{m,0}$ can end on the boundary. 
(See figure~\ref{fig:topological-boundary}.)
We say that  $W_{m,0}$ condenses on the boundary.
Such a boundary condition
corresponds to the even Lagrangian subgroup 
\begin{equation}
    \mathcal{C}_\text{(i)} 
        = 
        \{\,
\Gamma, e^1+\Gamma,\ldots,(k-1)e^1+\Gamma
        \,\}\subset
        \mathcal{D} 
\,.
\end{equation}
The basis $\{e^1,e^2\}$ of $\Gamma^*$ is related to the basis $\{e_1,e_2\}$ of $\Gamma$ as $e^1=(1/k)e_2$, $e^2=(1/k)e_1$.
According to the definition~(\ref{eq:Lambda-Gamma-mathcalC}), the lattice $\Lambda(\mathcal{C}_\text{(i)})$ is given by
\begin{equation}
\Lambda(\mathcal{C}_\text{(i)}) = 
\left\{
a e_1 + \left(b+\frac{m}{k}\right)e_2 \, \middle|\, a,b,m\in\mathbb{Z} 
\right\}
=\left \{ l_2 e_1 + \frac{l_1}{k} e_2 \, \middle| \, l_1,l_2\in\mathbb{Z} \right\}
\,,
\end{equation}
i.e., $(\lambda^{I=1} ,\lambda^{I=2})= (l_2, l_1/k)$.
The vector $(p_L,p_R) = (\lambda^I M_I{}^i)$ is given by
\begin{equation}\label{toric_momenta_i}
    \text{(i):} \quad
        p_L 
            = 
            \frac{1}{\sqrt{2}} \left( l_1 \frac{r}{\sqrt{k}} + l_2 \frac{\sqrt{k}}{r} \right) \,,
        \qquad
        p_R 
            = 
            \frac{1}{\sqrt{2}} \left( - l_1 \frac{r}{\sqrt{k}} + l_2 \frac{\sqrt{k}}{r} \right) \,.
\end{equation}
The partition function is given by~(\ref{eq:Z-pL-pR}) with $(p_L,p_R)$ given by~(\ref{toric_momenta_i}).
Similarly, the boundary condition (ii) corresponds to the even Lagrangian subgroup
\begin{equation}
    \mathcal{C}_\text{(ii)} 
        = 
        \{\,
\Gamma,e^2+\Gamma,\ldots,(k-1)e^2+\Gamma
        \,\} \subset
\mathcal{D} 
\,,
\end{equation} 
which labels the Wilson lines $W_{0,m}$ that condense.
The partition function is given by~(\ref{eq:Z-pL-pR}) with
\begin{equation}
    \text{(ii):} \quad
        p_L 
            = 
            \frac{1}{\sqrt{2}} \left( l_1  \sqrt{k}\, r +  \frac{l_2}{\sqrt{k}\,r} \right) \,,
    \qquad
        p_R 
            = 
            \frac{1}{\sqrt{2}} \left( - l_1  \sqrt{k}\, r +  \frac{l_2}{\sqrt{k}\,r}  \right) \,.
\end{equation}
Thus we have
\begin{equation}
\label{eq:partition-functions-i-ii}
    \begin{aligned}
        Z_\text{(i)}
            &= 
            \sum_{m=0}^{k-1} \Psi_{(m,0)} 
                = 
                    Z\left(R=\sqrt{\frac2k}\,r\right) 
                = 
                    Z\left(R=\frac{\sqrt{2k}}{r}\right) \,,
                                    \\
        Z_\text{(ii)} 
            &= 
                \sum_{m=0}^{k-1} \Psi_{(0,m)} 
            = 
                Z\left(R=\sqrt{2k}\,r\right) 
            = 
                Z\left(R=\sqrt{\frac{2}{k}}\,\frac{1}{r}\right) \,,
    \end{aligned}
\end{equation}
where $R$ is the radius of the compact boson $X\sim X+2\pi R$ in the $\alpha'=2$ convention~\cite{Polchinski:1998rq}.
We thus see that for any value of $k$ and for either choice of the boundary conditions (i) and~(ii), the resulting Narain CFT is the compact boson CFT of some radius~$R$.
The second equality in each line of~(\ref{eq:partition-functions-i-ii}) signifies the T-duality of the compact boson CFT.
Moreover, the two Narain CFTs defined by the boundary conditions (i) and (ii) are related by gauging (orbifolding by) the $\mathbb{Z}_k$ symmetries generated by the translation $X\rightarrow X + (2\pi/k) R$ on the circle and the T-dual circle.

\subsection{CFTs from the $\U(1)_{k}\times \U(1)_{-k}$ Chern-Simons theory}

Let us consider another $\U(1)\times \U(1)$ Chern-Simons theory, this time with the level matrix
\begin{equation}
K = 
\begin{pmatrix}
~k~ & ~0~ \\
~0~ & -k~
\end{pmatrix} 
\end{equation}
with $k$ an even positive integer~\cite{Kapustin:2010hk,Kapustin:2010if}.
The discriminant is $\mathcal{D} = \Gamma^*/\Gamma \simeq \mathbb{Z}_k\times\mathbb{Z}_k $ with the inner product $\oslash$ given by
\begin{equation}
(m_1,m_2) \oslash (m'_1,m'_2) =  
\frac{m_1\, m'_1 -  m_2\, m'_2}{k}
\quad\text{ mod } \mathbb{Z}
\end{equation}
for $(m_1,m_2)\in \mathbb{Z}_k \times \mathbb{Z}_k$.
We parameterize the matrix $M_I{}^i$ in~(\ref{eq:KMepsilon}) as
\begin{equation}
    (M_I{}^i) = \frac{\sqrt{k}}{2} 
    \begin{pmatrix}
        \frac{1}{r}+r & \frac{1}{r}-r \\
        \frac{1}{r}-r & \frac{1}{r}+r
    \end{pmatrix}
\end{equation}
with $r\in\BR$.

For any even $k$, the theory has at least two topological boundary conditions:
\begin{equation}
    (\text{i}) \  (A^1+A^2)| = \text{pure gauge} \,,
\qquad
    (\text{ii}) \ (A^1-A^2)|= \text{pure gauge} \,,
\end{equation}
which correspond to the even Lagrangian subgroups\footnote{In this subsection, we regard the Lagrangian subgroup as a subgroup of $\BZ_k\times \BZ_k$ through the identification $m_1e^1+m_2e^2+\Gamma\in \CD$ with $(m_1,m_2)\in \BZ_k\times \BZ_k$.
We will use similar notation in other parts of the paper.}
\begin{align}
    \mathcal{C}_\text{(i)} 
        &= 
            \{\,(0,0),\, (1,1),\,\ldots,\,(k-1,k-1)\,\}\subset\mathbb{Z}_k\times\mathbb{Z}_k \,, \\
    \mathcal{C}_\text{(ii)} 
        &= 
            \{\,(0,0),\, (k-1,1),\,\ldots,\,(1,k-1)\,\}\subset\mathbb{Z}_k\times\mathbb{Z}_k \,.
\end{align}
The partition functions are given by~(\ref{eq:Z-pL-pR}) with
\begin{align}
    \text{(i):} & &
        p_L &= l_1 \frac{1}{\sqrt{k}\,r} + l_2 \frac{\sqrt{k}\,r}{2} \,, &
        p_R &= l_1 \frac{1}{\sqrt{k}\,r} - l_2 \frac{\sqrt{k}\,r}{2} \,, \\
    \text{(ii):} & &
        p_L &= l_1 \frac{\sqrt{k}}{2r} + l_2 \frac{r}{\sqrt{k}} \,, &
        p_R &= l_1 \frac{\sqrt{k}}{2r} - l_2 \frac{r}{\sqrt{k}} \,,
\end{align}
and thus
\begin{align}
    Z_\text{(i)} 
        &= 
            \sum_{m=0}^{k-1} \Psi_{(m,m)} = Z\left(R=\sqrt{k}\,r\right) = 
            Z\left(R=\frac{2}{\sqrt{k}\,r}\right) \,, \\
    Z_\text{(ii)} 
        &= 
            \sum_{m=0}^{k-1} \Psi_{(-m,m)} = Z\left(R=\frac{2r}{\sqrt{k}}\right) = Z\left(R=\frac{\sqrt{k}}{r}\right) \,.
\end{align}

In general, a subgroup
$\mathcal{C}\subset\BZ_{k}\times\BZ_{k}$ is an even Lagrangian subgroup with respect to $\oslash$ if and only if $\mathcal{C}$ can be written as
\begin{equation}
    \mathcal{C} = \left\{\, (m_1,m_2) \mid m_1+m_2\in 2w\, \BZ ,~ m_1-m_2\in 2v\, \BZ\, \right\} \ ,
\end{equation}
where $w,v\in\BZ$ such that $k=2wv$.
When we regard this as the topological boundary condition, from
\begin{align}
    \begin{aligned}
        \Lambda(\mathcal{C}) 
            &= 
                \{ t_1\,e^1 + t_2\,e^2 \mid (t_1,t_2)\equiv (m_1,m_2) \,\bmod k ,\, (m_1,m_2)\in\mathcal{C} \} \\
            &= 
                \{ t_1\,e^1 + t_2\,e^2 \mid t_1+t_2\in 2w\, \BZ ,\, t_1-t_2\in 2v\, \BZ \} \\
            &= 
                \left\{ \frac{1}{\sqrt{k}}\left(\frac{1}{r}\,w\,l_1 + r\,v\,l_2,\, \frac{1}{r}\,w\,l_1 - r\,v\,l_2 \right) \;\middle|\; l_1,l_2\in\BZ \right\} \,,
    \end{aligned}
\end{align}
the partition function is given with
\begin{equation}
    p_L = l_1 \frac{w}{\sqrt{k}\,r} + l_2 \frac{\sqrt{k}\,r}{2w} \,, \quad
    p_R = l_1 \frac{w}{\sqrt{k}\,r} - l_2 \frac{\sqrt{k}\,r}{2w} \,,
\end{equation}
and thus
\begin{equation}
    Z 
        = 
            \sum_{(m_1,m_2)\in\mathcal{C}} \Psi_{(m_1,m_2)} 
        = 
            Z\left(R=\frac{\sqrt{k}\,r}{w}\right) 
        = 
            Z\left(R=\frac{2w}{\sqrt{k}\,r}\right) \,.
\end{equation}
The Lagrangian subgroups $\mathcal{C}_\text{(i)}$ and $\mathcal{C}_\text{(ii)}$ correspond to $w=1,\,v=\frac{k}{2}$ and $w=\frac{k}{2},\,v=1$ respectively.
For any even Lagrangian subgroup $\mathcal{C}$ with the parameter $w$, the corresponding CFT can be obtained by the $\BZ_w$-gauging from that of $\mathcal{C}_\text{(i)}$ since the radius as the compact boson is multiplied by $\frac{1}{w}$.

When $r=1$, the matrix $M_I{}^i$ is diagonal and thus any $\Psi_{(m_1,m_2)}$ can be factorized into holomorphic and anti-holomorphic parts. This is consistent with the fact that for any $w\in\BZ$ the radius $R=\frac{\sqrt{k}}{w}$ satisfies the rationality condition $R^2\in\BQ$.

Now, we consider the orbifold theory $T[\mathcal{C}]/(\mathcal{C}' / (\mathcal{C} \cap \mathcal{C}'))$  for two even Lagrangian subgroups
\begin{align}
\begin{aligned}
    \mathcal{C} 
        &= 
            \left\{\, (m_1,m_2) \mid m_1+m_2\in 2w\, \BZ ,\, m_1-m_2\in 2v\, \BZ\, \right\} \,,\\
    \mathcal{C}' 
        &= 
            \left\{ (m_1,m_2) \mid m_1+m_2\in 2w' \,\BZ ,\, m_1-m_2\in 2v' \,\BZ \right\} \,, \\
\end{aligned}
\end{align}
where $k=2wv=2w'v'$.
The intersection of the two subgroups is
\begin{equation}
    \mathcal{C} \cap \mathcal{C}' = \left\{\, (m_1,m_2) \mid m_1+m_2\in 2w_0\, \BZ ,\, m_1-m_2\in 2v_0\, \BZ \,\right\}
\end{equation}
where $w_0=\mathrm{lcm}(w,w'),\, v_0=\mathrm{lcm}(v,v')$. 
Since there exist coprime integers $s,t$ such that $w_0=sw=tw',\, v_0=tv=sv'$, we find
\begin{align}
    \begin{aligned}
        \mathcal{C}' / (\mathcal{C} \cap \mathcal{C}')
        &\cong \BZ_t \times \BZ_s    \ ,
    \end{aligned}
\end{align}  
where the cosets $\left\{ (m_1,m_2) \mid m_1+m_2\in 2aw' + 2w_0\, \BZ ,\, m_1-m_2\in 2bv' + 2v_0\, \BZ \right\} \in \mathcal{C}' / (\mathcal{C} \cap \mathcal{C}') $ are parametrized by  $(a,b)\in \BZ_t\times \BZ_s$.

The vertex operators in the untwisted and twisted sectors can be regarded as $V_\lambda$ such that $\lambda +\Gamma \in  \mathcal{C}+\mathcal{C}'$.
It follows from
\begin{equation}
(m_1,m_2) \oslash (m'_1,m'_2) 
= \frac{1}{2k} \left( (m_1+m_2)(m_1'-m_2') + (m_1-m_2)(m_1'+m_2') \right) \text{ mod } \mathbb{Z} \,,
\end{equation}
that an element $(a,b)\in\BZ_t\times\BZ_s \cong \mathcal{C}' / (\mathcal{C} \cap \mathcal{C}')$ acts on the vertex operator $V_\lambda$ ($\lambda \in (u_1e^1+u_2e^2)+\Gamma,\, 
 (u_1,u_2)\in\mathcal{C}+\mathcal{C}'$) as
\begin{equation}
    V_\lambda \mapsto \exp\left[\frac{2\pi\i}{k}\left((u_1+u_2)bv'+(u_1-u_2)aw'\right)\right] V_\lambda \,.
\end{equation}
Since $V_\lambda$ is invariant for all $(a,b)\in\BZ_t\times\BZ_s$ if and only if $(u_1,u_2)\in\mathcal{C}'$, we can conclude that
\begin{equation}
    \frac{T[\mathcal{C}]}{\mathcal{C}' / (\mathcal{C} \cap \mathcal{C}')} = T[\mathcal{C'}] \,.
\end{equation}

\subsection{CFTs from the $\U(1)_{k_1}\times \U(1)_{-k_2}$ Chern-Simons theory}

As a generalization of the previous example, let us consider the $\U(1)\times \U(1)$ Chern-Simons theory with the level matrix
\begin{equation}
    K = 
    \begin{pmatrix}
        k_1 & 0 \\
        0 & -k_2
    \end{pmatrix} \,,
\end{equation}
where $k_1$ and $k_2$ are even positive integers.
We parameterize the matrix $M_I{}^i$ as
\begin{equation}
    (M_I{}^i) = \frac{1}{2} 
    \begin{pmatrix}
        \sqrt{k_1}(\frac{1}{r}+r) & \sqrt{k_1}(\frac{1}{r}-r) \\
        \sqrt{k_2}(\frac{1}{r}-r) & \sqrt{k_2}(\frac{1}{r}+r)
    \end{pmatrix}
\end{equation}
with $r\in\BR$.
A subgroup $\mathcal{C}\subset\BZ_{k_1}\times\BZ_{k_2}$ is even and Lagrangian with respect to the inner product $\oslash$ given by
\begin{equation}
    (m_1,m_2) \oslash (m_1',m_2') = \frac{1}{k_1}m_1m_1' - \frac{1}{k_2} m_2m_2' \mod \BZ
\end{equation}
if and only if $k_1k_2$ is a square number and $\mathcal{C}$ can be written as
\begin{equation}
   \mathcal{C} = \left\{ (m_1,m_2) \;\middle|\; \frac{m_1}{u_1}+\frac{m_2}{u_2}\in 2w \BZ ,\, \frac{m_1}{u_1}-\frac{m_2}{u_2}\in 2v \BZ \right\}
\end{equation}
where $2wv=\gcd(k_1,k_2)=:k$, $k_1=ku_1^2,\, k_2=ku_2^2$.
We will prove this claim below.
Since $u_1\frac{1}{\sqrt{k_1}}=u_2\frac{1}{\sqrt{k_2}}=\frac{1}{\sqrt{k}}$, the lattice constructed from such a code is the same as the case $\U(1)_{k}\times \U(1)_{-k}$ with the same parameters $w,v$. Thus, the partition function is
\begin{equation}
    Z = \sum_{(m_1,m_2)\in\mathcal{C}} \Psi_{(m_1,m_2)} = Z\left(R=\frac{\sqrt{k}r}{w}\right) = Z\left(R=\frac{2w}{\sqrt{k}r}\right) \,.
\end{equation}

Let us prove the statement made below.
We assume that $\mathcal{C}\subset\BZ_{k_1}\times\BZ_{k_2}$ is a Lagrangian subgroup. 
Since generally the number of elements in a Lagrangian subgroup is the square root of that in the discriminant group, $k_1 k_2$ must be a square number. In that case, letting $k = \gcd(k_1, k_2)$, both $k_1/k$ and $k_2/k$ are also square numbers, as $k_1 k_2/k^2$ is a square and $k_1/k$ and $k_2/k$ share no common prime factors by the definition of $k$. Then, by using $u_i=\sqrt{k_i/k}\in\BZ$, the inner product with itself can be written as
\begin{equation}
\label{eq:m1m2-self}
    (m_1,m_2) \oslash (m_1,m_2) = \frac{1}{k} \left( \frac{m_1^2}{u_1^2} - \frac{m_2^2}{u_2^2} \right) \,,
\end{equation}
which can be integer only if $m_1^2/u_1^2$ and $m_2^2/u_2^2$ are both integers since $u_1$ and $u_2$ are coprime. Thus, the element of $\mathcal{C}$ can be written as $(m_1,m_2) = (l_1 u_1, l_2 u_2)$ where $l_i\in\BZ_{ku_i}$ and the inner product with itself becomes
\begin{equation}
\label{eq:m1m2-l1l2}
    (m_1,m_2) \oslash (m_1,m_2) = \frac{1}{k}(l_1+l_2)(l_1-l_2) \,.
\end{equation}
Since $k$ is even and $l_1+l_2$ and $l_1-l_2$ have the same parity, $l_1+l_2$ and $l_1-l_2$ must both be even for (\ref{eq:m1m2-l1l2}) to be an integer. Therefore, we can parametrize the element of $\mathcal{C}$ by $t_\pm\in\BZ$ with $t_\pm=(l_1\pm l_2)/2$ and the inner product is
\begin{equation}
    (m_1,m_2) \oslash (m_1',m_2') = \frac{2}{k} (t_+t_-' + t_-t_+') \,.
\end{equation}
Since it is evident from this expression that $(t_+,t_-)=(k/2,0),(0,k/2)$ are orthogonal to all elements, there exists a Lagrangian subgroup $\mathcal{C}' \subset (\BZ_{k/2})^2$ with the inner product
\begin{equation}
    (s_+,s_-) \circ (s_+',s_-') = \frac{2}{k} (s_+s_-' + s_-s_+')
\end{equation}
and $\mathcal{C}$ can be written as
\begin{align}
\begin{aligned}
    \mathcal{C} &= \left\{ ((t_++t_-)u_1, (t_+-t_-)u_2) \mid \exists (s_+,s_-)\in\mathcal{C}',\, (t_+,t_-) \equiv (s_+,s_-) \mod k/2  \right\} \\
    &= \left\{ (m_1,m_2) \;\middle|\; \exists (s_+,s_-)\in\mathcal{C}',\, \left(\frac{m_1}{u_1}+\frac{m_2}{u_2},\frac{m_1}{u_1}-\frac{m_2}{u_2}\right) \equiv 2 (s_+,s_-) \mod k  \right\} 
\end{aligned}
\end{align}

Let $w\in\BZ_{k/2}$ be the smallest nonzero $s_+$ among $(s_+,s_-)\in\mathcal{C}'$. By the linearity of $\mathcal{C}'$, $w$ is a divisor of $k/2$ and any $s_+$ is a multiple of $w$. Then, $(0,v)\in\mathcal{C}'$ where $v=k/(2w)\in\BZ$ from $\mathcal{C}'^\perp \subset \mathcal{C}'$, and if $(0,s_-)\in\mathcal{C}'$ then $s_-$ is a multiple of $v$ from $\mathcal{C}' \subset \mathcal{C}'^\perp$. Thus, there exists $m\in\BZ_{k/2}$ such that $\mathcal{C}' = \{ a(w,m) + b(0,v) \mid a,b\in\BZ\}$. By the orthogonality of $(w,m)$, $m\in(v/2)\BZ$ and then $m$ can be reduced to either $0$ or $v/2$ (if $v$ is even), i.e.,
\begin{equation}
    \mathcal{C}' = \{ (aw, bv) \mid a\in\BZ_v,\, b\in\BZ_w \}
\end{equation}
where $k/2 = wv$ or
\begin{align} \label{eq:odd_Lag_leng2}
\begin{aligned}
    \mathcal{C}' &= \{ (aw, (a+2b)v') \mid a\in\BZ_{2v'},\, b\in\BZ_w \} \\
    &= \{ (2xw, 2yv') \mid x\in\BZ_{v'},\, y\in\BZ_w \} \sqcup \{ ((2x+1)w, (2y+1)v') \mid x\in\BZ_{v'},\, y\in\BZ_w \} 
\end{aligned}
\end{align}
where $k/2 = 2wv'$ (this can be taken only if $k_1,k_2\in4\BZ$). 
It is easy to verify that the former is even and the latter is odd.
When we revert the former expression to $\mathcal{C}$,
\begin{equation}
    \mathcal{C} = \left\{ (m_1,m_2) \;\middle|\; \frac{m_1}{u_1}+\frac{m_2}{u_2} \in 2w\BZ,\, \frac{m_1}{u_1}-\frac{m_2}{u_2} \in 2v\BZ \right\} 
\end{equation}
where $k=2wv$.

\subsection{Examples with discrete torsion}
\label{eq:ex-dis-tor}

Here we provide explicit examples of discrete torsion discussed in section~\ref{sec:orb-bos}.
Let us consider the $\U(1)^4$ Chern-Simons theory with level matrix
\begin{equation}
    K
        = 
        \begin{pmatrix}
            ~0~&~0~&~k~&~0~\\
            ~0~&~0~&~0~&~k~\\
            ~k~&~0~&~0~&~0~\\
            ~0~&~k~&~0~&~0~
        \end{pmatrix}  \,.
\end{equation}
Although the bulk Chern-Simons theory is the product of two decoupled theories each described by the level matrix~(\ref{eq:Zk-toric-K}), some  boundary conditions will lead to a 2d CFT that is not the product of two decoupled CFTs.
We denote the basis of $\Gamma^*$ by $\{e^I\}$ and that of $\Gamma$ by $\{e_I\}$ ($I=1,2,3,4$), such that $K_{IJ} = e_I\oslash e_J$.
The discriminant group is $\mathcal{D}=\Gamma^*/\Gamma\simeq (\mathbb{Z}_k)^4$, with the inner product
\begin{equation}
(m_1,\ldots,m_4)\oslash(m'_1,\ldots,m'_4) = 
\frac{m_1 m'_3 +m_2 m'_4 + m_3 m'_1 + m_4 m'_2}{k} 
\ \text{ mod } \mathbb{Z}     
\end{equation}
for $m_i,m'_i\in\{0,1,\ldots,k-1 \text{ mod } k\}$.

Let us consider the even Lagrangian subgroup
\begin{equation}
\mathcal{C} := \{ (m_1,m_2,0,0) \, |
\,
m_1,m_2 =0,1,\ldots, k-1 \}
\subset (\mathbb{Z}_k)^4
\end{equation}
and the corresponding CFT $T[\mathcal{C}]$.
Recall the projection $\varphi: \mathcal{D}\rightarrow \mathcal{D}/\mathcal{C}$.
Let us take another even Lagrangian subgroup
\begin{equation}
\mathcal{C}' := \{ (0,0,m_3,m_4) \, |
\,
m_3,m_4 =0,1,\ldots, k-1 \}
\subset (\mathbb{Z}_k)^4 \,.
\end{equation}
Note that $\mathcal{C}\cap\mathcal{C}'=\{0\}$.
We are interested in orbifolding $T[\mathcal{C}]$ by $H:=\varphi(\mathcal{C}')\simeq (\mathbb{Z}_k)^2$.
The most natural orbifolding action (\ref{eq:orb-action-C'}) leads to the CFT $T[\mathcal{C}']$ as we saw in section~\ref{sec:orb-bos}.
As is well known~\cite{Vafa:1986wx}, possible choices of discrete torsion are labeled by the second group cohomology~$H^2\left(H,\U(1)\right)$, which is in this case isomorphic to $\mathbb{Z}_k$.
A cocycle $\omega: H\times H\rightarrow \U(1)$ that represents the generator of the cohomology is given by $\omega\left(\,(a_1,b_1),\,(a_2,b_2)\,\right) = \exp\left( \frac{2\pi\i}{k}\, b_1 a_2\right)$.%
\footnote{
This is the phase that appears in the projective representation of $\mathbb{Z}_k\times\mathbb{Z}_k$ 
by the $k$-dimensional qudit generalizations $X$ and $Z$ of the Pauli matrices $\sigma_x$ and~$\sigma_z$, $X^{a_1}Z^{b_1} X^{a_2} Z^{b_2} = \omega\left(\,(a_1,b_1),\,(a_2,b_2)\,\right)   X^{a_1+a_2} Z^{b_1+b_2}$.
}
We expect to find $k$ even Lagrangian subgroups $\mathcal{C}_{s}$ corresponding to the cocycles $\omega_s:=\omega^s$ for $s=0,1,\ldots, k-1$, such that $\mathcal{C}_{0}=\mathcal{C}'$.
Indeed, taking the even Lagrangian subgroup
\begin{equation}
\mathcal{C}_{s} 
:= \{ (sm',-s m,m,m') \, | \,
m,m'=0,1,\ldots,k-1 \} \subset (\mathbb{Z}_k)^4
\end{equation}
 as $\mathcal{C}''$ in section~\ref{sec:orb-bos} gives the discrete torsion phase, via (\ref{eq:disc-tor-rewite}), as the ratio of cycles, 
\begin{align}
    \begin{aligned}
    \epsilon_s\left(\,(a_1,b_1),\,(a_2,b_2)\,\right) 
        &= 
        \exp\left( 
        \frac{2\pi\i }{k} \,
        (s b_1,-s a_1,a_1,b_1)\oslash (0,0,a_2,b_2)
        \right)
        \\
        &=
            \frac{ \omega_s\left(\,(a_1,b_1),\,(a_2,b_2)\,\right) }{ \omega_s\left(\,(a_2,b_2),\,(a_1,b_1)\,\right) } \,,
    \end{aligned}
\end{align}
 which is the standard relation between the discrete torsion and the cocycle~\cite{Vafa:1986wx}.

\subsection{Chiral code CFTs}

In this subsection, we revisit the well-studied construction of the chiral CFTs through Euclidean lattices from classical codes~~\cite{frenkel1984natural,frenkel1989vertex,Dolan:1994st,Gaiotto:2018ypj,Kawabata:2023nlt,Kawabata:2023rlt,Kawabata:2024gek,Okada:2024imk}, which can be regarded as a special case of our construction based on Chern-Simons theory.

To see this, let us consider the $n\times n$ level matrix~$K=(K_{IJ})$ given by
\begin{equation}
  K_{IJ} = k\, \delta_{IJ} \,,
\end{equation}
where $k$ is an even positive integer.
In our construction, the lattice~$\Gamma$ is spanned by a basis $\{e_I\}$ with an inner product $\oslash$ given by $e_I\oslash e_J = k\, \delta_{IJ}$.
Choosing $x = (x_1,\ldots,x_n) \in \mathbb{R}^n$ as the coordinates of the vector space spanned by $\{e_I\}$ such that $e_I = (\sqrt{k}\,\delta_I^i)_{i=1}^n \in \mathbb{R}^{n}$, the inner product is simply
\begin{equation}
x \oslash x' 
= 
\sum_{i=1}^n x_i\, x'_i   \,.
\end{equation}
(There is no factor of $1/k$ here because we have absorbed $\sqrt{k}$ in $e_I$.)
As lattices embedded in the real vector space~$\mathbb{R}^n$, we have
\begin{equation}
\Gamma = \sqrt{k}\, \mathbb{Z}^{n}\,,
\qquad
\Gamma^* = \frac{1}{\sqrt{k}}\, \mathbb{Z}^{n}\,.
\end{equation}
A topological boundary condition is specified by a Lagrangian subgroup~$\mathcal{C} $ of $\mathcal{D} = \Gamma^*/\Gamma = (\mathbb{Z}_k)^n$.
In the literature on code CFTs, such $\mathcal{C}$ is called a self-dual code.
Then the momentum lattice
\begin{equation}\label{eq:Lambda-Gamma-C}
    \Lambda(\mathcal{C}) = \left\{\frac{c}{\sqrt{k}}+\sqrt k\, m \, \bigg|\, \frac{c}{\sqrt{k}}+\Gamma \in\mathcal{C}  \,,m\in\mathbb{Z}^n     \right\}
\end{equation}
is nothing but the Construction A lattice for the self-dual code $\mathcal{C}$.
The self-dual lattice $\Lambda(\mathcal{C}) $ is even if and only if the Lagrangian subgroup $\mathcal{C}$ is even in the sense defined below~(\ref{eq:evenness-condition}).
The resulting CFT is a bosonic chiral Narain code CFT studied, for example, in~\cite{Dolan:1994st} for $k=2$.

\subsection{Non-chiral code CFTs}
\label{sec:non-spin-CS-non-chiral-bosonic-CFT}

Over the last few years, there has been much study of the non-chiral CFTs constructed through Lorentzian lattices~\cite{Dymarsky:2020bps,Dymarsky:2020qom,Dymarsky:2020pzc,Dymarsky:2021xfc,Henriksson:2021qkt,Buican:2021uyp,Yahagi:2022idq,Furuta:2022ykh,Henriksson:2022dnu,Angelinos:2022umf,Henriksson:2022dml,Dymarsky:2022kwb,Kawabata:2022jxt,Furuta:2023xwl,Alam:2023qac,Kawabata:2023usr,Kawabata:2023iss,Aharony:2023zit,Barbar:2023ncl,Singh:2023mom,Ando:2024gcf,Singh:2024qjm,Mizoguchi:2024ahp} from classical and quantum codes.
Such CFTs can also be regarded as a special case of the Chern-Simons construction, as already pointed out in reference~\cite{Kawabata:2023iss}.

The case of non-chiral code CFTs corresponds to the choice
\begin{equation}
    (K_{IJ}) = 
    \begin{pmatrix}
        0 & k\, 1_{n\times n} \\
        k\, 1_{n\times n} & 0
    \end{pmatrix} \,.
\end{equation}
Choosing $x = (x_1,\ldots,x_{2n}) \in \mathbb{R}^{2n}$ as the coordinates of the vector space spanned by $\{e_I\}$ such that $e_I = (\sqrt{k}\,\delta_I^i)_{i=1}^{2n} \in \mathbb{R}^{2n}$, the inner product is given by
\begin{equation}
x \oslash x' 
        = 
        \sum_{i=1}^{2n} (x_i\, x'_{n+i} + x_{n+i}\, x'_i) \,.
\end{equation}
for $ x ,x' \in \mathbb{R}^{2n}$.
As lattices embedded in $\mathbb{R}^{2n}$, we have
\begin{equation}
\Gamma = \sqrt{k}\, \mathbb{Z}^{2n}\,,
\qquad
\Gamma^* = \frac{1}{\sqrt{k}}\, \mathbb{Z}^{2n}
\,.
\end{equation}
A topological boundary condition is specified by a Lagrangian subgroup (self-dual code) $\mathcal{C} $.
The momentum lattice
\begin{equation}
    \Lambda(\mathcal{C}) 
        = 
        \left\{
        \frac{c}{\sqrt{k}}+
        \sqrt k\, m \, \bigg|\, \frac{c}{\sqrt{k}} +\Gamma \in\mathcal{C} \,,m\in\mathbb{Z}^{2n}     \right\} \,,
\end{equation}
obtained again by Construction A, gives rise to a non-chiral Narain code CFT.
We assume that $\mathcal{C}$ is even so that the lattice~$\Lambda(\mathcal{C})$ is even.
The resulting non-chiral Narain code CFT is bosonic.
Such CFTs were introduced in~\cite{Dymarsky:2020bps,Dymarsky:2020qom} (for $k=2$) and studied in subsequent works for $k\geq 3$.

\section{Fermionic code CFTs}
\label{sec:fermionic-code-CFTs}

In this section, we extend the construction of bosonic CFTs from even self-dual codes in \cite{Dymarsky:2020qom,Kawabata:2022jxt} to that of fermionic CFTs from odd self-dual codes.
The construction of fermionic CFTs in the present section is a special case of the one given in section \ref{sec:ferm-cft}.%
\footnote{\label{footnote:relation-to-sec-general-1}%
We identify  $\mathbb{Z}_{2k}^{2n}$ with the discriminant group $\mathcal{D}=\Gamma^*/\Gamma$ and a self-dual code $C$ with a Lagrangian subgroup $\mathcal{C}$.
The inner product $c\odot c'$ defined mod $2k$ is related to the one $c\oslash c'$ defined mod $\mathbb{Z}$  as $ c \odot c' = 2k\, c\oslash c' $.
}
However, this section is written independently of the previous one, making it more accessible to those who are familiar with the bosonic counterpart.

In section \ref{ss:odd_sd_codes_shadows}, we consider odd self-dual codes over $\BZ_{2k}$ and introduce their shadows, specializing the discussion in section~\ref{sec:shadow-lagrangian}.
Then, in section \ref{ss:conA_odd}, we define the Construction A lattices for the codes and shadows and identify them as the momentum lattices for the vertex operators in fermionic CFTs.
In section \ref{ss:torus_PF_ferm}, we express the torus partition function of the fermionic CFTs by using the weight enumerator polynomials of the associated codes.
In section \ref{ss:example_ferm_CFT}, we apply our construction to an odd self-dual code of length two and derive $c=1$ fermionic CFTs, which turn out to be the massless Thirring model with the coupling constant tuned to particular values.

\subsection{Odd self-dual codes over $\BZ_{2k}$ and their shadows}\label{ss:odd_sd_codes_shadows}

In what follows, we consider a code $C$ of length $2n$
over $\BZ_{2k}$, i.e., an additive subgroup $C$ of $\BZ_{2k}^{2n}$.
We assume that $C$ is equipped with an inner product~$c\odot c' =  h_{ij} c_i c_j$, where $h=(h_{ij})$ is a symmetric integral matrix with determinant $\det h =\pm 1$ and $c=(c_i), c'= (c'_i) \in \BZ_{2k}^{2n}$.
The dual $C^\perp$ of $C$ with respect to the inner product $\odot$ over $\BZ_{2k}$ is defined by
\begin{align}
    C^\perp 
        :=
            \left\{\, c \in \BZ_{2k}^{2n}\, \big|\, c\odot c' \in 2k\,\BZ \quad {}\forall c'\in C\, \right\} \ .
\end{align}
A code is \emph{self-orthogonal} if $C \subset C^\perp$ and \emph{self-dual} if $C = C^\perp$.
Moreover, we call a code $C$ \emph{even} if all codewords satisfy $c\odot c \in 4k\,\BZ$.
Otherwise, we call it \emph{odd}.\footnote{Even and odd self-dual codes in our terminology are called Type II and Type I in \cite{bannai1999type}, respectively.}

Let $C$ be an odd self-dual code of length $2n$ over $\BZ_{2k}$ with respect to an inner product~$\odot$.
We then define the subset $C_0$ and $C_2$ as follows:
\begin{align}
    C_0
        &:=
            \left\{\, c\in C\,\big|\, c\odot c \in 4k\,\BZ \,\right\} \ , \\
    C_2
        &:=
            \left\{\, c\in C\,\big|\, c\odot c \in 2k(2\BZ +1) \, \right\} \ . 
\end{align}
Note that $C_0$ is closed under addition and an even self-orthogonal code by itself while $C_2$ is not closed and does not form a code.

Let us choose an element $s$ in $C_0^\perp\backslash C$.
We claim that
\begin{equation}
\label{eq:ccs-claim}
c \odot c = c \odot (2s) \text{ mod } 4 k \mathbb{Z} 
\end{equation}
for $c\in C$.
This can be shown following the discussion below~\eqref{eq:shadow_grad}.
By~(\ref{eq:ccs-claim}), we can use $s$ to characterize $C_0$ and $C_2$ as
\begin{align}\label{C0_C2}
    \begin{aligned}
        C_0
        &=
            \left\{\, c\in C\,\big|\, c\odot (2s) \in 4k\,\BZ \,\right\} \ , \\
        C_2
            &=
                \left\{\, c\in C\,\big|\, c\odot (2s) \in 2k\,(2\BZ +1) \, \right\} \ .
    \end{aligned} 
\end{align}

Let $t$ be an element of $C_2$.
We can write the decomposition of the code $C$ as
\begin{align}
    C = C_0 \cup (C_0 + t) \ .
\end{align}
The shadow $S$ of $C$ is defined by $S = C + s$ \cite{dougherty2000shadow,dougherty2001shadow,bonnecaze2003splitting,dougherty2003generalized}, or equivalently
\begin{align}
    S = C_1 \cup C_3 \ ,
\end{align}
where
\begin{align}
    C_1
        &:=
            C_0 + s \ , \\
    C_3
        &:=
            C_0 + s + t \ .
\end{align}
One can show $S= C_0^\perp\backslash C$ as follows.
The definition of $S$ implies $S\subset C_0^\perp\backslash C$.
To show the converse $S\supset C_0^\perp\backslash C$, suppose $s' \in C_0^\perp\backslash C$. 
We would like to show $\delta s: = s' - s \in C$.
To this end, we calculate the inner product between $\delta s$ and $c\in C$:
\begin{align}
    \delta s \odot c
        =
            (s' - s) \odot c
        =
            s' \odot c - s \odot c
        \in 2k \BZ \ ,
\end{align}
where we used the fact that \eqref{C0_C2} holds both for $s$ and $s'$.
Thus, we find $\delta s \in C^\perp = C$, which implies $s' \in C + s$ for any $s' \in C_0^\perp\backslash C$.

The definition of the shadow $S$ here is closely related to the one for lattices in section \ref{sec:ferm-cft} as we will see in the next subsection.

\subsection{Construction A lattices}\label{ss:conA_odd}
For a code $C$ of length $2n$ over $\BZ_{2k}$, we define the Construction A lattice $\Lambda(C)$ as%
\footnote{
To relate to section~\ref{sec:general}, 
we take the basis vectors of $\Gamma$ to be $e_I=(\sqrt{2k}\delta_{Ii})_{i=1}^{2n}$ so that $K_{IJ} = e_I\oslash e_J = 2k h_{IJ}$.  For lattice vectors $\lambda,\lambda'$, we have $\lambda\oslash\lambda'=\lambda\odot\lambda'= h_{ij} \lambda_i\lambda_j$.
}
\begin{align}
    \Lambda(C)
        :=
            \left\{\,
                \frac{c + 2k\,m}{\sqrt{2k}} ~ \bigg|~ c\in C\ , ~ m\in \BZ^{2n}
            \,\right\} \ .
\end{align}\\
For any pair of lattice vectors $\lambda, \lambda'$, their inner product is 
\begin{align}
    \begin{aligned}
        \lambda\odot \lambda'
            &=
                \frac{c + 2k\,m}{\sqrt{2k}} \odot  \frac{c' + 2k\,m'}{\sqrt{2k}} \\
            &=
                \frac{c\odot c'}{2k} + c\odot m' + m\odot c' + 2k\,m\odot m' \,.
    \end{aligned}
\end{align}
Thus $\Lambda(C)$ is an integral lattice if $C$ is a self-orthogonal code.
Furthermore, by using the assumption $\det h = \pm 1$, one can show that $\Lambda(C)$ is self-dual if and only if $C$ is self-dual.
We note that $\Lambda(C)$ is an odd self-dual lattice if and only if $C$ is odd self-dual.

Next, consider a lattice vector $\chi$ that corresponds to the codeword $2s\in C$ used to define the shadow $S$ of $C$ via the Construction A:
\begin{align}\label{choice_of_characteristic_vector}
    \chi
        :=
            \frac{2s}{\sqrt{2k}} \ .
\end{align}
Then, for any $\lambda = \dfrac{c + 2k\,m}{\sqrt{2k}}\in \Lambda(C)$,
\begin{align}
    \begin{aligned}
        \lambda \odot \lambda
            &=
                \frac{c\odot c}{2k} + 2\, c\odot m + 2k\, m\odot m \\
            &=
                \begin{cases}
                    ~0 & \quad c\in C_0 \\
                    ~1 & \quad c\in C_2 
                \end{cases}
                \qquad (\text{mod}~2) \ .
    \end{aligned}
\end{align}
On the other hand, it follows from \eqref{C0_C2} that
\begin{align}
    \begin{aligned}
        \chi \odot \lambda
            &=
                \frac{c\odot (2s)}{2k} + (2s)\odot m \\
            &=
                \begin{cases}
                    ~0 & \quad c\in C_0 \\
                    ~1 & \quad c\in C_2 
                \end{cases}
                \qquad (\text{mod}~2) \ ,
    \end{aligned}
\end{align}
where we used the fact that $s$ is a vector over $\BZ_{2k}$.
Hence, we show that any choice of $\chi$ satisfies the relation:
\begin{align}\label{characteristic_vector_def}
    \chi \odot \lambda
        =
        \lambda \odot \lambda
        \quad
        \text{mod}~2 \qquad {}^\forall \lambda \in \Lambda(C)\ .
\end{align}
A lattice vector $\chi$ satisfying \eqref{characteristic_vector_def} is called a characteristic vector, which has been already defined for a general lattice in \eqref{eq:characteristic-vector}.

The shadow of the lattice can be defined in a similar manner to that of the code as follows.
First, we introduce the sets $\Lambda_i~(i=0,1,2,3)$ by
\begin{align}
    \Lambda_i
        :=
            \Lambda(C_i)
            \qquad (i=0,1,2,3) \ .
\end{align}
Clearly, $\Lambda_0$ and $\Lambda_2$ can be written as
\begin{align}
    \begin{aligned}
        \Lambda_0
            &=
                \left\{\, \lambda\in \Lambda(C)\, \big|\, \chi\odot \lambda \in 2\BZ\,\right\} \ , \\
        \Lambda_2
            &=
                \left\{\, \lambda\in \Lambda(C)\, \big|\, \chi\odot \lambda \in 2\BZ+1\,\right\} \ .
    \end{aligned}
\end{align}
On the other hand, $\Lambda_1$ and $\Lambda_3$ are expressed as
\begin{align}
    \begin{aligned}
        \Lambda_1
            &=
                \Lambda_0 + \frac{\chi}{2} \ , \\
        \Lambda_3
            &=
                \Lambda_2 + \frac{\chi}{2} \ .
    \end{aligned}
\end{align}

\subsection{Torus partition functions}\label{ss:torus_PF_ferm}
The sets $\Lambda_i$ specify the spectrum of the vertex operators in the fermionic CFT constructed from the odd self-dual code $C$ once we set $\odot$ to be the off-diagonal Lorentzian inner product defined by
\begin{align}
    \lambda \odot \lambda'
        :=
            \sum_{i=1}^n \left(\lambda_i\,
            \lambda_{n+i}' + \lambda_{n+i}\,
            \lambda_{i}'\right) \ .
\end{align}

Let $\tilde\Lambda$ be the momentum lattice associated with $\Lambda$ by
\begin{align}
    \tilde\Lambda
        :=
            \left\{\, (p_L, p_R) = \left( \frac{\lambda_1 + \lambda_2}{\sqrt{2}},\, \frac{\lambda_1 - \lambda_2}{\sqrt{2}}\right) \, \bigg|\, (\lambda_1, \lambda_2) \in \Lambda\,\right\} \ .
\end{align}
The torus partition functions \eqref{eq:ZC0011} in the fermionic CFTs can be expressed as follows:
\begin{align}
    Z_{\text{NS}}
        &:=
            \frac{1}{|\eta(\tau)|^{2n}}\,\left[ \Theta_{\tilde\Lambda_0}(\tau, \bar\tau) + \Theta_{\tilde\Lambda_2}(\tau, \bar\tau)\right]\ , \\
    Z_{\widetilde{\text{NS}}}
        &:=
            \frac{1}{|\eta(\tau)|^{2n}}\,\left[ \Theta_{\tilde\Lambda_0}(\tau, \bar\tau) - \Theta_{\tilde\Lambda_2}(\tau, \bar\tau)\right] \ , \\
    Z_{\text{R}}
        &:=
            \frac{1}{|\eta(\tau)|^{2n}}\,\left[\Theta_{\tilde\Lambda_1}(\tau, \bar\tau) + \Theta_{\tilde\Lambda_3}(\tau, \bar\tau) \right]\ , \\
    Z_{\widetilde{\text{R}}}
        &:=
            \frac{1}{|\eta(\tau)|^{2n}}\,\left[\Theta_{\tilde\Lambda_1}(\tau, \bar\tau) - \Theta_{\tilde\Lambda_3}(\tau, \bar\tau) \right] \ ,
\end{align}
where $\Theta_{\tilde\Lambda}(\tau, \bar\tau)$ is the lattice theta function
\begin{align}
    \Theta_{\tilde\Lambda}(\tau, \bar\tau)
        :=
            \sum_{(p_L, p_R)\in \Tilde\Lambda} q^{\frac{p_L^2}{2}}\,\bar q^{\frac{p_R^2}{2}} \ ,
\end{align}
where $p_{L}^2 := \sum_{i=1}^n (p_{L\,i})^2$ and similarly for $p_R^2$.

Note that these lattice theta functions satisfy
\begin{align}
    \Theta_{\tilde\Lambda_0}(\tau, \bar\tau)
        +
        \Theta_{\tilde\Lambda_2}(\tau, \bar\tau)
            &=
               \sum_{(\lambda_1, \lambda_2)\in \Lambda(C)} q^{\frac{1}{4}(\lambda_1 + \lambda_2)^2}\,\bar q^{\frac{1}{4}(\lambda_1 - \lambda_2)^2}  \ , \\
    \Theta_{\tilde\Lambda_0}(\tau, \bar\tau)
        -
        \Theta_{\tilde\Lambda_2}(\tau, \bar\tau)
            &=
               \sum_{(\lambda_1, \lambda_2)\in \Lambda(C)} (-1)^{\chi\odot\lambda}\,q^{\frac{1}{4}(\lambda_1 + \lambda_2)^2}\,\bar q^{\frac{1}{4}(\lambda_1 - \lambda_2)^2}   \ ,           
\end{align}
and
\begin{align}
    \Theta_{\tilde\Lambda_1}(\tau, \bar\tau)
        +
        \Theta_{\tilde\Lambda_3}(\tau, \bar\tau)
            &=
               \sum_{(\lambda_1, \lambda_2)\in \Lambda(C)} q^{\frac{1}{4}\left(\lambda_1 + \lambda_2 + \frac{\chi_1 + \chi_2}{2}\right)^2}\,\bar q^{\frac{1}{4}\left(\lambda_1 - \lambda_2 + \frac{\chi_1 - \chi_2}{2}\right)^2}   \ , \\
    \Theta_{\tilde\Lambda_1}(\tau, \bar\tau)
        -
        \Theta_{\tilde\Lambda_3}(\tau, \bar\tau)
            &=
               \sum_{(\lambda_1, \lambda_2)\in \Lambda(C)} (-1)^{\chi\odot\lambda}\,q^{\frac{1}{4}\left(\lambda_1 + \lambda_2 + \frac{\chi_1 + \chi_2}{2}\right)^2}\,\bar q^{\frac{1}{4}\left(\lambda_1 - \lambda_2 + \frac{\chi_1 - \chi_2}{2}\right)^2}   \ .         
\end{align}
It follows from these expressions with the choice of the characteristic vector \eqref{choice_of_characteristic_vector} that the lattice theta function can be written as
\begin{align}
    \Theta_{\tilde\Lambda_0}(\tau, \bar\tau)
        &=
            \frac{1}{2}\left[ W_C\left( \left\{ \psi_{\alpha_i\beta_i}^+\right\}\right) + W_C\left( \left\{ \psi_{\alpha_i\beta_i}^{-, \,(s^{(i)})}\right\}\right)\right] \ , \\
    \Theta_{\tilde\Lambda_2}(\tau, \bar\tau)
        &=
            \frac{1}{2}\left[ W_C\left( \left\{ \psi_{\alpha_i\beta_i}^+\right\}\right) - W_C\left( \left\{ \psi_{\alpha_i\beta_i}^{-, \,(s^{(i)})}\right\}\right)\right] \ , \\
    \Theta_{\tilde\Lambda_1}(\tau, \bar\tau)
        &=
            \frac{1}{2}\left[ W_C\left( \left\{ \tilde\psi_{\alpha_i\beta_i}^{+,(s^{(i)})}\right\}\right) + W_C\left( \left\{ \tilde\psi_{\alpha_i\beta_i}^{-, \,(s^{(i)})}\right\}\right)\right] \ , \\
    \Theta_{\tilde\Lambda_3}(\tau, \bar\tau)
        &=
            \frac{1}{2}\left[ W_C\left( \left\{ \tilde\psi_{\alpha_i\beta_i}^{+,(s^{(i)})}\right\}\right) - W_C\left( \left\{ \tilde\psi_{\alpha_i\beta_i}^{-, \,(s^{(i)})}\right\}\right)\right] \ ,
\end{align}
where we denote the pair of $i$-th and $(n+i)$-th components of a $2n$-dimensional vector $v=(v_1, v_2)$ by $v^{(i)} := (v_{1i}, v_{2i})$ and $W_C$ be the full enumerator polynomial of a code $C$:
\begin{align}
    W_C\left( \left\{ x_{\alpha_i\beta_i}^{(i)} \right\}\right)
        :=
            \sum_{(\alpha,\beta)\in C}\,\prod_{i=1}^n x_{\alpha_i\beta_i}^{(i)}  \ .
\end{align}
For a two-dimensional vector $u=(u_1, u_2)$, we also introduce the functions as follows:
\begin{align}
    \psi_{ab}^+
        &:=
            \sum_{m_1,m_2\in\BZ}
                q^{\frac{k}{2}\left(\frac{a+b}{2k} + m_1 + m_2\right)^2}\,\bar q^{\frac{k}{2}\left(\frac{a-b}{2k} + m_1 - m_2\right)^2} \nonumber\\
        &\,=
            \Theta_{a+b,2k}
            \,\bar\Theta_{a-b,2k}
            + \Theta_{a+b-2k,2k}
            \,\bar\Theta_{a-b-2k,2k}
                \ , \\
    
    \psi_{ab}^{-,\, (u)}
        &:=
            \sum_{m_1,m_2\in\BZ}
                (-1)^{\frac{1}{k}(u_{2}a + u_1b)}\,q^{\frac{k}{2}\left(\frac{a+b}{2k} + m_1 + m_2\right)^2}\,\bar q^{\frac{k}{2}\left(\frac{a-b}{2k} + m_1 - m_2\right)^2} \nonumber\\
        &\,=
            (-1)^{\frac{1}{k}(u_{2}a + u_1b)}\,\psi_{ab}^+
            \ , \\
    \tilde\psi_{ab}^{+,\, (u)}
        &:=
            \sum_{m_1,m_2\in\BZ}
                q^{\frac{k}{2}\left(\frac{a+b+u_1+u_2}{2k} + m_1 + m_2\right)^2}\,\bar q^{\frac{k}{2}\left(\frac{a-b+u_1-u_2}{2k} + m_1 - m_2\right)^2} \nonumber\\
        &\,=
            \Theta_{a+b+u_1+u_2,2k}
            \,\bar\Theta_{a-b+u_1-u_2,2k}
            + \Theta_{a+b+u_1+u_2-2k,2k}
            \,\bar\Theta_{a-b+u_1-u_2-2k,2k}
            \ , \\
    \tilde\psi_{ab}^{-,\, (u)}
        &:=
            \sum_{m_1,m_2\in\BZ}
                (-1)^{\frac{1}{k}(u_{2}a + u_1b)}\,q^{\frac{k}{2}\left(\frac{a+b+u_1+u_2}{2k} + m_1 + m_2\right)^2}\,\bar q^{\frac{k}{2}\left(\frac{a-b+u_1-u_2}{2k} + m_1 - m_2\right)^2} \nonumber\\
        &\,=
            (-1)^{\frac{1}{k}(u_{2}a + u_1b)}\,\tilde\psi_{ab}^{+,\, (u)}
            \ ,
\end{align}
where $\Theta_{l,k}:=\Theta_{l,k}(\tau)$ is the theta function
\begin{align}
    \Theta_{l,k}(\tau)
        :=
            \sum_{v\in\BZ}\,q^{k\left(v +\frac{l}{2k}\right)^2} 
\end{align}
and $\bar\Theta_{l,k}:=\overline{\Theta_{l,k}(\tau)}$.
We can represent the torus partition function with these functions as follows:
\begin{align}\label{fermionic_torus_PF}
    \begin{aligned}
    Z_{\text{NS}}
        &=
            \frac{1}{|\eta(\tau)|^{2n}}\, W_C\left( \left\{ \psi_{\alpha_i\beta_i}^+\right\}\right) \ , \\
    Z_{\widetilde{\text{NS}}}
        &=
            \frac{1}{|\eta(\tau)|^{2n}}\, W_C\left( \left\{ \psi_{\alpha_i\beta_i}^{-, \,(s^{(i)})}\right\}\right)\ , \\
    Z_{\text{R}}
        &=
            \frac{1}{|\eta(\tau)|^{2n}}\, W_C\left( \left\{ \tilde\psi_{\alpha_i\beta_i}^{+,(s^{(i)})}\right\}\right) \ , \\
    Z_{\widetilde{\text{R}}}
        &=
            \frac{1}{|\eta(\tau)|^{2n}}\,W_C\left( \left\{ \tilde\psi_{\alpha_i\beta_i}^{-, \,(s^{(i)})}\right\}\right)  \ .
    \end{aligned}
\end{align}

\subsection{Examples: $c=1$ fermionic CFTs}\label{ss:example_ferm_CFT}

We consider an odd self-dual code of length $2$ over $\BZ_{2k}$ defined by
\begin{align}
    C :=
        \left\{\, (2m + a, k\,a)\,\big|\, m\in \BZ_k,\, a\in \BZ_2\,\right\} \ .
\end{align}
The subsets $C_0$ and $C_2$ become
\begin{align}
    \begin{aligned}
        C_0
            &=
                \left\{\, (2m , 0)\,\big|\, m \in \BZ_{2k}\,\right\} \ , \\
        C_2
            &=
                \left\{\, (2m + 1 , k)\,\big|\, m \in \BZ_{2k}\,\right\} \ ,
    \end{aligned}
\end{align}
and the dual $C_0^\perp$ of $C_0$ is
\begin{align}
    C_0^\perp
        =
             \left\{\, (l , ka)\,\big|\, l \in \BZ_{2k}\,,a\in\BZ_2\right\}
\end{align}

To define the shadow, we may choose a vector $s\in C_0^\perp\backslash C = C_1\cup C_3$ as
\begin{align}
    s = (1, 0) \ .
\end{align}
Using the weight enumerator polynomial
\begin{align}
    W_C
        =
            \sum_{m\in \BZ_k}(x_{2m, 0} + x_{2m+1, 1})  \ .
\end{align}
and set $\tau = \i\,\ell$, we find
\begin{align}\label{Thirring_code}
    \begin{aligned}
        W_C\left( \left\{ \psi_{\alpha_i\beta_i}^+\right\}\right)
            &=
                \theta_3\left( \i\,\frac{2\ell}{k}\right)\, \theta_3\left( \i\,2\ell\,k\right) + \theta_2\left( \i\,\frac{2\ell}{k}\right)\, \theta_2\left( \i\,2\ell\,k\right)\ , \\
        W_C\left( \left\{ \psi_{\alpha_i\beta_i}^{-, \,(s)}\right\}\right)
            &=
                \theta_3\left( \i\,\frac{2\ell}{k}\right)\, \theta_3\left( \i\,2\ell\,k\right) - \theta_2\left( \i\,\frac{2\ell}{k}\right)\, \theta_2\left( \i\,2\ell\,k\right)\ , \\
        W_C\left( \left\{ \tilde\psi_{\alpha_i\beta_i}^{+,(s)}\right\}\right)
            &= 
                \theta_2\left( \i\,\frac{2\ell}{k}\right)\, \theta_3\left( \i\,2\ell\,k\right) + \theta_3\left( \i\,\frac{2\ell}{k}\right)\, \theta_2\left( \i\,2\ell\,k\right)\ , \\
        W_C\left( \left\{ \tilde\psi_{\alpha_i\beta_i}^{-, \,(s)}\right\}\right)
            &=
                \theta_2\left( \i\,\frac{2\ell}{k}\right)\, \theta_3\left( \i\,2\ell\,k\right) - \theta_3\left( \i\,\frac{2\ell}{k}\right)\, \theta_2\left( \i\,2\ell\,k\right)\ ,
    \end{aligned}
\end{align}
where
\begin{align}
    \begin{aligned}\label{eq: Jacobi theta function}
        \theta_{2}(\tau)
            =
                \sum_{n \in \BZ +\frac{1}{2}}  q^{\frac{n^2}{2}}  \ , \qquad
        \theta_{3}(\tau) 
            =
        \sum_{n \in \BZ }\,  q^{\frac{n^2}{2}} \ ,\qquad
        \theta_{4}(\tau)
            =
            \sum_{n \in \BZ }\, (-1)^n \,q^{\frac{n^2}{2}}\ . 
    \end{aligned}
\end{align}
With the help of the identities of the theta functions,
the fermionic partition functions \eqref{fermionic_torus_PF} can be written as
\begin{align}\label{Thirring_PF}
    \begin{aligned}
        Z_\text{NS}
            &=
                \frac{1}{2|\eta(\i\,\ell)|^2}\left[       
                    \theta_3\left( \i\,\frac{2\ell}{k}\right)\, \theta_3\left( \i\,\frac{k\,\ell}{2}\right) 
                + 
                    \theta_3\left( \i\,\frac{2\ell}{k}\right)\, \theta_4\left( \i\,\frac{k\,\ell}{2}\right)  
                    \right.\\
        &\qquad \qquad \qquad \qquad 
                \left. + 
                    \theta_2\left( \i\,\frac{2\ell}{k}\right)\, \theta_3\left( \i\,\frac{k\,\ell}{2}\right)
                -
                    \theta_2\left( \i\,\frac{2\ell}{k}\right)\, \theta_4\left( \i\,\frac{k\,\ell}{2}\right)
                \right]\ , \\
        Z_{\widetilde{\text{NS}}}
            &=
                                \frac{1}{2|\eta(\i\,\ell)|^2}\left[       
                    \theta_3\left( \i\,\frac{2\ell}{k}\right)\, \theta_3\left( \i\,\frac{k\,\ell}{2}\right) 
                + 
                    \theta_3\left( \i\,\frac{2\ell}{k}\right)\, \theta_4\left( \i\,\frac{k\,\ell}{2}\right)  
                    \right.\\
        &\qquad \qquad \qquad \qquad 
                \left. - 
                    \theta_2\left( \i\,\frac{2\ell}{k}\right)\, \theta_3\left( \i\,\frac{k\,\ell}{2}\right)
                +
                    \theta_2\left( \i\,\frac{2\ell}{k}\right)\, \theta_4\left( \i\,\frac{k\,\ell}{2}\right)
                \right]\ , \\
        Z_\text{R}        
            &= 
                \frac{1}{2|\eta(\i\,\ell)|^2}\left[   
                    \theta_3\left( \i\,\frac{2\ell}{k}\right)\, \theta_3\left( \i\,\frac{k\,\ell}{2}\right)
                -
                    \theta_3\left( \i\,\frac{2\ell}{k}\right)\, \theta_4\left( \i\,\frac{k\,\ell}{2}\right)
                    \right.\\
        &\qquad \qquad \qquad \qquad 
                \left. + 
                    \theta_2\left( \i\,\frac{2\ell}{k}\right)\, \theta_3\left( \i\,\frac{k\,\ell}{2}\right) 
                + 
                    \theta_2\left( \i\,\frac{2\ell}{k}\right)\, \theta_4\left( \i\,\frac{k\,\ell}{2}\right)
                \right]\ , \\
        Z_{\widetilde{\text{R}}}      
            &= 
                \frac{1}{2|\eta(\i\,\ell)|^2}\left[    
                - 
                    \theta_3\left( \i\,\frac{2\ell}{k}\right)\, \theta_3\left( \i\,\frac{k\,\ell}{2}\right)
                +
                    \theta_3\left( \i\,\frac{2\ell}{k}\right)\, \theta_4\left( \i\,\frac{k\,\ell}{2}\right)
                    \right.\\
        &\qquad \qquad \qquad \qquad 
                \left. 
                +
                    \theta_2\left( \i\,\frac{2\ell}{k}\right)\, \theta_3\left( \i\,\frac{k\,\ell}{2}\right) 
                + 
                    \theta_2\left( \i\,\frac{2\ell}{k}\right)\, \theta_4\left( \i\,\frac{k\,\ell}{2}\right)  
                \right]\ , 
    \end{aligned}
\end{align}

These expressions agree with the partition functions of the massless Thirring model (see e.g., \cite{Fujimura:2023esr}) with the coupling $\lambda = k - 1$, which can be obtained by fermionizing the compact boson theory of the radius $R = 2/\sqrt{k}$~\cite{Karch:2019lnn}.%
\footnote{
The formal transformation $k\rightarrow 1/k$, corresponding to $R\rightarrow 4/R$, leaves invariant the first three expressions in~(\ref{Thirring_code}) and flips the sign of the fourth expression.
Via~(\ref{fermionic_torus_PF}), the only effect of the transformation on the torus partition functions is to change the sign of $Z_{\widetilde{\mathrm{R}}}$, i.e., the stacking of the Arf theory~\cite{Karch:2019lnn,Ji:2019ugf}.}

\paragraph{Cases with even $k$}
When $k$ is even, there exists an odd self-dual code defined by\footnote{We have already classified odd self-dual codes (referred to as Lagrangian subgroups in the terminology of the previous sections) of length 2 with respect to the off-diagonal Lorentzian inner product around \eqref{eq:odd_Lag_leng2}. This code is obtained by substituting $k/2, w, v'$ with $2k, k/2, 2$ and swapping the components.}
\begin{align}\label{eq:odd-self-dual-code}
    C :=
        \left\{\, \left( 2(2m + a), \frac{k}{2}\,a\right) \in \BZ_{2k} \times \BZ_{2k}\,\bigg|\, m\in \BZ_\frac{k}{2},\, a\in \BZ_4\,\right\} \ .
\end{align}
The subsets $C_0$ and $C_2$ become
\begin{align}
    \begin{aligned}
        C_0
            &=
                \left\{\, (4m , ka)\,\bigg|\, m \in \BZ_{\frac{k}{2}},\, a\in \BZ_2\right\} \ , \\
        C_2
            &= C_0 + t = \left\{\, \left(4m+2 , ka+\frac{k}{2}\right)\,\bigg|\, m \in \BZ_{\frac{k}{2}},\, a\in \BZ_2\right\}\,,
    \end{aligned}
\end{align}
where $t = (2,\tfrac{k}{2})$.
The dual $C_0^\perp$ of $C_0$ is
\begin{align}
    C_0^\perp
        =
             \left\{\, \left( 2l ,\, \frac{k}{2}\,a\right)\,\bigg|\, l \in \BZ_{k},\, a\in \BZ_4\,\right\}
\end{align}
To define the shadow, we may choose $s\in C_0^\perp\backslash C$ as
\begin{align}
    s = (2, 0) \ .
\end{align}
Then, $C_1$ and $C_3$ are defined as
\begin{align}
\begin{aligned}
    C_1 &= C_0 + s = \left\{\, (4m+2 , ka)\,\bigg|\, m \in \BZ_{\frac{k}{2}},\, a\in \BZ_2\right\} \ ,\\
    C_3 &= C_0 + s + t = \left\{\, \left(4m , ka+\frac{k}{2}\right)\,\bigg|\, m \in \BZ_{\frac{k}{2}},\, a\in \BZ_2\right\}\,.
\end{aligned}
\end{align}

The weight enumerator polynomial of $C$ becomes
\begin{align}
    W_C
        =
            \sum_{m\in \BZ_\frac{k}{2}}\left( x_{4m, 0} + x_{4m, k} + x_{4m+2, \frac{k}{2}} + x_{4m+2, \frac{3k}{2}}\right) \ .
\end{align}
Then, for $\tau = \i\,\ell$, we obtain
\begin{align}
    \begin{aligned}
        W_C\left( \left\{ \psi_{\alpha_i\beta_i}^+\right\}\right)
            &=
                    
                    \theta_3\left( \i\,\frac{8\ell}{k}\right)\, \theta_3\left( \i\,2k\,\ell\right) 
                + 
                    \theta_3\left( \i\,\frac{8\ell}{k}\right)\, \theta_2\left( \i\,2k\,\ell\right)  
                +
                    \theta_2\left( \i\,\frac{8\ell}{k}\right)\, \theta_2\left( \i\,\frac{k\,\ell}{2}\right) \ ,
                 \\
        W_C\left( \left\{ \psi_{\alpha_i\beta_i}^{-, \,(s)}\right\}\right)
            &=
                    \theta_3\left( \i\,\frac{8\ell}{k}\right)\, \theta_3\left( \i\,2k\,\ell\right) 
                + 
                    \theta_3\left( \i\,\frac{8\ell}{k}\right)\, \theta_2\left( \i\,2k\,\ell\right)  
                -
                    \theta_2\left( \i\,\frac{8\ell}{k}\right)\, \theta_2\left( \i\,\frac{k\,\ell}{2}\right) \ ,
                \\
        W_C\left( \left\{ \tilde\psi_{\alpha_i\beta_i}^{+,(s)}\right\}\right)
            &= 
                    \theta_2\left( \i\,\frac{8\ell}{k}\right)\, \theta_3\left( \i\,2k\,\ell\right) 
                + 
                    \theta_2\left( \i\,\frac{8\ell}{k}\right)\, \theta_2\left( \i\,2k\,\ell\right)  
                +
                    \theta_3\left( \i\,\frac{8\ell}{k}\right)\, \theta_2\left( \i\,\frac{k\,\ell}{2}\right) \ ,
                \\
        W_C\left( \left\{ \tilde\psi_{\alpha_i\beta_i}^{-, \,(s)}\right\}\right)
            &= 
                    \theta_2\left( \i\,\frac{8\ell}{k}\right)\, \theta_3\left( \i\,2k\,\ell\right) 
                + 
                    \theta_2\left( \i\,\frac{8\ell}{k}\right)\, \theta_2\left( \i\,2k\,\ell\right)  
                -
                    \theta_3\left( \i\,\frac{8\ell}{k}\right)\, \theta_2\left( \i\,\frac{k\,\ell}{2}\right) \ .
    \end{aligned}
\end{align}
Using the identities for the theta functions, we find
\begin{align}\label{Thirring_PF2}
    \begin{aligned}
        Z_\text{NS}
            &=
                \frac{1}{2|\eta(\i\,\ell)|^2}\left[       
                    \theta_3\left( \i\,\frac{2\ell}{k}\right)\, \theta_3\left( \i\,\frac{k\,\ell}{2}\right) 
                + 
                    \theta_4\left( \i\,\frac{2\ell}{k}\right)\, \theta_3\left( \i\,\frac{k\,\ell}{2}\right)  
                    \right.\\
        &\qquad \qquad \qquad 
                \left. + 
                    \theta_3\left( \i\,\frac{2\ell}{k}\right)\, \theta_2\left( \i\,\frac{k\,\ell}{2}\right)
                -
                    \theta_4\left( \i\,\frac{2\ell}{k}\right)\, \theta_2\left( \i\,\frac{k\,\ell}{2}\right)
                \right]\ , \\
        Z_{\widetilde{\text{NS}}}
            &=
                    \frac{1}{2|\eta(\i\,\ell)|^2}\left[       
                    \theta_3\left( \i\,\frac{2\ell}{k}\right)\, \theta_3\left( \i\,\frac{k\,\ell}{2}\right) 
                + 
                    \theta_4\left( \i\,\frac{2\ell}{k}\right)\, \theta_3\left( \i\,\frac{k\,\ell}{2}\right)  
                    \right.\\
        &\qquad \qquad \qquad 
                \left. - 
                    \theta_3\left( \i\,\frac{2\ell}{k}\right)\, \theta_2\left( \i\,\frac{k\,\ell}{2}\right)
                +
                    \theta_4\left( \i\,\frac{2\ell}{k}\right)\, \theta_2\left( \i\,\frac{k\,\ell}{2}\right)
                \right]\ , \\
        Z_\text{R}        
            &= 
                \frac{1}{2|\eta(\i\,\ell)|^2}\left[   
                    \theta_3\left( \i\,\frac{2\ell}{k}\right)\, \theta_3\left( \i\,\frac{k\,\ell}{2}\right)
                -
                    \theta_4\left( \i\,\frac{2\ell}{k}\right)\, \theta_3\left( \i\,\frac{k\,\ell}{2}\right)
                    \right.\\
        &\qquad \qquad \qquad 
                \left. + 
                    \theta_3\left( \i\,\frac{2\ell}{k}\right)\, \theta_2\left( \i\,\frac{k\,\ell}{2}\right) 
                + 
                    \theta_4\left( \i\,\frac{2\ell}{k}\right)\, \theta_2\left( \i\,\frac{k\,\ell}{2}\right)
                \right]\ , \\
        Z_{\widetilde{\text{R}}}      
            &= 
                -\frac{1}{2|\eta(\i\,\ell)|^2}\left[    
                - 
                    \theta_3\left( \i\,\frac{2\ell}{k}\right)\, \theta_3\left( \i\,\frac{k\,\ell}{2}\right)
                +
                    \theta_4\left( \i\,\frac{2\ell}{k}\right)\, \theta_3\left( \i\,\frac{k\,\ell}{2}\right)
                    \right.\\
        &\qquad \qquad \qquad \quad
                \left. 
                +
                    \theta_3\left( \i\,\frac{2\ell}{k}\right)\, \theta_2\left( \i\,\frac{k\,\ell}{2}\right) 
                + 
                    \theta_4\left( \i\,\frac{2\ell}{k}\right)\, \theta_2\left( \i\,\frac{k\,\ell}{2}\right)  
                \right]\ .
    \end{aligned}
\end{align}
Since the partition functions~\eqref{Thirring_PF2} are related to~\eqref{Thirring_PF} via $k\rightarrow 4/k$, the fermionic CFT defined by the odd self-dual code~\eqref{eq:odd-self-dual-code} is the fermionization of the compact boson CFT with radius~$R=\sqrt{k}$.
We note that T-duality $R\rightarrow 2/R$ does not commute with fermionization~\cite{Karch:2019lnn}.
Thus, while the parent bosonic theories (with radii $2/\sqrt{k}$ and $\sqrt{k}$) are related via T-duality, the resulting fermionic theories~\eqref{Thirring_PF} and~\eqref{Thirring_PF2} are distinct.

\section{CFTs with level-one affine Lie algebra symmetries}
\label{sec:level-one-affine}
In this section, we construct CFTs with level-one affine Lie algebra symmetries from topological boundary conditions in the associated Chern-Simons theories based on the results in section~\ref{sec:general}.
The important ingredients include the realization of affine Lie algebras in terms of vertex operators~\cite{Frenkel:1980rn,Segal1981UnitaryRO} and the fact that self-dual lattices can be constructed from Lagrangian subgroups.
We will also use the results by Johnson-Freyd~\cite{Johnson-Freyd:2019wgb} who provided super vertex operator algebras (SVOAs) with affine Lie algebra symmetries and supersymmetry.
While the constructions of~\cite{Frenkel:1980rn,Segal1981UnitaryRO,Johnson-Freyd:2019wgb} realize the algebra of operators, to specify the full bosonic (spin structure-independent) or fermionic (spin structure-dependent) CFT we need additional data, which is precisely a topological boundary condition of the Chern-Simons theory.\footnote{
In the terminology of~\cite{Freed:2012bs}, the constructions of~\cite{Frenkel:1980rn,Segal1981UnitaryRO,Johnson-Freyd:2019wgb} define a CFT as a relative theory~\cite{Freed:2012bs} while a topological boundary condition specifies an absolute theory.}
In section~\ref{sec:affine-construcion}, we explain how to obtain CFTs with level-one affine Lie algebra symmetries by combining the construction of~\cite{Frenkel:1980rn,Segal1981UnitaryRO} with our Chern-Simons construction discussed in section~\ref{sec:general}.
In section~\ref{sec:affine-chiral}, as a warm-up, we classify bosonic and fermionic chiral CFTs with level-one affine Lie algebra symmetries for simply laced Lie algebras. 
In sections~\ref{sec:susy_chiral} and~\ref{sec:affine-nonchiral}, we construct fermionic chiral and non-chiral CFTs with supersymmetry using the classification results by Johnson-Freyd~\cite{Johnson-Freyd:2019wgb}.

\subsection{Construction of CFTs with level-one affine Lie algebra symmetries}
\label{sec:affine-construcion}

It has been known for a long time that there is an intimate relation~\cite{Frenkel:1980rn,Segal1981UnitaryRO} between the lattices associated with Lie algebras and chiral CFTs with level-one affine Lie algebra symmetries for simply laced Lie algebras.
In this subsection, we review a Chern-Simons interpretation of this relation and discuss the role of topological boundary conditions.
We assume a basic knowledge of Lie algebras.

Let $\mathfrak{g}$ be a semisimple Lie algebra and $T^a$ ($a=1,\ldots,\dim\mathfrak{g}$) form a basis of $\mathfrak{g}$.
We will denote by ``$\cdot$'' the Killing metric such that the long roots have length $\sqrt{2}$, and assume that $T^a\cdot T^b=\delta^{ab}$.
The affine Lie algebra is the algebra generated by holomorphic currents $j^a(z)$ ($a=1,\ldots,\dim\mathfrak{g}$) having the operator product expansion (OPE)
\begin{equation}
j^a(z) j^b(0) \sim \frac{k\,\delta^{ab}}{z^2}     + \frac{\i f^{ab}{}_c}{z} j^c(0) \,,
\end{equation}
where $f^{ab}{}_c$ are the structure constants, $[T^a,T^b]=\i f^{ab}{}_c T^c$.
The constant $k$ is called the level.

Let us assume that $\mathfrak{g}$ is simply laced but not necessarily simple.
We let $\mathfrak{h}$ be the Cartan subalgebra and identify it with its dual by the metric.
Suppose that $H^i$ ($i=1,\ldots,\dim\mathfrak{h}=:\mathrm{rank}\,\mathfrak{g}$) form a basis of $\mathfrak{h}$ such that $H^i\cdot H^j=\delta^{ij}$.
A root $\alpha$ can be expanded as $\alpha=\sum_i\alpha^i H^i$.
Let $\varphi^i(z)$ be chiral bosons with the OPE
\begin{equation}
\varphi^i(z)\varphi^j(0) \sim -\delta^{ij} \ln z \,.
\end{equation} 
With a suitable choice of the operators $c_\alpha$ (cocycle factors) that involve only the zero-modes of~$\varphi$, the vertex operators
$E^\alpha(z) = c_\alpha\, e^{\i\,\alpha^i \varphi^i}$ and the currents $H^i(z)=\i\, \partial\varphi^i$ have the OPE defining the affine Lie algebra at level one ($k=1$) for $\mathfrak{g}$~\cite{Frenkel:1980rn,Segal1981UnitaryRO}.
We note that the chiral bosons~$\varphi^i$ alone do not define a full modular invariant (or covariant) chiral CFT, which would require additional data.

We now argue that the extra data are precisely the topological boundary conditions in the construction of the CFT from an abelian Chern-Simons theory via dimensional reduction, as formulated in section~\ref{sec:general}.
We recall that the weight lattice $\Gamma_\text{weight}$, which we can identify with the coweight lattice because $\mathfrak{g}$ is simply laced, is the dual of the root lattice with respect to the metric: $\Gamma_\text{weight}=\Gamma_\text{root}^*$.
We consider the abelian Chern-Simons theory defined by taking
\begin{equation}
\Gamma = \Gamma_\text{root}
\,,
\qquad
e_I = \alpha_I \,,
\qquad
K_{IJ} = \alpha_I \cdot \alpha_J \,,
\end{equation}
where $\alpha_I\in \Gamma_\text{root}$ ($I=1,\ldots,\mathrm{rank}\,\mathfrak{g}$) are the simple roots of $\mathfrak{g}$.
On the lattice, the inner product~``$\oslash$'' coincides with the one ``$\cdot$'' determined by the metric.
Because $K_{II}=2$, the Chern-Simons theory is bosonic.
The discriminant group 
\begin{equation}
\label{eq:D-weight-root}
\mathcal{D} =\Gamma^*/\Gamma= \Gamma_\text{weight} /\Gamma_\text{root}
\end{equation}
is isomorphic to the center $Z(G)$ of the simply connected Lie group $G$ whose Lie algebra is $\mathfrak{g}$.
It is equipped with a quadratic form $Q:\mathcal{D}\rightarrow\mathbb{Q}/\mathbb{Z}$ and the induced
inner product $\oslash: \mathcal{D}\times \mathcal{D} \rightarrow \mathbb{Q}/\mathbb{Z}$ such that 
\begin{equation}
Q([w])= \frac{1}{2}w \cdot w \text{ mod } \mathbb{Z}
\end{equation}
and $[w]\oslash [w'] = w \cdot w' \text{ mod } \mathbb{Z}$, where $w$ and $w'$ are elements of $\Gamma^*$ and the bracket $[w]=w+\Gamma$ denotes the equivalence class.
One can check that there is a one-to-one correspondence between the elements of $\mathcal{D} =  \Gamma_\text{weight}/\Gamma_\text{root}$ and the integrable highest-weight representations of the level-one affine Lie algebra for $\mathfrak{g}$.
Canonical quantization of the abelian Chern-Simons theory discussed in section~\ref{sec:canonical-quantization} gives rise to the wave functions~(\ref{eq:wave-function}), which are nothing but the level-one affine characters for $\mathfrak{g}$.
As we explained in section~\ref{sec:topological-boundary}, to fully reduce the three-dimensional system to a two-dimensional CFT, we need to specify a topological boundary condition, which is specified by a Lagrangian subgroup~$\mathcal{C}$ of the discriminant group~$\mathcal{D}$.
The upshot is that the Lagrangian subgroup gives rise to a self-dual lattice, which then defines a CFT.
The resulting CFT is bosonic if $\mathcal{C}$ is even as defined in (\ref{eq:evenness-condition}), and is fermionic if $\mathcal{C}$ is odd as defined in~(\ref{eq:oddness-condition}).
Their torus partition functions, discussed in section~\ref{sec:torus}, are given as~(\ref{eq:torus-partition-function}) in the bosonic case and as~(\ref{eq:ZC0011}) in the fermionic case.
We will study such CFTs in sections~\ref{sec:affine-chiral} and~\ref{sec:susy_chiral}.

We can also consider a non-chiral CFT based on two simply laced Lie algebras $\mathfrak{g}$ and $\mathfrak{g}'$, both of which are not necessarily simple.
We assume that their ranks are equal.
Let $\Gamma_\text{root}$ and $\Gamma_\text{root}'$ be their root lattices, and $\{\alpha_i\}$ and $\{\alpha'_i\}$ be their simple roots, respectively.
To apply our construction in section~\ref{sec:general}, we take
\begin{equation}
\Gamma = \Gamma_\text{root} \times \Gamma_\text{root}' \,,
\quad
K
= 
\begin{pmatrix}
\alpha_i \cdot \alpha_j & 0 \\
0 & -\alpha'_i \cdot \alpha'_j
\end{pmatrix}  \,.
\end{equation}
The discriminant group 
\begin{equation}
\mathcal{D} =\Gamma^*/\Gamma= (\Gamma_\text{weight} /\Gamma_\text{root})
\times
(\Gamma_\text{weight}' /\Gamma_\text{root}')
\end{equation}
is isomorphic to $Z(G)\times Z(G')$, where Lie groups $G$ and $G'$ are the simply connected Lie groups whose Lie algebras are $\mathfrak{g}$ and $\mathfrak{g}'$, respectively.
It is equipped with a quadratic form $Q:\mathcal{D}\rightarrow\mathbb{Q}/\mathbb{Z}$ and the induced
inner product $\oslash: \mathcal{D}\times \mathcal{D} \rightarrow \mathbb{Q}/\mathbb{Z}$ such that 
\begin{equation}
Q([w],[w'])= \frac{1}{2}(w \cdot w - w'\cdot w') \quad \text{ mod } \mathbb{Z}
\end{equation}
and $([w_1],[w_1'])\oslash ([w_2'],[w_2']) = w_1 \cdot w_1' - w_2 \cdot w_2' \text{ mod } \mathbb{Z}$.
A Lagrangian subgroup of $\mathcal{D}$ gives rise to a non-chiral CFT with level-one affine Lie algebra symmetries, corresponding to $\mathfrak{g}$ in the left-moving sector and to $\mathfrak{g}'$ in the right-moving sector.%
\footnote{%
When $G=G'=\mathrm{SU}(2)^n$ and $\mathcal{C}$ is even, this construction reduces to the one studied in~\cite{Henriksson:2022dml}.}
We will study such CFTs with supersymmetry in section~\ref{sec:affine-nonchiral}.

\begin{table}[t]
\begin{center}
\begin{tblr}{c|cccccc}
\toprule
 $\mathfrak{g}$ & $A_{m-1}$ & $D_{2m}$ & $D_{2m+1}$   & $E_6$ & $E_7$ & $E_8$
\\
\hline
$G$ & $\mathrm{SU}(m)$ & $\mathrm{Spin}(4m)$ &$\mathrm{Spin}(4m+2)$ & $E_6$ & $E_7$ & $E_8$
\\
$Z(G)$ & $\mathbb{Z}_m$ & $\mathbb{Z}_2\times\mathbb{Z}_2$ & $\mathbb{Z}_4$ &$\mathbb{Z}_3$ & $\mathbb{Z}_2$ & $\{0\}$\\
\bottomrule
\end{tblr}
\end{center}
\caption{Simply laced simple Lie algebras, the corresponding simply connected Lie groups, and their centers.}
\label{table:centers}

\end{table}

\subsection{Chiral CFTs with affine Lie algebra symmetries}
\label{sec:affine-chiral}

In this subsection, we classify the bosonic and fermionic chiral CFT (modular invariant or covariant) completions of the level-one affine Lie algebra for a simply laced simple Lie algebra by classifying the Lagrangian subgroups of~(\ref{eq:D-weight-root}) that provide topological boundary conditions for the Chern-Simons theory.
We do this only as a warm-up and preparation for later subsections.
No result here is new since even and odd self-dual Euclidean lattices are already classified and their relations to Lie algebra lattices are well understood.
\paragraph{\underline{$\SU(m)$}}
Let us begin with $\mathfrak{g}=A_{m-1}$ ($m\geq 2$).
In this case, the discriminant group $\mathcal{D}\simeq\mathbb{Z}_m$ is generated by $[w^1] = w^1+\Gamma$ with 
\begin{equation}
w^1\cdot w^1 = \frac{m-1}{m}\,,    
\end{equation} 
where $w^1$ is the highest weight of the fundamental representation of $\SU(m)$.
A Lagrangian subgroup exists if and only if $m$ is a square number: $m= (m')^2$ with $m'\in\mathbb{Z}_{>0}$.
For such $m$, the Lagrangian subgroup~$\mathcal{C}$ is unique
and generated by $m'\, [w^1]$ so that $\mathcal{C}\simeq\mathbb{Z}_{m'}$.
The corresponding self-dual lattice is
\begin{equation}\label{eq:Lambda_C_A}
\Lambda(\mathcal{C}) =  \Gamma \sqcup (\Gamma + m' w^1) \sqcup (\Gamma + 2m' w^1) \sqcup\ldots\sqcup (\Gamma + (m'-1)m' w^1) \,.
\end{equation}

Thus, we get the following classification for the chiral CFT completion of the level-one affine Lie algebra for $\SU(m)$.
\begin{itemize}
    \item When $m$ is the square of an odd integer~$m'$ ($m=9,25,49,81,\ldots$), there is a unique chiral CFT completion, which is bosonic and has central charge $c_L=m-1\in 8\mathbb{Z}$.
    The spectrum is given by the even self-dual lattice $\Lambda(\mathcal{C})$ in~(\ref{eq:Lambda_C_A}).
    The partition function is
    \begin{equation}
    Z(\tau) = \chi_0+\chi_{m'}+\chi_{2m'}+\ldots+\chi_{(m'-1)m'} \,,
    \end{equation}
    where $\chi_j$ denotes the affine character corresponding to the highest weight $j\, w^1$.
    
    \item When $m$ is the square of an even integer~$m'$ ($m=4,16,36,64,\ldots$), there is a unique chiral CFT completion, which is fermionic.
    The odd self-dual lattice $\Lambda(\mathcal{C})$ in~(\ref{eq:Lambda_C_A}) constitutes the NS spectrum.
    The general analysis in section~\ref{sec:ferm-cft} determines the R spectrum as well.
    Let us choose the vector $s=(m'/2)w^1$ and define the fermion number for the R sector accordingly (see footnote~\ref{footnote:R-F}).
    Then the partition functions are given as
    \begin{equation}
        \begin{aligned}
            Z_\mathrm{NS}(\tau)
                &=
                    \sum_{j=0}^{m'-1}\chi_{jm'} \,, 
                &
            Z_\mathrm{\widetilde{NS}}(\tau)
                &=
                    \sum_{j=0}^{m'-1}(-1)^j\,\chi_{jm'} \,,\\
            Z_\mathrm{R}(\tau)
                &=
                    \sum_{j=0}^{m'-1}\chi_{j m' +\frac{m'}{2}} \,, 
                &
            Z_\mathrm{\widetilde{R}}(\tau)
                &=
                    \sum_{j=0}^{m'-1}(-1)^j\, \chi_{j m'+\frac{m'}{2}} \,.
        \end{aligned}    
    \end{equation}
    
    \item When $m$ is not a square number, there is no chiral CFT completion.
\end{itemize}

\paragraph{\underline{$\Spin(4m)$}}
In the case $\mathfrak{g}=D_{2m}$ ($m\geq 2$), the discriminant 
$\mathcal{D}\simeq\mathbb{Z}_2\times\mathbb{Z}_2$ is generated by $[w_S]$ and $[w_C]$, where $w_S$ and $w_C$ are respectively the highest weights of the spinor and conjugate spinor representations.
The quadratic form $Q$ on $\mathcal{D}$ is determined by
\begin{equation}
w_S \cdot w_C
 = \frac{m-1}{2}
  \,, \qquad
w_S \cdot w_S = w_C \cdot w_C   =  \frac{m}{2} 
 \,.
\end{equation}
When $m \in 4\mathbb{Z}$, there are two even Lagrangian subgroups, each generated by $[w_S]$ or $[w_C]$, and one odd Lagrangian subgroup generated by $[w_S+w_C]$.
When $m\in 4\mathbb{Z}+2$, we get three odd Lagrangian subgroups, each generated by $[w_S]$, $[w_C]$ or $[w_S+w_C]$.
 When $m\in 2\mathbb{Z}+1$, the only Lagrangian subgroup is odd and generated by $[w_S+w_C]$.
We note that $w_S$ and $w_C$ are exchanged by an outer automorphism of $D_{2m}$ swapping $w_S$ and $w_C$.
Thus, we get the following results for the classification of the chiral CFT completions of the level-one affine Lie algebra for $\Spin(4m)$.
\begin{itemize}
\item When $m \in 4\mathbb{Z}$, there are one bosonic (up to automorphism) and one fermionic chiral CFT completions, all with central charge $c_L=2m \in 8\mathbb{Z}$.
The bosonic CFT has the partition function
\begin{equation}
    Z(\tau) = \chi_{00} + \chi_{10}
\,,
\end{equation}
where $\chi_{ij}$ denotes the affine character corresponding to the highest weight $i\,w_S+j\,w_C$ ($i,j \in \{0,1\}$).%
\footnote{For $m=4$, the bosonic CFT coincides with the $E_8$ CFT.}
If we choose $s=w_S$ to define the fermion number, the fermionic CFT has the partition functions
\begin{equation}\label{eq:MW-ferm-pfs}
    \begin{aligned}
        Z_\mathrm{NS}(\tau) 
            &= 
                \chi_{00}+\chi_{11}  \,, 
            &
        \qquad 
        Z_\mathrm{\widetilde{NS}}(\tau) 
            &= 
                \chi_{00}-\chi_{11}  \,,
        \\
        Z_\mathrm{R}(\tau) 
            &= 
                \chi_{10}+\chi_{01}  \,, 
            &
        \qquad
        Z_\mathrm{\widetilde{R}}(\tau) 
            &= 
                \chi_{10}-\chi_{01} \,,
    \end{aligned}    
\end{equation}
which implies that the CFT is the theory of $4m$ free Majorana-Weyl fermions.%
\footnote{%
For $\mathfrak{g}=D_{2m}\times D_{2m}$ with $m \in 4\mathbb{Z}$, $\mathcal{D}\simeq \mathbb{Z}_2^4$.
The subgroup generated by $([w_S,w_S])$ and $([w_S+w_C],[w_C])$ is Lagrangian and odd.
The corresponding CFT appears as a fermionic $c=16$ entry in the classification in \cite{BoyleSmith:2023xkd,Rayhaun:2023pgc,Hohn:2023auw}.
}
\item When $m\in 4\mathbb{Z}+2$ ($m\geq 6$), we get two fermionic chiral CFT completions up to automorphism.
The CFT corresponding to $[w_S]$ has, if we choose $s=w_C$,
the partition functions
\begin{equation}
    \begin{aligned}
        Z_\mathrm{NS}(\tau) 
            &= 
                \chi_{00}+\chi_{10}  \,, 
            &\qquad
        Z_\mathrm{\widetilde{NS}}(\tau) 
            &= 
                \chi_{00}-\chi_{10}  \,,
        \\
        Z_\mathrm{R}(\tau) 
            &= 
                \chi_{01}+\chi_{11}  \,, 
            &\qquad
        Z_\mathrm{\widetilde{R}}(\tau) 
            &= 
                \chi_{01}-\chi_{11} \,.
    \end{aligned}        
\end{equation}
The CFT corresponding to $[w_S+w_C]$ is the theory of $4m$ free Majorana-Weyl fermions with the partition functions given in~(\ref{eq:MW-ferm-pfs}).
When $m=2$, the three weights $w_S$, $w_C$, and $w_S+w_C$ are related by  triality so that, up to automorphism, there is a single fermionic chiral CFT, which is the theory of 8 free Majorana-Weyl fermions.

\item When $m\in 2\mathbb{Z}+1$, the only chiral CFT completion is the theory of $4m$ free Majorana-Weyl fermions with the partition functions~(\ref{eq:MW-ferm-pfs}).

\end{itemize}

\paragraph{\underline{$\Spin(4m+2)$}}
In the case $\mathfrak{g}=D_{2m+1}$ ($m\geq 1$), the discriminant group $\mathcal{D}\simeq \mathbb{Z}_4$ is generated by $[w_S]=w_S+\Gamma$, where $w_S$ is the highest weight of the spinor representation.
The quadratic form $Q$ is determined by
\begin{equation}
w_S \cdot w_S =
\frac{2m+1}{4}
\,.
\end{equation}
The unique Lagrangian subgroup is odd and generated by $2[w_S]$, which corresponds to the vector representation.
This is the theory of $4m+2$ free Majorana fermions with the partition functions
\begin{equation}\label{eq:MW-ferm-pfs-4m+2}
    \begin{aligned}
        Z_\mathrm{NS}(\tau) 
            &= 
                \chi_{0}+\chi_{2}  \,, 
            &
        \qquad 
        Z_\mathrm{\widetilde{NS}}(\tau) 
            &= 
                \chi_{0}-\chi_{2}  \,,
        \\
        Z_\mathrm{R}(\tau) 
            &= 
                \chi_{1}+\chi_{3}  \,, 
            &
        \qquad
        Z_\mathrm{\widetilde{R}}(\tau) 
            &= 
                \chi_{1}-\chi_{3} \,,
    \end{aligned}    
\end{equation}
where $\chi_{i}$ denotes the affine character corresponding to the highest weight $i\,w_S$ ($i\in \{0,1,2,3\}$).
(We chose $s=w_S$.)

\paragraph{\underline{$E_6$}}
In this case, $\mathcal{D}\simeq \mathbb{Z}_3$ is generated by $[w_\mathbf{27}]$, where $w_\mathbf{27}$ is the highest weight of the 27-dimensional irreducible representation.
The quadratic form $Q$ is determined by $w_\mathbf{27}\cdot w_\mathbf{27} = 4/3$.
There is no Lagrangian subgroup.
\paragraph{\underline{$E_7$}}
In this case, $\mathcal{D}\simeq \mathbb{Z}_2$ is generated by $[w_\mathbf{56}]$, where $w_\mathbf{56}$ is the highest weight of the 56-dimensional irreducible representation.
The quadratic form $Q$ is determined by $w_\mathbf{56}\cdot w_\mathbf{56} = 3/2$.
There is no Lagrangian subgroup.%
\footnote{\label{footnote:E7xE7}%
If we take $\mathfrak{g}=E_7\times E_7$, then $\mathcal{D}\simeq \mathbb{Z}_2\times\mathbb{Z}_2$ and
there is a unique Lagrangian subgroup which is odd and generated by $([w_\mathbf{56}],[w_\mathbf{56}])$.
The corresponding CFT appears as a $c=14$ fermionic entry in the classification in~\cite{BoyleSmith:2023xkd,Rayhaun:2023pgc,Hohn:2023auw}. In the next subsection, we will see that this theory has supersymmetry.}

\paragraph{\underline{$E_8$}}
In this case, $\mathcal{D} = \{0\}$, and there is a unique Lagrangian subgroup $\{0\}$ which is even.
This corresponds to the well-known $E_8$ bosonic chiral CFT with central charge $c=8$.

\subsection{Chiral superconformal field theories}
\label{sec:susy_chiral}

\begin{table}[t]
    \centering
    \begin{tblr}{c|ccc}
    \toprule
        Group $G$ & \;center $Z(G)$ & $c$ & supercurrent \\ \hline
        $\Spin(4m)^3$ & $(\BZ_2\times\BZ_2)^3$ & $6m$ & $(1,1,1,1,1,1)$ \\
        $\Spin(12)^2$ & $(\BZ_2\times\BZ_2)^2$ & $12$ & $(1,0,1,0)$ \\
        $\Spin(16) \times \Spin(8)$ & $(\BZ_2\times\BZ_2)^2$ & $12$ & $(1,0,1,0)$ \\
        $\Spin(24)$ & $\BZ_2\times\BZ_2$ & $12$ & $(1,0)$ \\
        $\Spin(4m+2)^3$ & $(\BZ_4)^3$ & $6m+3$ & $(2,2,2)$ \\
        $\SU(6)^2$ & $(\BZ_6)^2$ & $10$ & $(3,3)$ \\
        $\SU(12)$ & $\BZ_{12}$ & $11$ & $(6)$ \\
        $E_7^2$ & $\BZ_2^2$ & $14$ & $(1,1)$\\
    \bottomrule
    \end{tblr}
    \caption{The simply connected Lie groups for which the corresponding chiral CFTs can exhibit supersymmetry, with their centers, ranks (equivalently central charge), and the elements of the centers associated with the representations that contain supercurrents.
    This summarizes the result of~\cite{Johnson-Freyd:2019wgb} for simply laced Lie algebras.}
    \label{tab:susy_lie_groups}
\end{table}

This subsection constructs chiral supersymmetric CFTs with affine Lie algebra symmetry at level one.
Notably, the author of~\cite{Johnson-Freyd:2019wgb} classified $\CN=1$ supersymmetric SVOAs without free fermions and with the bosonic subalgebra given by a product of affine Lie algebras and identified which representation contains supercurrents.
The result of~\cite{Johnson-Freyd:2019wgb} for the level-one simply laced cases is summarized in table~\ref{tab:susy_lie_groups}. 
In our framework, the work classifies the Chern-Simons theories that contain an anyon corresponding to supercurrents.
In this subsection, we leverage the result~\cite{Johnson-Freyd:2019wgb} to construct supersymmetric chiral CFTs by adding the data of global structure specified by topological boundaries.
Our method is based on the enumeration of odd Lagrangian subgroups containing the representations with supercurrents.

\paragraph{\underline{$\Spin(4m)^3$}}

The discriminant group is $\CD\cong \BZ_2^6$.
Table~\ref{tab:susy_lie_groups} shows that the supercurrents are in the representation $(1,1,1,1,1,1)$.
There are three odd Lagrangian subgroups containing $(1,1,1,1,1,1)$.
The enumeration has been performed up to the equivalence transformations:
\begin{itemize}
    \item Swapping $w_S$ and $w_C$ in each $\mathrm{Spin}(4m)$
    \item Permuting the three copies of $\mathrm{Spin}(4m)$ in the whole $\mathrm{Spin}(4m)^3$
\end{itemize}
By the first equivalence, we regard the spinor and conjugate spinor representations as equivalent.
In terms of characters, the transformation exchanges $\chi_{10}$ and $\chi_{01}$.
The second transformation permutes the three blocks $\BZ_2^2$ in the discriminant group $\CD\cong\BZ_2^6$.

When $m$ is even, each odd Lagrangian subgroup is generated by
\begin{align}
    G_1 = \begin{bmatrix}
    1 & 1 & 0 & 0 & 0 & 0 \\
    0 & 0 & 1 & 1 & 0 & 0 \\
    0 & 0 & 0 & 0 & 1 & 1 
\end{bmatrix}\,,\quad
    G_2 = \begin{bmatrix}
    1 & 1 & 0 & 0 & 0 & 0 \\
    0 & 0 & 1 & 0 & 0 & 1 \\
    0 & 0 & 0 & 1 & 1 & 0 
\end{bmatrix}\,,\quad
    G_3 = \begin{bmatrix}
    1 & 0 & 0 & 0 & 0 & 1 \\
    0 & 1 & 0 & 1 & 1 & 1 \\
    0 & 0 & 1 & 0 & 0 & 1 
\end{bmatrix}\,.
\end{align}
The NS partition functions are
\begin{align} \label{eq:spin4m^3_NS}
    \begin{aligned}
        Z_{1,\mathrm{NS}}(\tau) &= (\chi_{00}+\chi_{11})^3\,,\\
        Z_{2,\mathrm{NS}}(\tau) &= (\chi_{00}+\chi_{11}) (\chi_{00}^2 + 2\,\chi_{01}\,\chi_{10} + \chi_{11}^2)\,,\\
        Z_{3,\mathrm{NS}}(\tau) &= \chi_{00}^3 + 2\chi_{00}\,\chi_{01}\,\chi_{10}+\chi_{00}\,\chi_{10}^2+\chi_{01}^2\,\chi_{11}+2\chi_{01}\,\chi_{10}\,\chi_{11}+\chi_{11}^3\,.
    \end{aligned}
\end{align}
For each odd Lagrangian subgroup, an element of its shadow is
\begin{equation}
    s_1 = (0,1,0,1,0,1)\,,\quad
    s_2 = (0,1,0,0,0,0)\,,\quad
    s_3 = (0,0,0,0,0,1)\,.
\end{equation}
From these expressions, one can obtain the R and $\widetilde{\mathrm{R}}$ partition functions
\begin{align}
\begin{aligned}
    Z_{1,\mathrm{R}}(\tau) &= (\chi_{01}+\chi_{10})^3\,,\\
    Z_{2,\mathrm{R}}(\tau) &= (\chi_{01}+\chi_{10})(\chi_{00}^2 + 2\,\chi_{01}\,\chi_{10} + \chi_{11}^2)\,,\\
    Z_{3,\mathrm{R}}(\tau) &= \chi_{00}^{2} \chi_{01} + 2 \chi_{00}^{2} \chi_{10} + \chi_{01}^{2} \chi_{10} + \chi_{01} \chi_{10}^{2} + 2 \chi_{01} \chi_{11}^{2} + \chi_{10} \chi_{11}^{2}\,,
\end{aligned}
\end{align}
and $Z_{1,\widetilde{\mathrm{R}}}(\tau)=Z_{2,\widetilde{\mathrm{R}}}(\tau) =0$, $Z_{3,\widetilde{\mathrm{R}}}(\tau) = 3\cdot 2^{2m-1}$ where we used $q$-expansion for the $\widetilde{\mathrm{R}}$ partition functions.
Since these two-dimensional theories are supersymmetric by construction, the $\widetilde{\mathrm{R}}$ partition function is constant, i.e., the Witten index~\cite{Witten:1982df}.
The non-zero Witten indices indicate the existence of the supersymmetric Ramond vacua, the Ramond ground states satisfying the BPS bound $h=c/24$.
The existence of supersymmetric Ramond vacua is equivalent to the fact that supersymmetry is unbroken.
Thus, the non-zero Witten index implies that supersymmetry is unbroken.
However, when the Witten index is zero, supersymmetry may be broken.
One can confirm whether supersymmetry is broken or not by seeing the $q$-expansion of the R partition functions
\begin{align}
    \begin{aligned}
        Z_{1,\mathrm{R}}(\tau) = 0 + O(q^\frac{m}{2})\,,\qquad
        Z_{2,\mathrm{R}}(\tau) = 4^m + O(q)\,.
    \end{aligned}
\end{align}
The constant terms count the supersymmetric R vacua satisfying $h = c/24$. From the absence of such vacua, supersymmetry is broken in the first theory. In the second theory, supersymmetry is not broken.

The first theory consists of $12m$ Majorana-Weyl fermions.
The second theory is a direct product of two subsectors: $\chi_{00}+\chi_{11}$ and $\chi_{00}^2 + 2\chi_{01}\chi_{10} + \chi_{11}^2$.
The former corresponds to $4m$ Majorana-Weyl fermions with the central charge $c=2m$.
The latter part with $c=4m$ is bosonic for even $m$ since $\chi_{00}^2 + 2\chi_{01}\chi_{10} + \chi_{11}^2$ is invariant under modular $T$ transformation up to a phase by a gravitational anomaly, which implies that its spectrum consists of bosonic states with integer spins.
The third theory generically does not contain free Majorana-Weyl fermions.

When $m$ is odd, the three odd Lagrangian subgroups including supercurrents are generated by
\begin{align}
    G_1 = \begin{bmatrix}
    1 & 1 & 0 & 0 & 0 & 0 \\
    0 & 0 & 1 & 1 & 0 & 0 \\
    0 & 0 & 0 & 0 & 1 & 1 
\end{bmatrix}\,,\quad
    G_2 = \begin{bmatrix}
    1 & 1 & 0 & 0 & 0 & 0 \\
    0 & 0 & 1 & 0 & 0 & 1 \\
    0 & 0 & 0 & 1 & 1 & 0 
\end{bmatrix}\,,\quad
    G_3 = \begin{bmatrix}
    1 & 0 & 0 & 0 & 0 & 1 \\
    0 & 1 & 0 & 1 & 0 & 0 \\
    0 & 0 & 1 & 0 & 1 & 0 
\end{bmatrix}\,.
\end{align}
The NS partition functions are
\begin{align}
    \begin{aligned}
        Z_{1,\mathrm{NS}}(\tau) &= (\chi_{00}+\chi_{11})^3\,,\\
        Z_{2,\mathrm{NS}}(\tau) &= (\chi_{00}+\chi_{11}) (\chi_{00}^2 + 2\,\chi_{01}\,\chi_{10} + \chi_{11}^2)\,,\\
        Z_{3,\mathrm{NS}}(\tau) &= \chi_{00}^3 +\chi_{00}\,\chi_{01}^2+ \chi_{00}\,\chi_{01}\,\chi_{10}+\chi_{00}\,\chi_{10}^2+\chi_{01}^2\,\chi_{11}+\chi_{01}\,\chi_{10}\,\chi_{11}+\chi_{10}^2\,\chi_{11}+\chi_{11}^3\,.
    \end{aligned}
\end{align}
For each odd Lagrangian subgroup, an element of its shadow is
\begin{equation}
    s_1 = (0,1,0,1,0,1)\,,\quad
    s_2 = (0,1,0,0,1,1)\,,\quad
    s_3 = (0,0,0,1,1,1)\,.
\end{equation}
Correspondingly, the R partition functions are
\begin{align}
\begin{aligned}
    Z_{1,\mathrm{R}}(\tau) &= (\chi_{01}+\chi_{10})^3\,,\\
    Z_{2,\mathrm{R}}(\tau) &= (\chi_{01}+\chi_{10})(\chi_{01}^2 + 2\,\chi_{00}\,\chi_{11} + \chi_{10}^2)\,,\\
    Z_{3,\mathrm{R}}(\tau) &= (\chi_{01}+\chi_{10}) (\chi_{01}\chi_{10} + 3\chi_{00}\chi_{11})\,,
\end{aligned}
\end{align}
and the $\widetilde{\mathrm{R}}$ partition functions are all zero.
From $q$-expansion of the R partition functions, all these theories do not have the supersymmetric Ramond vacua, and supersymmetry is broken.

As for even $m$, the first theory always consists of $12m$ Majorana-Weyl fermions.
The second theory is a direct product of two parts: $\chi_{00}+\chi_{11}$ and $\chi_{00}^2 + 2\chi_{01}\chi_{10} + \chi_{11}^2$.
The former part corresponds to $4m$ Majorana-Weyl fermions.
The latter part with central charge $c=4m$ is fermionic for odd $m$.
This is because for odd $m$, $\chi_{00}^2 + 2\chi_{01}\chi_{10} + \chi_{11}^2$ is not modular $T$ invariant but modular $T^2$ invariant, which shows the existence of fermionic states with half-integer spins.

For each $m\in \BZ$, there are three types of odd Lagrangian subgroups.
In general, the corresponding fermionic CFTs are distinct (see table~\ref{tab:susy4m}).
The first theory is $12m$ Majorana-Weyl fermions.
The second theory is a direct product of $4m$ Majorana-Weyl fermions and a lattice CFT based on a $4m$-dimensional self-dual lattice.
This type of product theory is known to have $\CN=1$ supersymmetry~\cite{heluani2007susy,Gaiotto:2018ypj}.
The third theory is a lattice CFT based on a $6m$-dimensional odd self-dual lattice.
When $m=1$, the three theories are degenerate: $Z_{\mathrm{NS}}(\tau) = (\chi_{00}+\chi_{11})^3$ and only 12 Majorana-Weyl fermions appear.
When $m=2$, the second theory is a direct product of $E_8$ lattice CFT and eight Majorana-Weyl fermions, and the third theory is the supermoonshine module $V^{f\natural}$~\cite{duncan2007super}.
When $m=3$, the third theory has the NS partition function
\begin{equation}
    Z_\mathrm{NS}(\tau) = q^{-\frac{3}{4}} + 198 \,q^\frac{1}{4} + 4800\, q^\frac{3}{4} + 51849\, q^\frac{5}{4} + O(q^{\frac{7}{4}})\,.
\end{equation}
This is the lattice CFT based on the $18$-dimensional odd self-dual lattice $\Lambda_{18}$ containing the root lattice $D_6^3$, the third $18$-dimensional lattice in the classification list~\cite[Table 16.7]{conway2013sphere}.
When $m=4$, the second theory is a direct product of $16$ Majorana-Weyl fermions and the bosonic lattice CFT based on the lattice $D_{16}^+$.
The third theory for $m=4$ is based on the $24$-dimensional odd self-dual lattice containing the root lattice $D_8^3$. The lattice has the theta function
\begin{equation}
\label{eq:lambda_24}
    \Theta_{\Lambda_{24}}(q) = \sum_{\lambda\in\Lambda_{24}}q^{\frac{\lambda^2}{2}} = 1+336\,q+4096\,q^\frac{3}{2}+73808\,q^2+O(q^{\frac{5}{2}})\,.
\end{equation}
This lattice appears at 145th in the classification list~\cite[Table 17.1c]{conway2013sphere}.

\begin{table}[t]
    \centering
    \setlength{\tabcolsep}{10pt} 
    \begin{tabular*}{8.5cm}{rccc}
        \toprule
        $m$ & 1st theory & 2nd theory & 3rd theory \\
        \hline
        1 & $12\psi$ & $12\psi$ & $12\psi$ \\
        2 & $24\psi$& $E_8\times 8\psi$ & $V^{f\natural}$ \\
        3 & $36\psi$ & $V^{f\natural}\times 12\psi$ & {$\Lambda_{18}$}  \\
        4 & $48\psi$ & {$D_{16}^+\times 16\psi$} & {$\Lambda_{24}$}\\
        \bottomrule
    \end{tabular*}
    \caption{Chiral supersymmetric CFTs with affine Lie algebra symmetry corresponding to $G=\Spin(4m)^3$ at small $m$ (correspondingly, central charge is $c=6m$). Here, the symbols $\psi$, $E_8$, $V^{f\natural}$, and $D_{16}^+$ denote a free Majorana-Weyl fermion, the $E_8$ lattice CFT, the $\CN=1$ supermoonshine module, and the $D_{16}^+$ lattice CFT respectively. The symbol $\Lambda_{18}$ denotes the lattice CFT based on the $18$-dimensional odd self-dual lattice containing the root lattice $D_6^3$~(the third lattice of dimension $18$ in~\cite[Table 16.7]{conway2013sphere}). The symbol $\Lambda_{24}$ denotes the lattice CFT based on a $24$-dimensional odd self-dual lattice containing the root lattice $D_8^3$ (the 145th lattice in~\cite[Table 17.1c]{conway2013sphere}).}
    \label{tab:susy4m}
\end{table}

\paragraph{\underline{$\Spin(12)^2$}}
The discriminant group is $\CD\cong\BZ_2^4$.
As in the case of $\mathrm{Spin}(4m)^3$ group, we consider the classification of odd Lagrangian subgroups up to the following operations:
\begin{itemize}
    \item Swapping $w_S$ and $w_C$ in each $\mathrm{Spin}(12)$
    \item Permuting the two copies of $\mathrm{Spin}(12)$ in the whole $\mathrm{Spin}(12)^2$
\end{itemize}
We consider a supercurrent in the representation $(1,0,1,0)$ and search for odd Lagrangian subgroups containing it.
Note that this choice of a supercurrent is not invariant under the first operation above.
Thus, we only have to consider the equivalence by the second operation when fixing a supercurrent by $(1,0,1,0)$.
The odd Lagrangian subgroup containing the element $(1,0,1,0)$ is unique and is generated by
\begin{align}
    G = \begin{bmatrix}
    1 & 0 & 1 & 0 \\
    0 & 1 & 0 & 1
\end{bmatrix}\,.
\end{align}
Correspondingly, the NS partition function is
\begin{align} \label{eq:spin12^2_NS}
\begin{aligned}
    Z_{\mathrm{NS}}(\tau) &= \chi_{00}^2 + \chi_{01}^2 + \chi_{10}^2 + \chi_{11}^2\,,\\
    &= \frac{1}{\sqrt{q}}+276\, \sqrt{q}+2048 \,q+11202\, q^{3/2}+ O(q^2)\,.
\end{aligned}
\end{align}
From the $q$-expansion, this theory turns out to be the $\CN=1$ supermoonshine module at central charge $c=12$~\cite{duncan2007super}.

\paragraph{\underline{$\Spin(16)\times\Spin(8)$}}

The discriminant group is $\CD\cong \BZ_2^2\times \BZ_2^2$ where the first part corresponds to $\Spin(16)$ and the second does to $\Spin(8)$.
The supercurrents are in the representation corresponding to $(1,0,1,0)\in\BZ_2^2\times\BZ_2^2$. 
We enumerate all the odd Lagrangian subgroups containing $(1,0,1,0)$ up to the following operations:
\begin{itemize}
    \item Swapping $w_S$ and $w_C$ in $\mathrm{Spin}(16)$ or $\mathrm{Spin}(8)$
    \item Swapping $w_C$ and $w_S+w_C$ in $\mathrm{Spin}(8)$
\end{itemize}
The second equivalence is due to the triality of the three representations $w_S,w_C,w_S+w_C$.
Then, the odd Lagrangian subgroups containing $(1,0,1,0)$ are generated by
\begin{align}
    G_1 = \begin{bmatrix}
        1 & 0 & 0 & 0\\
        0 & 0 & 1 & 0
    \end{bmatrix}\,,\quad
    G_2 = \begin{bmatrix}
        1 & 0 & 1 & 0\\
        0 & 1 & 0 & 1
    \end{bmatrix}\,.
\end{align}
Each generator gives the NS partition function
\begin{align}
\begin{aligned}
    Z_{1,\mathrm{NS}}(\tau) &= (\chi_{00}+\chi_{10})(\psi_{00}+\psi_{10})\,,\\
    Z_{2,\mathrm{NS}}(\tau) &= \chi_{00}\,\psi_{00} +  \chi_{01}\,\psi_{01} +
    \chi_{10}\,\psi_{10} + \chi_{11}\,\psi_{11}\,,
\end{aligned}
\end{align}
where $\chi$ and $\psi$ denote the affine characters corresponding to $\Spin(16)$ and $\Spin(8)$, respectively.
The $q$-expansion of these partition functions are
\begin{align}
    \begin{aligned}
        Z_{1,\mathrm{NS}}(\tau) 
            &= 
                \frac{1}{\sqrt{q}}+8+276\, \sqrt{q}+2048\, q+11202\, q^{3/2}+ O(q^2)\,,\\
        Z_{2,\mathrm{NS}}(\tau) 
            &=  \frac{1}{\sqrt{q}}+276\, \sqrt{q}+2048\, q+11202\, q^{3/2}+ O(q^2)\,.
    \end{aligned}
\end{align}
Thus, the first theory is the $E_8$ lattice CFT plus four pairs of Majorana-Weyl fermions. The second theory is the $\CN=1$ supermoonshine module.

\paragraph{\underline{$\Spin(24)$}}

The discriminant group is $\CD \cong \BZ_2^2$.
There are three odd Lagrangian subgroups, each of which is generated by $G_1 =  [1,0]$, $G_2=[0,1]$, and $G_3=[1,1]$.
The first Lagrangian subgroup $\CC = \{00,10\}$ contains a supercurrent in the representation $[w_S]\simeq (1,0)$.
Thus, the NS partition function is
\begin{align}
    Z_{\mathrm{NS}}(\tau) = \chi_{00} + \chi_{10}=\frac{1}{\sqrt{q}}+276\, \sqrt{q}+2048\, q+11202\, q^{\frac{3}{2}}+ O(q^2)\,.
\end{align}
Again, this theory is the $\CN=1$ supermoonshine module~\cite{duncan2007super}.

\paragraph{\underline{$\Spin(4m+2)^3$}}
The discriminant group is $\CD\cong\BZ_4^3$.
The supercurrents are in the representation $(2,2,2)$.

The odd Lagrangian subgroup containing $(2,2,2)$ is unique and generated by $G=\mathrm{diag}(2,2,2)$.
The corresponding NS partition function is
\begin{align}
    Z_{\mathrm{NS}}(\tau) = (\chi_0+\chi_2)^3\,.
\end{align}
This theory can be described by $(12m+6)$ pairs of Majorana-Weyl fermions.

\paragraph{\underline{$\SU(6)^2$}}
The discriminant group $\CD$ is generated by $[(w^1,0)]$ and $[(0,w^1)]$, and isomorphic to $\BZ_6\times \BZ_6$.
From $w^1\oslash w^1=5/6$, the inner product is $(a,b)\oslash (a',b') = \frac{5}{6}(a\,a'+b\,b')$ for $(a,b),(a',b')\in\BZ_6^2$.
However, a straightforward computation shows the absence of any Lagrangian subgroup in the discriminant group $\CD$.

\paragraph{\underline{$\SU(12)$}}
The discriminant group is $\CD = \BZ_{12}$, but this does not have any Lagrangian subgroup as discussed in section~\ref{sec:affine-chiral}.
Thus, no chiral CFTs can be constructed in this case.

\paragraph{\underline{$E_7^2$}}

The discriminant group is $\mathcal{D}\simeq \mathbb{Z}_2\times\mathbb{Z}_2$. Each $\BZ_2$ is generated by $[w_\mathbf{56}]$.
A supercurrent is in the representation $([w_\mathbf{56}],[w_\mathbf{56}])\cong (1,1)$.
A unique odd Lagrangian subgroup is generated by $G = [1,1]$ and contains the supercurrent.
The NS partition function is
\begin{equation}
    Z_\mathrm{NS}(\tau) =  \chi_{0}^2 + \chi_{1}^2 = q^{-\frac{7}{12}} + 266\,q^{\frac{5}{12}} + 3136\,q^{\frac{11}{12}} + O(q^{\frac{17}{12}})\,.
\end{equation}
An element of the shadow is $s=(0,1)$.
The R and $\widetilde{\mathrm{R}}$ partition functions are
\begin{equation}
    Z_\mathrm{R}(\tau) = 2\chi_{0}\chi_1=112\, q^{\frac{1}{6}} + 16832\,q^{\frac{7}{6}} + O(q^{\frac{13}{6}})\,,\qquad Z_{\widetilde{\mathrm{R}}}(\tau) = 0\,.
\end{equation}
From the above expression, the theory does not have the supersymmetric Ramond vacua, and hence supersymmetry is broken.

\subsection{Non-chiral superconformal field theories}
\label{sec:affine-nonchiral}

In this subsection, we classify non-chiral CFTs in which both the left-moving and right-moving sectors are associated with level-one affine Lie algebras for simply laced Lie algebras and possess supersymmetry.\footnote{For an approach toward a similar problem based on the fermionization of WZW model, see~\cite{Bae:2021lvk}.}

For classification, in addition to the identification in the chiral case, we also identify theories that are transformed into each other by the inversion $w^1\to -w^1$ in each $\SU(m)$ or swapping the left-moving sector and the right-moving sector. In terms of the partition function, the latter corresponds to taking complex conjugates for all characters.
Summarizing the discussion from the previous section, we identify theories in the classification that are equivalent under the transformations listed below:
\begin{itemize}
    \item Swapping $w_S$ and $w_C$ in $\Spin(4m)$ ; $\chi_{10}\leftrightarrow\chi_{01}$
    \item Swapping $w_C$ and $w_S+w_C$ in $\Spin(8)$ ; $\chi_{01}\leftrightarrow\chi_{11}$
    \item Replacing $w^1$ with $-w^1$ in $\SU(m)$ ; $\chi_i\leftrightarrow\chi_{-i}$
    \item Permuting the $u$ copies of $H$ in $G=H^u$ ($H$: simple Lie group) ; $\chi_l^a\rightarrow\chi_l^{\sigma(a)}$
    \item Swapping the left-moving sector and the right-moving sector ; $\chi_l\leftrightarrow\bar{\chi}_l$
\end{itemize}
As mentioned in the previous section, $\Spin(8)$ has the triality, and by combining the first and second points, we can permute $w_S,w_C,$ and $w_S+w_C$.
For the fourth point, we explicitly indicated which $H$ the character originates from and represented the permutation by $\sigma\in \mathrm{Sym}(u)$.

First, we consider the case where the Lie algebras for both sectors are the same.

To illustrate the idea in the non-chiral case, let us consider an example where $G=G'=\SU(12)$. The discriminant group $\mathcal{D}$ is isomorphic to $\BZ_{12}^2$ with $[(w^1,0)]\simeq(1,0)$ and $[(0,w^1)]\simeq(0,1)$. From $w^1 \cdot w^1 = 11/12$, 
the quadratic form~$Q$ is given by $Q((x,y)) = \frac{11}{24}(x^2-y^2) \text{ mod } \BZ$ for $(x,y)\in \BZ_{12}^2$.
From table~\ref{tab:susy_lie_groups}, the supercurrents of the left-moving and right-moving sectors are in the representations corresponding to $(6,0)$ and $(0,6)$, respectively. A straightforward calculation shows that there exists two Lagrangian subgroups that contain both elements:
\begin{equation}
    \mathcal{C}^{(1)} \simeq \langle\,(2,2),(0,6)\,\rangle \,,\quad
    \mathcal{C}^{(2)} \simeq \langle\,(2,10),(0,6)\,\rangle \,.
\end{equation}
Note that since $(6,0)$ and $(0,6)$ are included, these must necessarily be odd.

For the first case, we can take $s=(1,1) \in S$ where $S$ is the shadow of $\mathcal{C}^{(1)}$ and then the subsets $\mathcal{C}_0$, $\mathcal{C}_2$ of $\mathcal{C}^{(1)}$ and $\mathcal{C}_1$, $\mathcal{C}_3$ of $S$
are
\begin{align}
\begin{aligned}
    \mathcal{C}_0 &= \{ (0,0), (2,2), (4,4), (6,6), (8,8), (10,10)\} \,, \\
    \mathcal{C}_2 &= \{ (0,6), (2,8), (4,10), (6,0), (8,2), (10,4) \} \,, \\
    \mathcal{C}_1 &= \{ (1,1), (3,3), (5,5), (7,7), (9,9), (11,11) \} \,, \\
    \mathcal{C}_3 &= \{ (1,7), (3,9), (5,11), (7,1), (9,3), (11,5) \} \,.
\end{aligned}
\end{align}
From \eqref{eq:Z_by_C0123},
the partition functions are
\begin{align}
\begin{aligned}
    Z_{\mathrm{NS}}(\tau,\bar{\tau}) &= |\chi_0 + \chi_6|^2 + |\chi_2 + \chi_8|^2 + |\chi_4 + \chi_{10}|^2 \,, \\
    Z_{\widetilde{\mathrm{NS}}}(\tau,\bar{\tau}) &= |\chi_0 - \chi_6|^2 + |\chi_2 - \chi_8|^2 + |\chi_4 - \chi_{10}|^2 \,, \\
    Z_{\mathrm{R}}(\tau,\bar{\tau}) &= |\chi_1 + \chi_7|^2 + |\chi_3 + \chi_9|^2 + |\chi_5 + \chi_{11}|^2 \,, \\
    Z_{\widetilde{\mathrm{R}}}(\tau,\bar{\tau}) &= |\chi_1 - \chi_7|^2 + |\chi_3 - \chi_9|^2 + |\chi_5 - \chi_{11}|^2 = 2 \cdot 12^2 \,.
\end{aligned}
\end{align}

Similarly, in the second case, the partition functions are
\begin{align}
\begin{aligned}
    Z_{\mathrm{NS}}(\tau,\bar{\tau}) &= |\chi_0 + \chi_6|^2 + \Bigl[ (\chi_2 + \chi_8) (\bar{\chi}_{10} + \bar{\chi}_4) + (\text{c.c.}) \Bigr] \,, \\
    Z_{\widetilde{\mathrm{NS}}}(\tau,\bar{\tau}) &=  |\chi_0 - \chi_6|^2 + \Bigl[ (\chi_2 - \chi_8) (\bar{\chi}_{10} - \bar{\chi}_4) + (\text{c.c.}) \Bigr] \,, \\
    Z_{\mathrm{R}}(\tau,\bar{\tau}) &= |\chi_3 + \chi_9|^2 + \Bigl[ (\chi_1 + \chi_7) (\bar{\chi}_{11} + \bar{\chi}_5) + (\text{c.c.}) \Bigr] \,, \\
    Z_{\widetilde{\mathrm{R}}}(\tau,\bar{\tau}) &= - |\chi_3 - \chi_9|^2 + \Bigl[ (\chi_1 - \chi_7) (\bar{\chi}_{11} - \bar{\chi}_5) + (\text{c.c.}) \Bigr] = 2 \cdot 12^2 \,.
\end{aligned}
\end{align}
It is clear from these expressions that two theories transform into each other by taking conjugate for the right-moving sector, i.e., $\bar{\chi}_i \leftrightarrow \bar{\chi}_{-i}$. As already mentioned, they are considered equivalent in our classification.

The classification for each Lie algebra is shown below, according to table~\ref{tab:susy_lie_groups}. We describe some features of the theories here, with the generator matrices of the Lagrangian subgroups, the element of the shadow to define the fermion parity in the R sector, and the NS partition functions provided in Appendix~\ref{app:gen_mat}.

\paragraph{\underline{$\Spin(4m)^3$}}
The supercurrents are in the representations corresponding to 
\begin{equation}
    (1,1,1,1,1,1,0,0,0,0,0,0),\, (0,0,0,0,0,0,1,1,1,1,1,1) \in (\BZ_2\times \BZ_2)^6 \,.    
\end{equation}
For any $m$, there are 15 Lagrangian subgroups that contain these elements.
The characteristics of the corresponding CFTs are summarized in table~\ref{tab:spin4m^3_CFTs}, depending on whether $m$ is even or odd.

\begin{table}[t]
    \begin{subtable}{\textwidth}
        \centering
        \begin{tblr}{c|ccccccccccccccc}
        \toprule
        label $i$ & $1$ & $2$ & $3$ & $4$ & $5$ & $6$ & $7$ & $8$ & $9$ & $10$ & $11$ & $12$ & $13$ & $14$ & $15$ \\ \hline
        $Z_{i,\widetilde{\mathrm{R}}}/4^{2m-1}$ & $0$ & $3$ & $3$ & $6$ & $6$ & $0$ & $0$  & $0$  & $0$  & $9$  & $0$  & $0$  & $0$  & $0$  & $0$ \\
        \#MW$/(4m)$ (left) & $0$ & $0$ & $0$ & $0$ & $0$ & $1$ & $1$ & $1$ & $1$ & $0$ & $1$ & $1$ & $3$ & $3$ & $3$ \\
        \#MW$/(4m)$ (right) & $0$ & $0$ & $0$ & $0$ & $0$ & $0$ & $0$ & $1$ & $1$ & $0$ & $0$ & $1$ & $0$ & $1$ & $3$ \\
        SSB (left) &&&&&&&&&&&&& $\checkmark$ & $\checkmark$ & $\checkmark$ \\
        SSB (right) &&&&&&&&&&&&&&& $\checkmark$ \\
        factorizable &&&&&&&&&& $\checkmark$ & $\checkmark$ & $\checkmark$ & $\checkmark$ & $\checkmark$ & $\checkmark$ \\
        \bottomrule
        \end{tblr}
        \caption{$m:\text{even}$}
    \end{subtable}
    \\ \\
    \begin{subtable}{\textwidth}
        \centering
        \begin{tblr}{c|ccccccccccccccc}
        \toprule
        label $i$ & $1$ & $2$ & $3$ & $4$ & $5$ & $6$ & $7$ & $8$ & $9$ & $10$ & $11$ & $12$ & $13$ & $14$ & $15$ \\ \hline
        $Z_{i,\widetilde{\mathrm{R}}}/4^{2m-1}$ & $6$ & $6$ & $6$ & $0$ & $0$ & $0$ & $0$  & $0$  & $0$  & $0$  & $0$  & $0$  & $0$  & $0$  & $0$ \\
        \#MW$/(4m)$ (left) $^\ast$ & $0$ & $0$ & $0$ & $0$ & $0$ & $1$ & $1$ & $1$ & $1$ & $0$ & $1$ & $1$ & $3$ & $3$ & $3$ \\
        \#MW$/(4m)$ (right) $^\ast$ & $0$ & $0$ & $0$ & $0$ & $0$ & $0$ & $0$ & $1$ & $1$ & $0$ & $0$ & $1$ & $0$ & $1$ & $3$ \\
        SSB (left) &&&&&&&&&& $\checkmark$ & $\checkmark$ & $\checkmark$ & $\checkmark$ & $\checkmark$ & $\checkmark$ \\
        SSB (right) &&&&&&&&&& $\checkmark$ & $\checkmark$ & $\checkmark$ & $\checkmark$ & $\checkmark$ & $\checkmark$ \\
        factorizable &&&&&&&&&& $\checkmark$ & $\checkmark$ & $\checkmark$ & $\checkmark$ & $\checkmark$ & $\checkmark$ \\
        \bottomrule
        \end{tblr}
        \caption{$m:\text{odd}$}
    \end{subtable}
    \caption{The characteristics of the CFTs corresponding to the Lagrangian subgroups associated with $\Spin(4m)^3$. $Z_{i,\widetilde{\mathrm{R}}}$: the $\widetilde{\mathrm{R}}$ partition function, \#MW: the number of Majorana-Weyl (MW) fermions, SSB: whether supersymmetry is spontaneously broken, and factorizable: whether the theory is the direct product of the left-moving and right-moving sectors. The asterisks indicate that the number of MW fermions is valid only when $m\neq 1$.}
    \label{tab:spin4m^3_CFTs}
\end{table}

When $m=1$, certain Lagrangian subgroups generate equivalent theories, resulting in three distinct CFTs. The first contains no Majorana-Weyl (MW) fermions (corresponding to labels 1, 2, and 3 in the table); the second contains 4 MW fermions in both the left-moving and right-moving sectors (labels 4, 5, 6, 7, 8, and 9); and the third contains 12 MW fermions in both sectors, i.e., it consists entirely of MW fermions (labels 10, 11, 12, 13, 14, and 15). For other values of $m$, all theories are distinct.

\paragraph{\underline{$\Spin(12)^2$}}
The supercurrents are in the representations corresponding to
\begin{equation}
    (1,0,1,0,0,0,0,0),\, (0,0,0,0,1,0,1,0) \in (\BZ_2\times \BZ_2)^4 \,.
\end{equation}
There are two Lagrangian subgroups that contain these elements. The NS and $\widetilde{\mathrm{R}}$ partition functions are
\begin{align}
\begin{aligned}
    Z_{1,\mathrm{NS}}(\tau,\bar{\tau}) &= |\chi_{00}^2 + \chi_{10}^2|^2 + 2|\chi_{00}\, \chi_{01} + \chi_{10}\, \chi_{11}|^2 + |\chi_{01}^2 + \chi_{11}^2|^2 \,, \\
    Z_{1,\widetilde{\mathrm{R}}}(\tau,\bar{\tau}) &= 2 |\chi_{00}\,\chi_{11} - \chi_{01}\,\chi_{10}|^2 = 2\cdot 12^2
\end{aligned}
\end{align}
in the first theory and
\begin{align}
\begin{aligned}
    Z_{2,\mathrm{NS}}(\tau,\bar{\tau}) &= |\chi_{00}^2 + \chi_{01}^2 + \chi_{10}^2 + \chi_{11}^2|^2 \,, \\
    Z_{2,\widetilde{\mathrm{R}}}(\tau,\bar{\tau}) &= 4 |\chi_{00}\,\chi_{11} - \chi_{01}\,\chi_{10}|^2 = 4\cdot 12^2
\end{aligned}
\end{align}
in the second theory.

In cases like this, for certain Lie groups, there are only one or two Lagrangian subgroups corresponding to supersymmetric CFTs. The characteristics of such theories are summarized in table~\ref{tab:susy_CFTs}.
\begin{table}[t]
    \centering
    \begin{tblr}{c||cc|c|c|c|c}
    \toprule
    $G$ & \SetCell[c=2]{} $\Spin(12)^2$ && $\Spin(24)$ & $\SU(6)^2$ & $\SU(12)$ & \;$E_7^2$\;  \\ \hline
    $c$ & $12$ & $12$ & $12$ & $10$ & $11$ & $14$ \\
    $Z_{\widetilde{\mathrm{R}}}$ & $288$ & $576$ & $576$ & $144$ & $288$ & $0$ \\
    \#MW (left) & $0$ & $0$ & $0$ & $0$ & $0$ & $0$ \\
    \#MW (right) & $0$ & $0$ & $0$ & $0$ & $0$ & $0$ \\
    SSB (left) &&&&&& $\checkmark$ \\
    SSB (right) &&&&&& $\checkmark$ \\
    factorizable && $\checkmark$ & $\checkmark$ &&& $\checkmark$ \\
    \bottomrule
    \end{tblr}
    \caption{The simply connected Lie groups that admit one or two Lagrangian subgroups corresponding to supersymmetric CFTs. The symbol $c$ denotes the rank, and all other notations for the corresponding CFTs are consistent with those used in table~\ref{tab:spin4m^3_CFTs}.}
    \label{tab:susy_CFTs}
\end{table}

\paragraph{\underline{$\Spin(16)\times \Spin(8)$}}
The supercurrents are in the representations corresponding to
\begin{equation}
    (1,0,1,0,0,0,0,0),\, (0,0,0,0,1,0,1,0) \in (\BZ_2\times \BZ_2)^4 \,.    
\end{equation}
There are five Lagrangian subgroups that contain these elements. The characteristics of the corresponding CFTs are summarized in table~\ref{tab:spin16spin8_CFTs}.

\begin{table}[t]
    \centering
    \begin{tblr}{c|ccccc}
    \toprule
    label $i$ & $1$ & $2$ & $3$ & $4$ & $5$ \\ \hline
    $Z_{i,\widetilde{\mathrm{R}}}$ & $384$ & $192$ & $576$ & $0$ & $0$ \\
    \#MW (left) & $0$ & $0$ & $0$ & $8$ & $8$ \\
    \#MW (right) & $0$ & $0$ & $0$ & $0$ & $8$ \\
    SSB (left) &&&&& \\
    SSB (right) &&&&& \\
    factorizable &&& $\checkmark$ & $\checkmark$ & $\checkmark$ \\
    \bottomrule
    \end{tblr}
    \caption{The characteristics of the CFTs corresponding to the Lagrangian subgroups associated with $\Spin(16)\times\Spin(8)$.}
    \label{tab:spin16spin8_CFTs}
\end{table}

\paragraph{\underline{$\Spin(24)$}}
The supercurrents are in the representations corresponding to
\begin{equation}
    (1,0,0,0),\, (0,0,1,0) \in (\BZ_2\times \BZ_2)^2 \,.
\end{equation}
There is only one Lagrangian subgroup that contains these elements. The NS and $\widetilde{\mathrm{R}}$ partition functions are
\begin{align}
    Z_{\mathrm{NS}}(\tau,\bar{\tau}) = |\chi_{00} + \chi_{10}|^2 \,, \qquad
    Z_{\widetilde{\mathrm{R}}}(\tau,\bar{\tau}) = |\chi_{01} - \chi_{11}|^2 = 24^2 \,.
\end{align}

\paragraph{\underline{$\Spin(4m+2)^3$}}
The supercurrents are in the representations corresponding to
\begin{equation}
    (2,2,2,0,0,0),\, (0,0,0,2,2,2) \in (\BZ_4)^6 \,.
\end{equation}
For any $m$, there are three Lagrangian subgroups that contain these elements. The characteristics of the corresponding CFTs are summarized in table~\ref{tab:spin4m+2^3_CFTs}.

\begin{table}
    \centering
    \begin{tblr}{c|ccc}
    \toprule
    label $i$ & $1$ & $2$ & $3$  \\ \hline
    $Z_{i,\widetilde{\mathrm{R}}}$ & $6\cdot 4^{2m}$ & $0$ & $0$ \\
    \#MW (left) & $0$ & $4m+2$ & $12m+6$ \\
    \#MW (right) & $0$ & $4m+2$ & $12m+6$ \\
    SSB (left) &&& $\checkmark$ \\
    SSB (right) &&& $\checkmark$ \\
    factorizable &&& $\checkmark$ \\
    \bottomrule
    \end{tblr}
    \caption{The characteristics of the CFTs corresponding to the Lagrangian subgroups associated with $\Spin(4m+2)^3$.}
    \label{tab:spin4m+2^3_CFTs}
\end{table}

\paragraph{\underline{$\SU(6)^2$}}
The supercurrents are in the representations corresponding to
\begin{equation}
    (3,3,0,0),\, (0,0,3,3) \in (\BZ_6)^4 \,.
\end{equation}
There is only one Lagrangian subgroup that contains these elements. The NS and $\widetilde{\mathrm{R}}$ partition functions are
\begin{align}
\begin{aligned}
    Z_{\mathrm{NS}}(\tau,\bar{\tau}) &= |\chi_0^2 + \chi_3^2|^2 + |\chi_1^2 + \chi_4^2|^2 + |\chi_2^2 + \chi_5^2|^2 \\
    &\quad + 2 |\chi_0\, \chi_2 + \chi_3\, \chi_5|^2 + 2 |\chi_0\, \chi_4 + \chi_3\, \chi_1|^2 + 2 |\chi_1\, \chi_5 + \chi_2\, \chi_4|^2 \,, \\
    Z_{\widetilde{\mathrm{R}}}(\tau,\bar{\tau}) &= 2 |\chi_0\,\chi_1 - \chi_3\,\chi_4|^2 + 2 |\chi_0\,\chi_5 - \chi_3\,\chi_2|^2 + 2 |\chi_1\,\chi_2 - \chi_4\,\chi_5|^2
    = 4\cdot 6^2 \,.
\end{aligned}
\end{align}

\paragraph{\underline{$\SU(12)$}}
As we have already seen in the example, the supercurrents are in the representations corresponding to
\begin{equation}
    (6,0),\, (0,6) \in (\BZ_{12})^2 \,.
\end{equation}
There is only one Lagrangian subgroup that contains these elements. The NS and $\widetilde{\mathrm{R}}$ partition functions are
\begin{align}
\begin{aligned}
    Z_{\mathrm{NS}}(\tau,\bar{\tau}) &= |\chi_0 + \chi_6|^2 + |\chi_2 + \chi_8|^2 + |\chi_4 + \chi_{10}|^2 \,, \\
    Z_{\widetilde{\mathrm{R}}}(\tau,\bar{\tau}) &= |\chi_1 - \chi_7|^2 + |\chi_3 - \chi_9|^2 + |\chi_5 - \chi_{11}|^2 = 2 \cdot 12^2 \,.
\end{aligned}
\end{align}

\paragraph{\underline{$E_7^2$}}
The supercurrents are in the representations corresponding to
\begin{equation}
    (1,1,0,0),\, (0,0,1,1) \in (\BZ_2)^4 \,.
\end{equation}
There is only one Lagrangian subgroup that contains these elements. The NS and $\widetilde{\mathrm{R}}$ partition functions are
\begin{align}
    Z_{\mathrm{NS}}(\tau,\bar{\tau}) = |\chi_0^2 + \chi_1^2|^2 \,, \qquad
    Z_{\widetilde{\mathrm{R}}}(\tau,\bar{\tau}) = 0 \,.
\end{align}

Next, we consider the case where the Lie algebras for the left-moving and right-moving sectors are different.
In table~\ref{tab:susy_lie_groups}, Lie algebras with the same rank exist only at $c=12$ and the candidate Lie groups are $\Spin(8)^3,\, \Spin(12)^2,\, \Spin(16)\times\Spin(8)$, and $\Spin(24)$. Among the six combinations, except for that of $\Spin(8)^3$ and $\Spin(16) \times \Spin(8)$, superconformal theories exist only in forms that can be expressed as the direct product of the left-moving and right-moving sectors, which are already classified in section \ref{sec:susy_chiral}. For example, when $G=\Spin(8)^3$ and $G'=\Spin(12)^2$, there exist three theories and the NS partition functions are the product of \eqref{eq:spin4m^3_NS} and the complex conjugate of \eqref{eq:spin12^2_NS}.
The classification for $G=\Spin(8)^3$ and $G'=\Spin(16) \times \Spin(8)$ is shown below.

\paragraph{\underline{$G=\Spin(8)^3,\, G'=\Spin(16) \times \Spin(8)$}}
The supercurrents are in the representation corresponding to
\begin{equation}
    (1,1,1,1,1,1,0,0,0,0),\, (0,0,0,0,0,0,1,0,1,0) \in (\BZ_2\times\BZ_2)^5 \,.
\end{equation}
There are ten Lagrangian subgroups that contain these elements. The characteristics of the corresponding CFTs are summarized in table~\ref{tab:LneqR_CFTs}.

\begin{table}
    \centering
    \begin{tblr}{c|cccccccccc}
    \toprule
    label $i$ & $1$ & $2$ & $3$ & $4$ & $5$ & $6$ & $7$ & $8$ & $9$ & $10$ \\ \hline
    $Z_{i,\widetilde{\mathrm{R}}}$ & $384$ & $192$ & $0$ & $0$ & $576$ & $0$  & $0$  & $0$  & $0$  & $0$ \\
    \#MW (left) & $0$ & $0$ & $8$ & $8$ & $0$ & $0$ & $8$ & $8$ & $24$ & $24$ \\
    \#MW (right) & $0$ & $0$ & $0$ & $0$ & $0$ & $8$ & $0$ & $8$ & $0$ & $8$ \\
    SSB (left) &&&&&&&&& $\checkmark$ & $\checkmark$ \\
    SSB (right) &&&&&&&&&& \\
    factorizable &&&&& $\checkmark$ & $\checkmark$ & $\checkmark$ & $\checkmark$ & $\checkmark$ & $\checkmark$ \\
    \bottomrule
    \end{tblr}
    \caption{The characteristics of the CFTs corresponding to the Lagrangian subgroups associated with $G=\Spin(8)^3,\, G'=\Spin(16) \times \Spin(8)$.}
    \label{tab:LneqR_CFTs}
\end{table}

\section{Discussion}
\label{sec:discussion}

In this work, we studied the two-dimensional CFTs that can be constructed from bosonic Chern-Simons theories by imposing topological boundary conditions.  We mainly focused on the fermionic CFTs, which correspond to the topological boundary conditions specified by odd Lagrangian subgroups of the discriminant group.  As an application of our construction, we studied the fermionic generalization of code CFTs.  Another application was the construction of fermionic CFTs with level-one affine Lie algebra symmetries for simply laced Lie algebras.  In particular, we considered the Chern-Simons theories corresponding to a class of SVOAs which have affine Lie algebras as their bosonic parts, and which Johnson-Freyd showed to possess supersymmetry.  We classified the fermionic topological boundary conditions that give rise to full supersymmetric fermionic CFTs with modular covariant combinations of characters.

Let us discuss future directions.
In the present paper, we focused on the torus partition functions of the CFTs.  It would be interesting to generalize our Chern-Simons construction to more general CFT correlators with higher genus or operator insertions.
The recent paper~\cite{Wang:2020nmz} classified all abelian anyon theories and gave an algorithm to construct the level matrix $K$ in abelian Chern-Simons theory.  A worthwhile open problem is to classify bosonic and fermionic topological boundaries for a general anyon theory based on the results of~\cite{Wang:2020nmz}.
On the other hand, reference~\cite{Gannon:1992nq} classified the modular invariant combinations of level-one affine characters, thus classifying the bosonic CFTs that can be obtained from level-one affine Lie algebras.  It would interesting to classify the fermionic CFTs that can be similarly constructed, by enumerating the fermionic topological boundary conditions.

Another direction would be to explore a possible connection with the work~\cite{Benini:2022hzx}.\footnote{We thank the JHEP referee for pointing out the potential relevance of~\cite{Benini:2022hzx}.}
There, 3d abelian Chern-Simons theories are treated as candidate gravitational theories, and their one-form symmetries given by a Lagrangian subgroup~$\mathcal{C}$ of the discriminant group~$\mathcal{D}$ are gauged so that the resulting gravitational theory has no global symmetry and that the path integral becomes independent of the bulk topology.
In our set-up where Chern-Simons theories are non-gravitational, the Lagrangian subgroup~$\mathcal{C}$ specifies a topological boundary condition.
In other words, anyon condensation for $\mathcal{C}$ occurs in the bulk in~\cite{Benini:2022hzx}, while it occurs along the topological boundary in this work.
As discussed in~\cite{Kaidi:2021gbs} for example, the topological boundary condition may be obtained by introducing a fictitious region that shares the same boundary with the physical region and by gauging the one-form symmetry~$\mathcal{C}$ in the bulk of the fictitious region.
If one shrinks the physical region to zero size, which can be done without affecting physics because the Chern-Simons theory is topological, the set-up of the current work seem to reduce to that of~\cite{Benini:2022hzx}.
It would be interesting to study this connection further.

\acknowledgments

We are grateful to Yoshiki~Fukusumi for valuable discussions.
The researches of K.\,K. and S.\,Y. are supported in part by FoPM, WINGS Program, the University of Tokyo.
The research of K.\,K. is also supported by JSPS KAKENHI Grant-in-Aid for JSPS fellows Grant No.\,23KJ0436.
The research of T.\,O. is supported in part by Grant-in-Aid for Transformative Research Areas (A) ``Extreme Universe'' No.\,21H05190 and by JST PRESTO Grant Number JPMJPR23F3.
The research of T.\,N. is supported in part by the JSPS Grant-in-Aid for Scientific Research (B) No.\,24K00629, Grant-in-Aid for Scientific Research (A) No.\,21H04469, and
Grant-in-Aid for Transformative Research Areas (A) ``Extreme Universe'' No.\,21H05182 and No.\,21H05190.

\newpage 
\appendix

\section{Explicit expressions for level-one affine characters}
\label{app:characters}

Here we give explicit expressions for some characters of level-one affine Lie algebras, taken for example from~\cite[Chap.4]{conway2013sphere}.
As we noted below~(\ref{eq:D-weight-root}), there is a one-to-one correspondence between the center $Z(G)$ of the simply connected and simply laced Lie group~$G$ (listed in Table~\ref{table:centers}) and the integrable highest-weight representations of the level-one affine Lie algebra for the corresponding Lie algebra.
We thus use the elements of $Z(G)$ to label the representations.
The expressions are used in section~\ref{sec:susy_chiral}.

For $G=\SU(m)$,
\begin{equation}
    \chi_l = \frac{\sum_{k=0}^{m-1} \zeta^{-kl} \left(\sum_{n\in\BZ}\zeta^{kn}q^{\frac{1}{2}n^2}\right)^m}{\eta^{m-1} m \sum_{n\in\BZ}q^{\frac{m}{2}(n+\frac{l}{m})^2}}
\end{equation}
where $\zeta=e^{2\pi\i/m}$ and $l=0,\dots,m-1$.

For $G=\Spin(4m)$,
\begin{equation}
    \chi_{00} = \frac{\theta_3^k + \theta_4^k}{2 \eta^k} \,,\quad
    \chi_{11} = \frac{\theta_3^k - \theta_4^k}{2 \eta^k} \,,\quad
    \chi_{01} = \chi_{10} = \frac{\theta_2^k}{2 \eta^k} \,,
\end{equation}
where $k=2m$.

For $G=\Spin(4m+2)$,
\begin{equation}
    \chi_0 = \frac{\theta_3^k + \theta_4^k}{2 \eta^k} \,,\quad
    \chi_2 = \frac{\theta_3^k - \theta_4^k}{2 \eta^k} \,,\quad
    \chi_1 = \chi_3 = \frac{\theta_2^k}{2 \eta^k} \,,
\end{equation}
where $k=2m+1$.

For $G=E_7$,
\begin{equation}
    \chi_0 = \frac{\theta_3(2\tau)^7+7\theta_3(2\tau)^3\theta_2(2\tau)^4}{\eta(\tau)^7} \,,\quad
    \chi_1 = \frac{\theta_2(2\tau)^7+7\theta_2(2\tau)^3\theta_3(2\tau)^4}{\eta(\tau)^7} \,.
\end{equation}

Table~\ref{tab:rep_weights} shows the conformal weights of the representations for the Lie groups examined in sections~\ref{sec:susy_chiral}, \ref{sec:affine-nonchiral}.

\begin{table}[t]
    \begin{subtable}{\textwidth}
        \centering
        \begin{tblr}{c|cccccc}
        \toprule
        label $i$ & $0$ & $1$ & $2$ & $3$ & $4$ & $5$ \\ \hline
        weight $h$ & $0$ & $\frac{5}{12}$ & $\frac{2}{3}$ & $\frac{3}{4}$ & $\frac{2}{3}$ & $\frac{5}{12}$ \\
        \bottomrule
        \end{tblr}
        \caption{$\SU(6)$}
    \end{subtable}
    \\ \\
    \begin{subtable}{\textwidth}
        \centering
        \begin{tblr}{c|cccccccccccc}
        \toprule
        label $i$ & $0$ & $1$ & $2$ & $3$ & $4$ & $5$ & $6$ & $7$ & $8$ & $9$ & $10$ & $11$ \\ \hline
        weight $h$ & $0$ & $\frac{11}{24}$ & $\frac{5}{6}$ & $\frac{9}{8}$ & $\frac{4}{3}$ & $\frac{35}{24}$ & $\frac{3}{2}$ & $\frac{35}{24}$ & $\frac{4}{3}$ & $\frac{9}{8}$ & $\frac{5}{6}$ & $\frac{11}{24}$ \\
        \bottomrule
        \end{tblr}
        \caption{$\SU(12)$}
    \end{subtable}
    \\ \\
    \begin{subtable}{0.45\textwidth}
        \centering
        \begin{tblr}{c|cccc}
        \toprule
        label $i$ & $(0,0)$ & $(0,1)$ & $(1,1)$ & $(1,0)$ \\ \hline
        weight $h$ & $0$ & $\frac{m}{4}$ & $\frac{1}{2}$ & $\frac{m}{4}$ \\
        \bottomrule
        \end{tblr}
        \caption{$\Spin(4m)$}
    \end{subtable}
    \hfill
    \begin{subtable}{0.45\textwidth}
        \centering
        \begin{tblr}{c|cccc}
        \toprule
        label $i$ & $0$ & $1$ & $2$ & $3$ \\ \hline
        weight $h$ & $0$ & $\frac{2m+1}{8}$ & $\frac{1}{2}$ & $\frac{2m+1}{8}$  \\
        \bottomrule
        \end{tblr}
        \caption{$\Spin(4m+2)$}
    \end{subtable}
    \\ \\
    \begin{subtable}{\textwidth}
        \centering
        \begin{tblr}{c|cc}
        \toprule
        label $i$ & $0$ & $1$ \\ \hline
        weight $h$ & $0$ & $\frac{3}{4}$ \\
        \bottomrule
        \end{tblr}
        \caption{$E_7$}
    \end{subtable}
    \caption{The conformal weights of the representations.}
    \label{tab:rep_weights}
\end{table}

\newpage

\section{Generator matrices of Lagrangian subgroups}
\label{app:gen_mat}

In section \ref{sec:affine-nonchiral}, we classified non-chiral CFTs where both the left-moving and right-moving sectors are associated with level-one affine Lie algebras and possess supersymmetry. Here, we provide the generator matrices of the Lagrangian subgroups, the elements of the shadows, and the NS partition functions for the cases where the Lie groups for both sectors are the same (listed in table~\ref{tab:susy_lie_groups}) and where the Lie groups are $\Spin(8)^3$ for the left-moving sector and $\Spin(16) \times \Spin(8)$ for the right-moving sector.

For the discriminant group $\CD \cong \BZ_l^n$, the $k\times n$ generator matrix $G$ over $\BZ_l$ generates the Lagrangian subgroup $\CC\subset\CD$ by $\CC \cong \{ x\, G \mid x\in \BZ_l^k \}$.
The element of $\BZ_l^n$ below the matrix corresponds to $[s]\in S(\CC)$, which is used to fix the ambiguity of the fermion parity in the R sector as in section \ref{sec:shadow-lagrangian}.
The labels of the partition functions and the theories in the tables in section~\ref{sec:affine-nonchiral} correspond to the positions of the generator matrices.
The character is generally denoted by $\chi$, except that $\psi$ is used for the $\Spin(8)$ inside $\Spin(16)\times \Spin(8)$.

\renewcommand{\arraystretch}{0.8}
\allowdisplaybreaks[1]

\paragraph{\underline{$\Spin(4m)^3$}}

When $m$ is even, the generator matrices are as follows.
\[
\begin{array}{ccc}

\begin{bmatrix}
1 & 0 & 0 & 0 & 0 & 1 & 0 & 0 & 0 & 0 & 1 & 1 \\
0 & 1 & 0 & 0 & 0 & 1 & 0 & 0 & 0 & 1 & 0 & 1 \\
0 & 0 & 1 & 0 & 0 & 1 & 0 & 1 & 0 & 1 & 0 & 0 \\
0 & 0 & 0 & 1 & 0 & 1 & 0 & 1 & 1 & 0 & 1 & 1 \\
0 & 0 & 0 & 0 & 1 & 1 & 0 & 0 & 1 & 0 & 0 & 1 \\
0 & 0 & 0 & 0 & 0 & 0 & 1 & 1 & 1 & 1 & 1 & 1
\end{bmatrix}
, &
\begin{bmatrix}
1 & 0 & 0 & 0 & 0 & 1 & 0 & 0 & 0 & 0 & 1 & 1 \\
0 & 1 & 0 & 0 & 0 & 1 & 0 & 0 & 0 & 1 & 0 & 1 \\
0 & 0 & 1 & 0 & 0 & 1 & 0 & 0 & 1 & 1 & 1 & 1 \\
0 & 0 & 0 & 1 & 0 & 1 & 0 & 1 & 0 & 1 & 0 & 0 \\
0 & 0 & 0 & 0 & 1 & 1 & 0 & 1 & 1 & 1 & 0 & 1 \\
0 & 0 & 0 & 0 & 0 & 0 & 1 & 1 & 1 & 1 & 1 & 1
\end{bmatrix}
, &
\begin{bmatrix}
1 & 0 & 0 & 0 & 0 & 1 & 0 & 0 & 0 & 0 & 0 & 0 \\
0 & 1 & 0 & 0 & 1 & 0 & 0 & 0 & 0 & 1 & 0 & 1 \\
0 & 0 & 1 & 0 & 0 & 1 & 0 & 0 & 0 & 0 & 1 & 1 \\
0 & 0 & 0 & 1 & 0 & 1 & 0 & 0 & 0 & 1 & 1 & 0 \\
0 & 0 & 0 & 0 & 0 & 0 & 1 & 0 & 0 & 1 & 0 & 0 \\
0 & 0 & 0 & 0 & 0 & 0 & 0 & 1 & 1 & 0 & 1 & 1
\end{bmatrix}
, \\ \\

(0, 0, 0, 0, 0, 1, 0, 0, 0, 0, 0, 1) & 
(0, 0, 0, 0, 0, 1, 0, 1, 0, 1, 0, 1) & 
(0, 0, 0, 0, 0, 1, 0, 0, 0, 0, 1, 0) \\ \\
\end{array}
\]

The NS partition functions are as follows.
%

When $m$ is odd, the generator matrices are as follows.
\[
\begin{array}{ccc}

\begin{bmatrix}
1 & 0 & 0 & 0 & 0 & 1 & 0 & 0 & 0 & 0 & 1 & 1 \\
0 & 1 & 0 & 0 & 0 & 1 & 0 & 0 & 0 & 1 & 0 & 1 \\
0 & 0 & 1 & 0 & 0 & 1 & 0 & 0 & 1 & 0 & 0 & 1 \\
0 & 0 & 0 & 1 & 0 & 1 & 0 & 1 & 0 & 0 & 0 & 1 \\
0 & 0 & 0 & 0 & 1 & 1 & 0 & 1 & 1 & 1 & 1 & 0 \\
0 & 0 & 0 & 0 & 0 & 0 & 1 & 1 & 1 & 1 & 1 & 1
\end{bmatrix}
, &
\begin{bmatrix}
1 & 0 & 0 & 0 & 0 & 1 & 0 & 0 & 0 & 0 & 1 & 1 \\
0 & 1 & 0 & 0 & 0 & 1 & 0 & 0 & 0 & 1 & 0 & 1 \\
0 & 0 & 1 & 0 & 0 & 1 & 0 & 1 & 0 & 0 & 0 & 1 \\
0 & 0 & 0 & 1 & 0 & 1 & 0 & 1 & 1 & 1 & 1 & 0 \\
0 & 0 & 0 & 0 & 1 & 1 & 0 & 0 & 1 & 0 & 0 & 1 \\
0 & 0 & 0 & 0 & 0 & 0 & 1 & 1 & 1 & 1 & 1 & 1
\end{bmatrix}
, &
\begin{bmatrix}
1 & 0 & 0 & 0 & 0 & 1 & 0 & 0 & 0 & 1 & 0 & 1 \\
0 & 1 & 0 & 0 & 0 & 1 & 0 & 0 & 0 & 1 & 1 & 0 \\
0 & 0 & 1 & 0 & 0 & 1 & 0 & 1 & 0 & 1 & 0 & 0 \\
0 & 0 & 0 & 1 & 0 & 1 & 0 & 1 & 1 & 0 & 1 & 1 \\
0 & 0 & 0 & 0 & 1 & 1 & 0 & 0 & 1 & 1 & 0 & 0 \\
0 & 0 & 0 & 0 & 0 & 0 & 1 & 1 & 1 & 1 & 1 & 1
\end{bmatrix}
, \\ \\

(0, 0, 0, 0, 0, 1, 0, 0, 0, 0, 0, 1) & 
(0, 0, 0, 0, 0, 1, 0, 0, 0, 0, 0, 1) & 
(0, 0, 0, 0, 0, 1, 0, 0, 0, 1, 0, 0) \\ \\
\end{array}
\]

The NS partition functions are as follows.
%

\paragraph{\underline{$\Spin(12)^2$}}
\[
\begin{array}{cc}

\begin{bmatrix}
1 & 0 & 1 & 0 & 0 & 0 & 0 & 0 \\
0 & 1 & 0 & 0 & 0 & 0 & 0 & 1 \\
0 & 0 & 0 & 1 & 0 & 1 & 0 & 0 \\
0 & 0 & 0 & 0 & 1 & 0 & 1 & 0
\end{bmatrix}
, &
\begin{bmatrix}
1 & 0 & 1 & 0 & 0 & 0 & 0 & 0 \\
0 & 1 & 0 & 1 & 0 & 0 & 0 & 0 \\
0 & 0 & 0 & 0 & 1 & 0 & 1 & 0 \\
0 & 0 & 0 & 0 & 0 & 1 & 0 & 1
\end{bmatrix}
\\ \\

(0, 0, 1, 0, 1, 0, 0, 0) &
(0, 0, 1, 1, 0, 0, 1, 1)

\end{array}
\]

\begin{align*}
    Z_{1,\mathrm{NS}} &= |\chi_{00}^2 + \chi_{10}^2|^2 + 2|\chi_{00}\, \chi_{01} + \chi_{10}\, \chi_{11}|^2 + |\chi_{01}^2 + \chi_{11}^2|^2 \\
    Z_{2,\mathrm{NS}} &= |\chi_{00}^2 + \chi_{01}^2 + \chi_{10}^2 + \chi_{11}^2|^2
\end{align*}

\paragraph{\underline{$\Spin(16)\times \Spin(8)$}}

\[
\begin{array}{cccc}

\begin{bmatrix}
1 & 0 & 0 & 0 & 0 & 0 & 1 & 0 \\
0 & 1 & 0 & 1 & 0 & 1 & 0 & 1 \\
0 & 0 & 1 & 0 & 0 & 0 & 1 & 0 \\
0 & 0 & 0 & 0 & 1 & 0 & 1 & 0
\end{bmatrix}
, &
\begin{bmatrix}
1 & 0 & 0 & 0 & 0 & 1 & 0 & 1 \\
0 & 1 & 0 & 1 & 0 & 0 & 1 & 0 \\
0 & 0 & 1 & 0 & 0 & 1 & 0 & 1 \\
0 & 0 & 0 & 0 & 1 & 0 & 1 & 0
\end{bmatrix}
, &
\begin{bmatrix}
1 & 0 & 1 & 0 & 0 & 0 & 0 & 0 \\
0 & 1 & 0 & 1 & 0 & 0 & 0 & 0 \\
0 & 0 & 0 & 0 & 1 & 0 & 1 & 0 \\
0 & 0 & 0 & 0 & 0 & 1 & 0 & 1
\end{bmatrix}
, &
\begin{bmatrix}
1 & 0 & 0 & 0 & 0 & 0 & 0 & 0 \\
0 & 0 & 1 & 0 & 0 & 0 & 0 & 0 \\
0 & 0 & 0 & 0 & 1 & 0 & 1 & 0 \\
0 & 0 & 0 & 0 & 0 & 1 & 0 & 1
\end{bmatrix}
, \\ \\

(0, 0, 0, 1, 0, 0, 0, 1) & 
(0, 0, 0, 1, 1, 1, 0, 0) & 
(0, 0, 1, 1, 0, 0, 1, 1) & 
(0, 0, 0, 1, 0, 0, 1, 1) \\ \\

\begin{bmatrix}
1 & 0 & 0 & 0 & 0 & 0 & 0 & 0 \\
0 & 0 & 1 & 0 & 0 & 0 & 0 & 0 \\
0 & 0 & 0 & 0 & 1 & 0 & 0 & 0 \\
0 & 0 & 0 & 0 & 0 & 0 & 1 & 0
\end{bmatrix}
\\ \\

(0, 0, 0, 1, 0, 0, 0, 1)

\end{array}
\]

\begin{align*}

\begin{autobreak}
Z_{1,\mathrm{NS}} =
\chi_{00} \psi_{00} \overline{\chi_{00}} \overline{\psi_{00}}
+ \chi_{00} \psi_{00} \overline{\chi_{10}} \overline{\psi_{10}}
+ \chi_{00} \psi_{10} \overline{\chi_{00}} \overline{\psi_{10}}
+ \chi_{00} \psi_{10} \overline{\chi_{10}} \overline{\psi_{00}}
+ \chi_{01} \psi_{01} \overline{\chi_{01}} \overline{\psi_{01}}
+ \chi_{01} \psi_{01} \overline{\chi_{11}} \overline{\psi_{11}}
+ \chi_{01} \psi_{11} \overline{\chi_{01}} \overline{\psi_{11}}
+ \chi_{01} \psi_{11} \overline{\chi_{11}} \overline{\psi_{01}}
+ \chi_{10} \psi_{00} \overline{\chi_{00}} \overline{\psi_{10}}
+ \chi_{10} \psi_{00} \overline{\chi_{10}} \overline{\psi_{00}}
+ \chi_{10} \psi_{10} \overline{\chi_{00}} \overline{\psi_{00}}
+ \chi_{10} \psi_{10} \overline{\chi_{10}} \overline{\psi_{10}}
+ \chi_{11} \psi_{01} \overline{\chi_{01}} \overline{\psi_{11}}
+ \chi_{11} \psi_{01} \overline{\chi_{11}} \overline{\psi_{01}}
+ \chi_{11} \psi_{11} \overline{\chi_{01}} \overline{\psi_{01}}
+ \chi_{11} \psi_{11} \overline{\chi_{11}} \overline{\psi_{11}}
\end{autobreak} \\

\begin{autobreak}
Z_{2,\mathrm{NS}} =
\chi_{00} \psi_{00} \overline{\chi_{00}} \overline{\psi_{00}}
+ \chi_{00} \psi_{00} \overline{\chi_{10}} \overline{\psi_{10}}
+ \chi_{00} \psi_{10} \overline{\chi_{01}} \overline{\psi_{01}}
+ \chi_{00} \psi_{10} \overline{\chi_{11}} \overline{\psi_{11}}
+ \chi_{01} \psi_{01} \overline{\chi_{00}} \overline{\psi_{10}}
+ \chi_{01} \psi_{01} \overline{\chi_{10}} \overline{\psi_{00}}
+ \chi_{01} \psi_{11} \overline{\chi_{01}} \overline{\psi_{11}}
+ \chi_{01} \psi_{11} \overline{\chi_{11}} \overline{\psi_{01}}
+ \chi_{10} \psi_{00} \overline{\chi_{01}} \overline{\psi_{01}}
+ \chi_{10} \psi_{00} \overline{\chi_{11}} \overline{\psi_{11}}
+ \chi_{10} \psi_{10} \overline{\chi_{00}} \overline{\psi_{00}}
+ \chi_{10} \psi_{10} \overline{\chi_{10}} \overline{\psi_{10}}
+ \chi_{11} \psi_{01} \overline{\chi_{01}} \overline{\psi_{11}}
+ \chi_{11} \psi_{01} \overline{\chi_{11}} \overline{\psi_{01}}
+ \chi_{11} \psi_{11} \overline{\chi_{00}} \overline{\psi_{10}}
+ \chi_{11} \psi_{11} \overline{\chi_{10}} \overline{\psi_{00}}
\end{autobreak} \\

\begin{autobreak}
Z_{3,\mathrm{NS}} =
|(\chi_{00} \psi_{00}
+ \chi_{01} \psi_{01}
+ \chi_{10} \psi_{10}
+ \chi_{11} \psi_{11})|^2
\end{autobreak} \\

\begin{autobreak}
Z_{4,\mathrm{NS}} =
(\chi_{00}
+ \chi_{10}) (\psi_{00}
+ \psi_{10}) (\overline{\chi_{00}} \overline{\psi_{00}}
+ \overline{\chi_{01}} \overline{\psi_{01}}
+ \overline{\chi_{10}} \overline{\psi_{10}}
+ \overline{\chi_{11}} \overline{\psi_{11}})
\end{autobreak} \\

\begin{autobreak}
Z_{5,\mathrm{NS}} =
|(\chi_{00}
+ \chi_{10}) (\psi_{00}
+ \psi_{10})|^2
\end{autobreak}

\end{align*}

\paragraph{\underline{$\Spin(24)$}}

\[
\begin{array}{c} 
\begin{bmatrix}
1 & 0 & 0 & 0 \\
0 & 0 & 1 & 0
\end{bmatrix}
\\ \\
(0, 1, 0, 1)
\end{array}
\]

\begin{align*}
    Z_{\mathrm{NS}} &= |\chi_{00} + \chi_{10}|^2
\end{align*}

\paragraph{\underline{$\Spin(4m+2)^3$}}

\[
\begin{array}{cccc}

\begin{bmatrix}
1 & 0 & 1 & 0 & 1 & 1 \\
0 & 1 & 1 & 1 & 0 & 1 \\
0 & 0 & 2 & 0 & 0 & 2 \\
0 & 0 & 0 & 2 & 2 & 2
\end{bmatrix}
, &
\begin{bmatrix}
1 & 0 & 1 & 0 & 1 & 1 \\
0 & 2 & 0 & 0 & 0 & 0 \\
0 & 0 & 2 & 0 & 0 & 2 \\
0 & 0 & 0 & 2 & 0 & 0 \\
0 & 0 & 0 & 0 & 2 & 2
\end{bmatrix}
, &
\begin{bmatrix}
2 & 0 & 0 & 0 & 0 & 0 \\
0 & 2 & 0 & 0 & 0 & 0 \\
0 & 0 & 2 & 0 & 0 & 0 \\
0 & 0 & 0 & 2 & 0 & 0 \\
0 & 0 & 0 & 0 & 2 & 0 \\
0 & 0 & 0 & 0 & 0 & 2
\end{bmatrix}
\\ \\

(0, 0, 1, 0, 0, 1) &
(0, 1, 0, 1, 0, 0) &
(1, 1, 1, 1, 1, 1)

\end{array}
\]

\begin{align*}

\begin{autobreak}
Z_{1,\mathrm{NS}} =
\chi_{0}^{3} \overline{\chi_{0}}^{3}
+ \chi_{0}^{3} \overline{\chi_{2}}^{3}
+ 3 \chi_{0}^{2} \chi_{2} \overline{\chi_{0}}^{2} \overline{\chi_{2}}
+ 3 \chi_{0}^{2} \chi_{2} \overline{\chi_{0}} \overline{\chi_{2}}^{2}
+ 3 \chi_{0} \chi_{1}^{2} \overline{\chi_{0}} \overline{\chi_{1}}^{2}
+ 3 \chi_{0} \chi_{1}^{2} \overline{\chi_{2}} \overline{\chi_{3}}^{2}
+ 6 \chi_{0} \chi_{1} \chi_{3} \overline{\chi_{0}} \overline{\chi_{1}} \overline{\chi_{3}}
+ 6 \chi_{0} \chi_{1} \chi_{3} \overline{\chi_{1}} \overline{\chi_{2}} \overline{\chi_{3}}
+ 3 \chi_{0} \chi_{2}^{2} \overline{\chi_{0}}^{2} \overline{\chi_{2}}
+ 3 \chi_{0} \chi_{2}^{2} \overline{\chi_{0}} \overline{\chi_{2}}^{2}
+ 3 \chi_{0} \chi_{3}^{2} \overline{\chi_{0}} \overline{\chi_{3}}^{2}
+ 3 \chi_{0} \chi_{3}^{2} \overline{\chi_{1}}^{2} \overline{\chi_{2}}
+ 3 \chi_{1}^{2} \chi_{2} \overline{\chi_{0}} \overline{\chi_{3}}^{2}
+ 3 \chi_{1}^{2} \chi_{2} \overline{\chi_{1}}^{2} \overline{\chi_{2}}
+ 6 \chi_{1} \chi_{2} \chi_{3} \overline{\chi_{0}} \overline{\chi_{1}} \overline{\chi_{3}}
+ 6 \chi_{1} \chi_{2} \chi_{3} \overline{\chi_{1}} \overline{\chi_{2}} \overline{\chi_{3}}
+ \chi_{2}^{3} \overline{\chi_{0}}^{3}
+ \chi_{2}^{3} \overline{\chi_{2}}^{3}
+ 3 \chi_{2} \chi_{3}^{2} \overline{\chi_{0}} \overline{\chi_{1}}^{2}
+ 3 \chi_{2} \chi_{3}^{2} \overline{\chi_{2}} \overline{\chi_{3}}^{2}
\end{autobreak} \\

\begin{autobreak}
Z_{2,\mathrm{NS}} =
(\chi_{0}
+ \chi_{2}) (\overline{\chi_{0}}
+ \overline{\chi_{2}}) (\chi_{0}^{2} \overline{\chi_{0}}^{2}
+ \chi_{0}^{2} \overline{\chi_{2}}^{2}
+ 4 \chi_{0} \chi_{2} \overline{\chi_{0}} \overline{\chi_{2}}
+ \chi_{1}^{2} \overline{\chi_{1}}^{2}
+ \chi_{1}^{2} \overline{\chi_{3}}^{2}
+ 4 \chi_{1} \chi_{3} \overline{\chi_{1}} \overline{\chi_{3}}
+ \chi_{2}^{2} \overline{\chi_{0}}^{2}
+ \chi_{2}^{2} \overline{\chi_{2}}^{2}
+ \chi_{3}^{2} \overline{\chi_{1}}^{2}
+ \chi_{3}^{2} \overline{\chi_{3}}^{2})
\end{autobreak} \\

\begin{autobreak}
Z_{3,\mathrm{NS}} =
|(\chi_{0}
+ \chi_{2})^{3}|^2
\end{autobreak}

\end{align*}

\paragraph{\underline{$\SU(6)^2$}}

\[
\begin{array}{c}
\begin{bmatrix}
1 & 1 & 1 & 1 \\
0 & 2 & 3 & 5
\end{bmatrix}
\\ \\
(0, 1, 0, 1)
\end{array}
\]

\begin{align*}
    Z_{\mathrm{NS}} &= |\chi_0^2 + \chi_3^2|^2 + |\chi_1^2 + \chi_4^2|^2 + |\chi_2^2 + \chi_5^2|^2 \\
    &\quad + 2 |\chi_0\, \chi_2 + \chi_3\, \chi_5|^2 + 2 |\chi_0\, \chi_4 + \chi_3\, \chi_1|^2 + 2 |\chi_1\, \chi_5 + \chi_2\, \chi_4|^2
\end{align*}

\paragraph{\underline{$\SU(12)$}}

\[
\begin{array}{cc}
\begin{bmatrix}
2 & 2 \\
0 & 6
\end{bmatrix}
\\ \\
(1, 1)
\end{array}
\]

\begin{align*}
    Z_{\mathrm{NS}} &= |\chi_0 + \chi_6|^2 + |\chi_2 + \chi_8|^2 + |\chi_4 + \chi_{10}|^2
\end{align*}

\paragraph{\underline{$E_7^2$}}

\[
\begin{array}{c}
\begin{bmatrix}
1 & 1 & 0 & 0 \\
0 & 0 & 1 & 1
\end{bmatrix}
\\ \\
(0, 1, 0, 1)
\end{array}
\]

\begin{align*}
    Z_{\mathrm{NS}} &= |\chi_0^2 + \chi_1^2|^2
\end{align*}

\paragraph{\underline{$G=\Spin(8)^3,\, G'=\Spin(16) \times \Spin(8)$}}

\[
\begin{array}{cccc}

\begin{bmatrix}
1 & 0 & 0 & 0 & 0 & 1 & 0 & 0 & 0 & 0 \\
0 & 1 & 0 & 0 & 1 & 0 & 0 & 0 & 1 & 0 \\
0 & 0 & 1 & 0 & 0 & 1 & 0 & 1 & 0 & 1 \\
0 & 0 & 0 & 1 & 0 & 1 & 0 & 1 & 1 & 1 \\
0 & 0 & 0 & 0 & 0 & 0 & 1 & 0 & 1 & 0
\end{bmatrix}
, &
\begin{bmatrix}
1 & 0 & 0 & 0 & 0 & 1 & 0 & 0 & 0 & 0 \\
0 & 1 & 0 & 0 & 1 & 0 & 0 & 1 & 0 & 1 \\
0 & 0 & 1 & 0 & 0 & 1 & 0 & 0 & 1 & 0 \\
0 & 0 & 0 & 1 & 0 & 1 & 0 & 1 & 1 & 1 \\
0 & 0 & 0 & 0 & 0 & 0 & 1 & 0 & 1 & 0
\end{bmatrix}
, &
\begin{bmatrix}
1 & 0 & 0 & 0 & 0 & 1 & 0 & 0 & 1 & 0 \\
0 & 1 & 0 & 0 & 0 & 1 & 0 & 1 & 0 & 1 \\
0 & 0 & 1 & 1 & 0 & 0 & 0 & 0 & 0 & 0 \\
0 & 0 & 0 & 0 & 1 & 1 & 0 & 1 & 1 & 1 \\
0 & 0 & 0 & 0 & 0 & 0 & 1 & 0 & 1 & 0
\end{bmatrix}
, &
\begin{bmatrix}
1 & 0 & 0 & 0 & 0 & 1 & 0 & 1 & 0 & 1 \\
0 & 1 & 0 & 0 & 0 & 1 & 0 & 1 & 1 & 1 \\
0 & 0 & 1 & 1 & 0 & 0 & 0 & 0 & 0 & 0 \\
0 & 0 & 0 & 0 & 1 & 1 & 0 & 0 & 1 & 0 \\
0 & 0 & 0 & 0 & 0 & 0 & 1 & 0 & 1 & 0
\end{bmatrix}
, \\ \\

(0, 0, 0, 0, 0, 1, 1, 1, 0, 0) & 
(0, 0, 0, 0, 0, 1, 0, 0, 0, 1) & 
(0, 0, 0, 1, 0, 0, 0, 0, 1, 1) & 
(0, 0, 0, 1, 0, 0, 0, 1, 1, 0) \\ \\

\begin{bmatrix}
1 & 0 & 0 & 0 & 0 & 1 & 0 & 0 & 0 & 0 \\
0 & 1 & 0 & 1 & 1 & 1 & 0 & 0 & 0 & 0 \\
0 & 0 & 1 & 0 & 0 & 1 & 0 & 0 & 0 & 0 \\
0 & 0 & 0 & 0 & 0 & 0 & 1 & 0 & 1 & 0 \\
0 & 0 & 0 & 0 & 0 & 0 & 0 & 1 & 0 & 1
\end{bmatrix}
, &
\begin{bmatrix}
1 & 0 & 0 & 0 & 0 & 1 & 0 & 0 & 0 & 0 \\
0 & 1 & 0 & 1 & 1 & 1 & 0 & 0 & 0 & 0 \\
0 & 0 & 1 & 0 & 0 & 1 & 0 & 0 & 0 & 0 \\
0 & 0 & 0 & 0 & 0 & 0 & 1 & 0 & 0 & 0 \\
0 & 0 & 0 & 0 & 0 & 0 & 0 & 0 & 1 & 0
\end{bmatrix}
, &
\begin{bmatrix}
1 & 0 & 0 & 0 & 0 & 1 & 0 & 0 & 0 & 0 \\
0 & 1 & 0 & 0 & 1 & 0 & 0 & 0 & 0 & 0 \\
0 & 0 & 1 & 1 & 0 & 0 & 0 & 0 & 0 & 0 \\
0 & 0 & 0 & 0 & 0 & 0 & 1 & 0 & 1 & 0 \\
0 & 0 & 0 & 0 & 0 & 0 & 0 & 1 & 0 & 1
\end{bmatrix}
, &
\begin{bmatrix}
1 & 0 & 0 & 0 & 0 & 1 & 0 & 0 & 0 & 0 \\
0 & 1 & 0 & 0 & 1 & 0 & 0 & 0 & 0 & 0 \\
0 & 0 & 1 & 1 & 0 & 0 & 0 & 0 & 0 & 0 \\
0 & 0 & 0 & 0 & 0 & 0 & 1 & 0 & 0 & 0 \\
0 & 0 & 0 & 0 & 0 & 0 & 0 & 0 & 1 & 0
\end{bmatrix}
, \\ \\

(0, 0, 0, 0, 0, 1, 0, 0, 1, 1) & 
(0, 0, 0, 0, 0, 1, 0, 0, 0, 1) & 
(0, 0, 0, 1, 0, 0, 0, 0, 1, 1) & 
(0, 0, 0, 1, 0, 0, 0, 0, 0, 1) \\ \\

\begin{bmatrix}
1 & 1 & 0 & 0 & 0 & 0 & 0 & 0 & 0 & 0 \\
0 & 0 & 1 & 1 & 0 & 0 & 0 & 0 & 0 & 0 \\
0 & 0 & 0 & 0 & 1 & 1 & 0 & 0 & 0 & 0 \\
0 & 0 & 0 & 0 & 0 & 0 & 1 & 0 & 1 & 0 \\
0 & 0 & 0 & 0 & 0 & 0 & 0 & 1 & 0 & 1
\end{bmatrix}
, &
\begin{bmatrix}
1 & 1 & 0 & 0 & 0 & 0 & 0 & 0 & 0 & 0 \\
0 & 0 & 1 & 1 & 0 & 0 & 0 & 0 & 0 & 0 \\
0 & 0 & 0 & 0 & 1 & 1 & 0 & 0 & 0 & 0 \\
0 & 0 & 0 & 0 & 0 & 0 & 1 & 0 & 0 & 0 \\
0 & 0 & 0 & 0 & 0 & 0 & 0 & 0 & 1 & 0
\end{bmatrix}
\\ \\

(0, 1, 0, 1, 0, 1, 0, 0, 1, 1) & 
(0, 1, 0, 1, 0, 1, 0, 0, 0, 1)

\end{array}
\]

\begin{align*}

\begin{autobreak}
Z_{1,\mathrm{NS}} =
\psi_{00}^{3} \overline{\chi_{00}} \overline{\psi_{00}}
+ \psi_{00}^{3} \overline{\chi_{10}} \overline{\psi_{10}}
+ \psi_{00}^{2} \psi_{11} \overline{\chi_{00}} \overline{\psi_{10}}
+ \psi_{00}^{2} \psi_{11} \overline{\chi_{10}} \overline{\psi_{00}}
+ \psi_{00} \psi_{01}^{2} \overline{\chi_{01}} \overline{\psi_{11}}
+ \psi_{00} \psi_{01}^{2} \overline{\chi_{11}} \overline{\psi_{01}}
+ \psi_{00} \psi_{01} \psi_{10} \overline{\chi_{00}} \overline{\psi_{00}}
+ \psi_{00} \psi_{01} \psi_{10} \overline{\chi_{00}} \overline{\psi_{10}}
+ \psi_{00} \psi_{01} \psi_{10} \overline{\chi_{01}} \overline{\psi_{01}}
+ \psi_{00} \psi_{01} \psi_{10} \overline{\chi_{01}} \overline{\psi_{11}}
+ \psi_{00} \psi_{01} \psi_{10} \overline{\chi_{10}} \overline{\psi_{00}}
+ \psi_{00} \psi_{01} \psi_{10} \overline{\chi_{10}} \overline{\psi_{10}}
+ \psi_{00} \psi_{01} \psi_{10} \overline{\chi_{11}} \overline{\psi_{01}}
+ \psi_{00} \psi_{01} \psi_{10} \overline{\chi_{11}} \overline{\psi_{11}}
+ \psi_{00} \psi_{10}^{2} \overline{\chi_{01}} \overline{\psi_{01}}
+ \psi_{00} \psi_{10}^{2} \overline{\chi_{11}} \overline{\psi_{11}}
+ \psi_{00} \psi_{11}^{2} \overline{\chi_{00}} \overline{\psi_{10}}
+ \psi_{00} \psi_{11}^{2} \overline{\chi_{10}} \overline{\psi_{00}}
+ \psi_{01}^{2} \psi_{11} \overline{\chi_{01}} \overline{\psi_{01}}
+ \psi_{01}^{2} \psi_{11} \overline{\chi_{11}} \overline{\psi_{11}}
+ \psi_{01} \psi_{10} \psi_{11} \overline{\chi_{00}} \overline{\psi_{00}}
+ \psi_{01} \psi_{10} \psi_{11} \overline{\chi_{00}} \overline{\psi_{10}}
+ \psi_{01} \psi_{10} \psi_{11} \overline{\chi_{01}} \overline{\psi_{01}}
+ \psi_{01} \psi_{10} \psi_{11} \overline{\chi_{01}} \overline{\psi_{11}}
+ \psi_{01} \psi_{10} \psi_{11} \overline{\chi_{10}} \overline{\psi_{00}}
+ \psi_{01} \psi_{10} \psi_{11} \overline{\chi_{10}} \overline{\psi_{10}}
+ \psi_{01} \psi_{10} \psi_{11} \overline{\chi_{11}} \overline{\psi_{01}}
+ \psi_{01} \psi_{10} \psi_{11} \overline{\chi_{11}} \overline{\psi_{11}}
+ \psi_{10}^{2} \psi_{11} \overline{\chi_{01}} \overline{\psi_{11}}
+ \psi_{10}^{2} \psi_{11} \overline{\chi_{11}} \overline{\psi_{01}}
+ \psi_{11}^{3} \overline{\chi_{00}} \overline{\psi_{00}}
+ \psi_{11}^{3} \overline{\chi_{10}} \overline{\psi_{10}}
\end{autobreak} \\

\begin{autobreak}
Z_{2,\mathrm{NS}} =
\psi_{00}^{3} \overline{\chi_{00}} \overline{\psi_{00}}
+ \psi_{00}^{3} \overline{\chi_{10}} \overline{\psi_{10}}
+ \psi_{00}^{2} \psi_{11} \overline{\chi_{01}} \overline{\psi_{01}}
+ \psi_{00}^{2} \psi_{11} \overline{\chi_{11}} \overline{\psi_{11}}
+ \psi_{00} \psi_{01}^{2} \overline{\chi_{01}} \overline{\psi_{11}}
+ \psi_{00} \psi_{01}^{2} \overline{\chi_{11}} \overline{\psi_{01}}
+ \psi_{00} \psi_{01} \psi_{10} \overline{\chi_{00}} \overline{\psi_{00}}
+ \psi_{00} \psi_{01} \psi_{10} \overline{\chi_{00}} \overline{\psi_{10}}
+ \psi_{00} \psi_{01} \psi_{10} \overline{\chi_{01}} \overline{\psi_{01}}
+ \psi_{00} \psi_{01} \psi_{10} \overline{\chi_{01}} \overline{\psi_{11}}
+ \psi_{00} \psi_{01} \psi_{10} \overline{\chi_{10}} \overline{\psi_{00}}
+ \psi_{00} \psi_{01} \psi_{10} \overline{\chi_{10}} \overline{\psi_{10}}
+ \psi_{00} \psi_{01} \psi_{10} \overline{\chi_{11}} \overline{\psi_{01}}
+ \psi_{00} \psi_{01} \psi_{10} \overline{\chi_{11}} \overline{\psi_{11}}
+ \psi_{00} \psi_{10}^{2} \overline{\chi_{00}} \overline{\psi_{10}}
+ \psi_{00} \psi_{10}^{2} \overline{\chi_{10}} \overline{\psi_{00}}
+ \psi_{00} \psi_{11}^{2} \overline{\chi_{01}} \overline{\psi_{01}}
+ \psi_{00} \psi_{11}^{2} \overline{\chi_{11}} \overline{\psi_{11}}
+ \psi_{01}^{2} \psi_{11} \overline{\chi_{00}} \overline{\psi_{10}}
+ \psi_{01}^{2} \psi_{11} \overline{\chi_{10}} \overline{\psi_{00}}
+ \psi_{01} \psi_{10} \psi_{11} \overline{\chi_{00}} \overline{\psi_{00}}
+ \psi_{01} \psi_{10} \psi_{11} \overline{\chi_{00}} \overline{\psi_{10}}
+ \psi_{01} \psi_{10} \psi_{11} \overline{\chi_{01}} \overline{\psi_{01}}
+ \psi_{01} \psi_{10} \psi_{11} \overline{\chi_{01}} \overline{\psi_{11}}
+ \psi_{01} \psi_{10} \psi_{11} \overline{\chi_{10}} \overline{\psi_{00}}
+ \psi_{01} \psi_{10} \psi_{11} \overline{\chi_{10}} \overline{\psi_{10}}
+ \psi_{01} \psi_{10} \psi_{11} \overline{\chi_{11}} \overline{\psi_{01}}
+ \psi_{01} \psi_{10} \psi_{11} \overline{\chi_{11}} \overline{\psi_{11}}
+ \psi_{10}^{2} \psi_{11} \overline{\chi_{01}} \overline{\psi_{11}}
+ \psi_{10}^{2} \psi_{11} \overline{\chi_{11}} \overline{\psi_{01}}
+ \psi_{11}^{3} \overline{\chi_{00}} \overline{\psi_{00}}
+ \psi_{11}^{3} \overline{\chi_{10}} \overline{\psi_{10}}
\end{autobreak} \\

\begin{autobreak}
Z_{3,\mathrm{NS}} =
(\psi_{00}
+ \psi_{11}) (\psi_{00}^{2} \overline{\chi_{00}} \overline{\psi_{00}}
+ \psi_{00}^{2} \overline{\chi_{10}} \overline{\psi_{10}}
+ 2 \psi_{00} \psi_{11} \overline{\chi_{01}} \overline{\psi_{11}}
+ 2 \psi_{00} \psi_{11} \overline{\chi_{11}} \overline{\psi_{01}}
+ \psi_{01}^{2} \overline{\chi_{01}} \overline{\psi_{01}}
+ \psi_{01}^{2} \overline{\chi_{11}} \overline{\psi_{11}}
+ 2 \psi_{01} \psi_{10} \overline{\chi_{00}} \overline{\psi_{10}}
+ 2 \psi_{01} \psi_{10} \overline{\chi_{10}} \overline{\psi_{00}}
+ \psi_{10}^{2} \overline{\chi_{01}} \overline{\psi_{01}}
+ \psi_{10}^{2} \overline{\chi_{11}} \overline{\psi_{11}}
+ \psi_{11}^{2} \overline{\chi_{00}} \overline{\psi_{00}}
+ \psi_{11}^{2} \overline{\chi_{10}} \overline{\psi_{10}})
\end{autobreak} \\

\begin{autobreak}
Z_{4,\mathrm{NS}} =
(\psi_{00}
+ \psi_{11}) (\psi_{00}^{2} \overline{\chi_{00}} \overline{\psi_{00}}
+ \psi_{00}^{2} \overline{\chi_{10}} \overline{\psi_{10}}
+ 2 \psi_{00} \psi_{11} \overline{\chi_{00}} \overline{\psi_{10}}
+ 2 \psi_{00} \psi_{11} \overline{\chi_{10}} \overline{\psi_{00}}
+ \psi_{01}^{2} \overline{\chi_{01}} \overline{\psi_{11}}
+ \psi_{01}^{2} \overline{\chi_{11}} \overline{\psi_{01}}
+ 2 \psi_{01} \psi_{10} \overline{\chi_{01}} \overline{\psi_{01}}
+ 2 \psi_{01} \psi_{10} \overline{\chi_{11}} \overline{\psi_{11}}
+ \psi_{10}^{2} \overline{\chi_{01}} \overline{\psi_{11}}
+ \psi_{10}^{2} \overline{\chi_{11}} \overline{\psi_{01}}
+ \psi_{11}^{2} \overline{\chi_{00}} \overline{\psi_{00}}
+ \psi_{11}^{2} \overline{\chi_{10}} \overline{\psi_{10}})
\end{autobreak} \\

\begin{autobreak}
Z_{5,\mathrm{NS}} =
(\psi_{00}^{3}
+ 2 \psi_{00} \psi_{01} \psi_{10}
+ \psi_{00} \psi_{10}^{2}
+ \psi_{01}^{2} \psi_{11}
+ 2 \psi_{01} \psi_{10} \psi_{11}
+ \psi_{11}^{3}) (\overline{\chi_{00}} \overline{\psi_{00}}
+ \overline{\chi_{01}} \overline{\psi_{01}}
+ \overline{\chi_{10}} \overline{\psi_{10}}
+ \overline{\chi_{11}} \overline{\psi_{11}})
\end{autobreak} \\

\begin{autobreak}
Z_{6,\mathrm{NS}} =
(\psi_{00}^{3}
+ 2 \psi_{00} \psi_{01} \psi_{10}
+ \psi_{00} \psi_{10}^{2}
+ \psi_{01}^{2} \psi_{11}
+ 2 \psi_{01} \psi_{10} \psi_{11}
+ \psi_{11}^{3}) (\overline{\chi_{00}}
+ \overline{\chi_{10}}) (\overline{\psi_{00}}
+ \overline{\psi_{10}})
\end{autobreak} \\

\begin{autobreak}
Z_{7,\mathrm{NS}} =
(\psi_{00}
+ \psi_{11}) (\psi_{00}^{2}
+ 2 \psi_{01} \psi_{10}
+ \psi_{11}^{2}) (\overline{\chi_{00}} \overline{\psi_{00}}
+ \overline{\chi_{01}} \overline{\psi_{01}}
+ \overline{\chi_{10}} \overline{\psi_{10}}
+ \overline{\chi_{11}} \overline{\psi_{11}})
\end{autobreak} \\

\begin{autobreak}
Z_{8,\mathrm{NS}} =
(\psi_{00}
+ \psi_{11}) (\psi_{00}^{2}
+ 2 \psi_{01} \psi_{10}
+ \psi_{11}^{2}) (\overline{\chi_{00}}
+ \overline{\chi_{10}}) (\overline{\psi_{00}}
+ \overline{\psi_{10}})
\end{autobreak} \\

\begin{autobreak}
Z_{9,\mathrm{NS}} =
(\psi_{00}
+ \psi_{11})^{3} (\overline{\chi_{00}} \overline{\psi_{00}}
+ \overline{\chi_{01}} \overline{\psi_{01}}
+ \overline{\chi_{10}} \overline{\psi_{10}}
+ \overline{\chi_{11}} \overline{\psi_{11}})
\end{autobreak} \\

\begin{autobreak}
Z_{10,\mathrm{NS}} =
(\psi_{00}
+ \psi_{11})^{3} (\overline{\chi_{00}}
+ \overline{\chi_{10}}) (\overline{\psi_{00}}
+ \overline{\psi_{10}})
\end{autobreak}

\end{align*}
\newpage 
\bibliographystyle{JHEP}
\bibliography{refs}
\end{document}